\documentclass[useAMS, usenatbib, usegraphicx]{mn2e}
\title[Debris from giant impacts at large orbital radii]{Debris from giant impacts between planetary embryos at large orbital radii}
\author[A.P. Jackson et al.]{Alan P. Jackson$^{1}$\thanks{E-mail: ajackson@ast.cam.ac.uk}, Mark C. Wyatt$^1$, Amy Bonsor$^{2,3}$, and Dimitri Veras$^{1,4}$\\%
$^1$Institute of Astronomy, University of Cambridge, Madingley Road, Cambridge, CB3 0HA, UK\\%
$^2$Institut de Plan\'etologie et d'Astrophysique de Grenoble, 414 Rue de la Piscine, Domaine Universitaire, 38400 St-Martin d'H\`eres, France\\%
$^3$School of Physics, H. H. Wills Physics Laboratory, University of Bristol, Tyndall Avenue, Bristol, BS8 1TL, UK\\%
$^4$Department of Physics, University of Warwick, Gibbet Hill Road, Coventry CV4 7AL}
\date{Submitted 2013}
\pagerange{\pageref{firstpage}--\pageref{lastpage}}
\pubyear{2013}
\usepackage[total={17.8cm,24.0cm},centering]{geometry}
\usepackage{times}
\usepackage[fleqn]{amsmath}
\usepackage[none]{hyphenat}
\usepackage[bookmarks=true, bookmarksopen=true]{hyperref}
\usepackage{color}
\usepackage[mnras, extd]{journalnames}
\usepackage{lscape}
\begin{document}
\renewcommand{\theenumi}{\arabic{enumi}.}

\label{firstpage}
\maketitle
\begin{abstract}
We consider the observational signatures of giant impacts between planetary embryos.  While the debris released in the impact remains in a clump for only a single orbit, there is a much longer lasting asymmetry caused by the fact that all debris must pass through the collision-point.  The resulting asymmetry is stationary, it does not orbit the star.  The debris is concentrated in a clump at the collision-point, with a more diffuse structure on the opposite side.  The asymmetry lasts for typically around 1000 orbital periods of the progenitor, which can be several Myr at distances of $\sim$50~AU.  We describe how the appearance of the asymmetric disc depends on the mass and eccentricity of the progenitor, as well as viewing orientation.  The wavelength of observation, which determines the grain sizes probed, is also important.  Notably, the increased collision rate of the debris at the collision-point makes this the dominant production site for any secondary dust and gas created.  For dust small enough to be removed by radiation pressure, and gas with a short lifetime, this causes their distribution to resemble a jet emanating from the (stationary) collision-point.  We suggest that the asymmetries seen at large separations in some debris discs, like Beta Pictoris, could be the result of giant impacts.  If so this would indicate that planetary embryos are present and continuing to grow at several tens of AU at ages of up to tens of Myr.

\end{abstract}
\begin{keywords}
celestial mechanics -- circumstellar matter -- planet-disc interactions -- planets and satellites: formation
\end{keywords}

\sloppy

\section{Introduction}
\label{intro}
The final stage of terrestrial planet formation is now widely believed to be one of chaotic growth, with the terrestrial planets built up through series of planetary scale impacts (commonly known as giant impacts) between planetary embryos \citep[e.g.][]{kenyon2006, raymond2009, kokubo2010}.  While giant impacts are perhaps most frequently discussed in relation to terrestrial planet formation they are certainly not limited to the terrestrial zone of a planetary system.  In our own solar system the Pluto-Charon system \citep{canup2005, canup2011, stern2006}, and Haumea and its collisional family \citep{brown2007}, are both proposed to have their origin in giant impacts.  It has also long been suggested that a large impact could explain Uranus' large obliquity \citep[e.g.][]{benz1989, slattery1992}.

Outside our own solar system there are numerous examples of planets and debris discs that can be found at substantial distances from their parent star.  Examples are the HR8799 system with 4 massive gas giant planets, the outermost at 68AU and a debris disc at 90-300AU (e.g. \citealt{marois2008, soummer2011}; Matthews et al. in press), or Fomalhaut with a massive debris disc at 140 AU \citep[e.g.][]{stapelfeldt2004, boley2012}.  Debris discs are belts of planetesimals and dust produced by destructive collisions amongst the larger planetesimals \citep[e.g.][]{wyatt2007, wyatt2007b, wyatt2008}.

The outer regions of planetary systems thus clearly provide an environment in which giant impacts can occur.  Indeed some extrasolar systems possess much more material at large distances than our solar system, such that more, and larger, impacts can be expected than in the solar system.

An essential property of giant impacts, wherever they occur, is that they produce substantial quantities of debris.  Giant impacts span a large range of collision scenarios and outcomes from catastrophic disruption to fairly efficient accretion dependent on impact velocity and geometry \citep[e.g.][]{leinhardt2012}.  As a result the quantity of debris produced can vary greatly, but even impacts that are apparently efficient accretion events release $\ga$1 per cent of the mass of the colliding bodies in debris.  Indeed it is such debris that is liberated from the progenitor but remains bound to it that is key to the models for the formation of our Moon \citep[e.g.][]{canup2004b} and Pluto-Charon \citep[e.g.][]{canup2005}, while models for the formation of Mercury require large quantities of unbound debris \citep{anic2006, benz2007}.

Debris produced by giant impacts, its properties, evolution and detectability, remains an understudied topic in comparison to the giant impacts that produce it, and those studies that have been done have generally focussed on the terrestrial planet region.  Previous work that has been conducted in this area include \citet{kenyon2004}, who investigated debris produced during terrestrial planet formation in a statistical way, and \citet{jackson2012} who studied the evolution of debris produced by the Moon-forming impact.  Impacts occurring at large orbital distances emphasise different aspects of the debris evolution.  In particular since orbital periods are longer and velocities are lower at larger distances the evolution of a debris disc is substantially slower, which means that features that are short-lived and thus unlikely to be seen in the terrestrial zone can be much longer lived and become important characteristics in the outer regions.  Furthermore since a debris disc at a large orbital distance is, by definition, much larger in spatial extent, and also less likely to be hidden by the star, there is a much greater possibility of obtaining spatially resolved images to observe structure within the disc.

Of particular interest are disc asymmetries.  There are a growing number of young systems with resolved debris discs that display asymmetries and other as yet poorly understood features, such as HD15115 \citep[e.g.][]{kalas2007,rodigas2012} and HD32297 \citep[e.g.][]{kalas2005b, schneider2005, currie2012}, with the 12 million year old Beta Pictoris system \citep[e.g.][]{telesco2005, li2012} probably the best known example.  If we believe that giant impacts are indeed common, particularly during the epoch of planet formation, then we might reasonably expect that some of these systems may have experienced a giant impact in the comparatively recent past.  If this is the case a question that we should be asking is; can a giant impact explain some of the features of these discs?  If giant impacts can explain some of these features, thus providing evidence for massive bodies at large orbital distances, this also provides us with important information about the process of planet formation.

In this work we discuss the morphologies (\S~\ref{sec:obsdisc}) and detectability (\S\S~\ref{sec:detectability} and \S~\ref{sec:collevol}) of debris discs produced by giant impacts and how these vary with parameters such as the mass of the progenitor body.  We also discuss the morphology of small dust grains influenced by radiation pressure (\S\S~\ref{sec:blowout}) and CO (\S\S~\ref{sec:CO}), created in the destruction of the collisional debris.  Finally we apply our models of giant impact debris to the debris disc of Beta Pictoris (\S~\ref{sec:betapic}).  First however we describe the analytics that underpin the determination of the orbits of the debris, and thus the shape, and features of the disc, in Section~\ref{orbeq}.

\section{Orbit equations}
\label{orbeq}

Consider a body that undergoes a collision by some impacting projectile.  Whether the collision is catastrophic, totally disrupting the target, or cratering, excavating a small region of the surface, if we want to describe the motion of any debris that escapes from the target our starting point is the orbit of the progenitor body.  The orbit of the escaping debris is clearly not going to be exactly that of the progenitor however, since it has a velocity relative to the progenitor.  We can however use this `kick' in velocity relative to the progenitor to relate the orbits of the escaping debris to that of the progenitor.

\citet{jackson2012} briefly outlined this concept of relating the orbits of the debris produced by an impact to that of the progenitor via a velocity kick for the single case of the Moon-forming impact, with a progenitor on a circular orbit.  Since both Pluto-Charon and Haumea, the two Kuiper belt bodies believed to have suffered giant impacts, are presently in eccentric orbits (though it is unknown whether Pluto was eccentric at the time of impact), and there is no reason to believe similar bodies in extrasolar systems would be circular either, an extension to eccentric progenitor orbits is desired.  The concept of a velocity kick relating two orbits is also applicable to many other problems, the scattering of small bodies by a planet for example.  As such here we extend these equations to include eccentric and inclined progenitors and present them in a general manner, before focussing our discussion on the debris produced by a giant impact.

In this Section we first present general equations relating the pre- and post-kick orbital elements (\S\S~\ref{orbeqder}, see also Appendix~\ref{geneq}).  We then consider the simpler case of initially circular orbits in Section~\ref{circparent} to aid understanding and intuition for the behaviour of the post-kick orbital elements, before returning to eccentric orbits in Section~\ref{eccorb}.  For interest, we also include a comparison to the Gauss planetary equations in Appendix~\ref{gausscomp}, with a brief discussion of where the planetary equations differ from the kick formalism.

\subsection{General orbits}
\label{orbeqder}

Throughout this work we will use a standard notation for the orbital elements in which $a$=semi-major axis, $e$=eccentricity, $r$=distance from central body, $I$=inclination, $\Omega$=longitude of ascending node, $\omega$=argument of pericentre, $f$=true anomaly.  We also make use of the ancillary variables $q$=pericentre distance, $\boldsymbol{v}$=particle velocity, and $v_{\rm k}$=circular speed at orbital distance $a$, $M$=mass of central body, $m$=mass of particle.  Primes are used to indicate the post-kick elements as opposed to the unprimed pre-kick elements.

The orientation of a general three-dimensional orbit is described with respect to a fixed reference plane by the inclination, longitude of ascending node and argument of pericentre.  Without loss of generality however we may define the $x-y$ plane to be the plane of the pre-kick orbit and the pre-kick argument of pericentre to be zero (i.e. the pericentre of the pre-kick orbit lies on the x-axis), such that $I=\Omega=\omega=0$.  The $z$-axis is then defined to point in the direction of the initial orbital angular momentum with the pre-kick particle orbiting in an anti-clockwise sense.

The standard relation between the semi-major axis of the particle and its velocity; 
\begin{equation}
\label{av}
\frac{1}{a}=\frac{2}{r}-\frac{v^2}{G(M+m)},
\end{equation}
will apply both before and after the kick with the addition of primes to $a$ and $v$ for the post-impact case, since the position of the particle, and thus $r$, has not changed.

\begin{figure}
\includegraphics[width=\columnwidth]{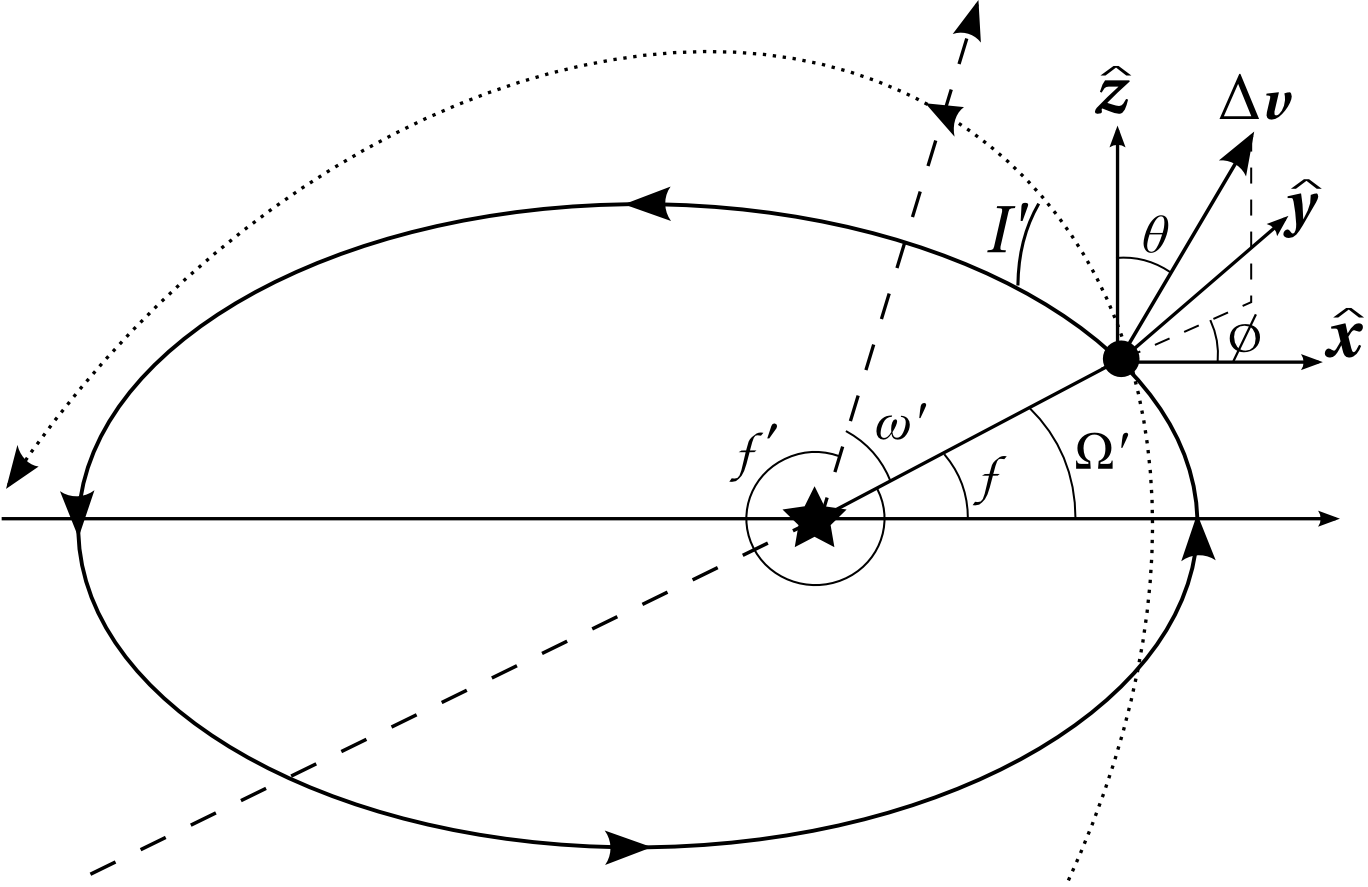}
\caption{Diagrammatic example of the effect of a single velocity kick $\boldsymbol{\Delta v}$ on the orbit of a particle.  The original orbit is shown as a solid black ellipse, while the new orbit after application of the velocity kick is the dotted ellipse.  The new orbit is constrained to pass through the kick-point, which sets the simple relations for $\Omega'$ and $\omega' + f'$.  Also shown are the orientation of the Cartesian axes and the definitions of $\theta$ and $\phi$.}
\label{orbdiag}
\end{figure}

The new velocity $\boldsymbol{v'}$ is simply the vector sum of the old velocity $\boldsymbol{v}$ and the kick velocity $\boldsymbol{\Delta v}$ and thus $v'^2=v^2+\Delta v^2+2\boldsymbol{v}\cdot\boldsymbol{\Delta v}$.  So we can combine the relations for $a$ and $a'$ from Eq.~\ref{av} to relate the semi-major axis before and after the kick in terms of the kick velocity by
\begin{equation}
\label{aaprimdeltav}
\frac{1}{a'}=\frac{1}{a}-\frac{\Delta v^2 + 2 \boldsymbol{v}\cdot\boldsymbol{\Delta v}}{G(M+m)}.
\end{equation}

To proceed it is convenient to represent $\boldsymbol{v}$ and $\boldsymbol{\Delta v}$ in a Cartesian coordinate system.  While $\boldsymbol{\Delta v}$ is best represented in a spherical polar coordinate system, there is no obvious preferred orientation for the associated Cartesian basis, and indeed in many cases the distribution will be spherically symmetric.  Thus we choose to use the same Cartesian system as that defined above based on the pre-kick orbit for simplicity, with $\theta$ as the angle between $\boldsymbol{\Delta v}$ and the $z$-axis and $\phi$ as the angle between the projection of $\boldsymbol{\Delta v}$ into the $x-y$ plane and the $x$-axis.  This definition of $\theta$ and $\phi$ is shown in Fig.~\ref{orbdiag}.  Defining the orientation of the kick solely in terms of the initial orbit of the particle also aids in allowing the kick formalism to be as general as possible, since if the problem to which the formalism is to be applied introduces a preferred direction it is a relatively simple matter to determine the appropriate values of $\theta$ and $\phi$ through rotations.

In this Cartesian coordinate system $\boldsymbol{v}=Q(-S_f, (e+C_f), 0)$ and $\boldsymbol{\Delta v}=\Delta v(S_{\theta}C_{\phi}, S_{\theta}S_{\phi}, C_{\theta})$.  Here $Q=v_{\rm k}/(1-e^2)^{\frac{1}{2}}$, and we introduce the common shorthand notation $S_x$ and $C_x$ for $\sin(x)$ and $\cos(x)$ respectively, which we will use throughout.

With these definitions
\begin{equation}
\label{v.deltav}
\boldsymbol{v}\cdot\boldsymbol{\Delta v}=Q \Delta v S_{\theta}[(e + C_f) S_{\phi} - S_f C_{\phi}].
\end{equation}
Using this with Eq.~\ref{aaprimdeltav} we obtain, after some condensing of trigonometric functions and de-dimensionalising using $v_{\rm k}^2=G(M+m)/a$;
\begin{equation}
\label{aaprimecc}
\frac{a}{a'}=1 - \left(\frac{\Delta v}{v_{\rm k}}\right)^2 - \frac{2}{\sqrt{1-e^2}}\left(\frac{\Delta v}{v_{\rm k}}\right) S_{\theta}(S_{(\phi-f)} + e S_{\phi}).
\end{equation}

In a similar manner the eccentricity relation,
\begin{equation}
\label{eccha}
e^2=1-\frac{h^2}{G(M+m)a},
\end{equation}
will hold both before and after impact, adding primes to $e$, $h$ and $a$ for the post-kick case, where $\boldsymbol{h}=\boldsymbol{R}\wedge\boldsymbol{v}=av_{\rm k}(1-e^2)^{\frac{1}{2}}\boldsymbol{\hat{z}}$ and $\boldsymbol{h'}=\boldsymbol{R}\wedge\boldsymbol{v'}$ with $\boldsymbol{R}=\frac{a(1-e^2)}{1+eC_f}(C_f, S_f, 0)$, the position vector of the particle at the time of the kick.  Combining pre- and post-kick eccentricity equations we obtain:
\begin{equation}
\label{ecceprim}
e'^2=1-(1-e^2)\left(\frac{h'^2}{h^2}\frac{a}{a'}\right).
\end{equation}

Expanding $\boldsymbol{h'}$ in terms of $\boldsymbol{h}$ and a `kick' term $\boldsymbol{h'}=\boldsymbol{h}+\boldsymbol{R}\wedge\boldsymbol{\Delta v}$, and from the form of $\boldsymbol{\Delta v}$ above, 
\begin{equation}
\label{Rcrossdeltav}
\boldsymbol{R}\wedge\boldsymbol{\Delta v}=a\frac{1-e^2}{1+e C_f} \Delta v \begin{pmatrix}S_f C_{\theta} \\
                                                                                        -C_f C_{\theta} \\
                                                                                         S_{\theta} S_{(\phi-f)}\end{pmatrix},
\end{equation}
and thus
\begin{equation}
\label{ecchprimh}
\begin{split}
\frac{h'^2}{h^2} =1 &+ 2 \frac{(1-e^2)^\frac{1}{2}}{1+e C_f} \left(\frac{\Delta v}{v_{\rm k}}\right) S_{\theta} S_{(\phi-f)} \\
& +\frac{1-e^2}{(1+e C_f)^2} \left(\frac{\Delta v}{v_{\rm k}}\right)^2 \left(C_{\theta}^2 + S_{\theta}^2 S_{(\phi-f)}^2\right).
\end{split}
\end{equation}

As in \citet{murray1999} $I'$ will be given by $\cos(I')=h'_z/h'$, which may be written as
\begin{equation}
\label{eccIprimh}
\cos I'=\left(\frac{h'_z}{h}\right)\left(\frac{h'^2}{h^2}\right)^{-\frac{1}{2}},
\end{equation}
to utilise the simple form of $h$ with $h'^2/h^2$ as already determined by Eq.(\ref{ecchprimh})
\begin{equation}
\label{eccIprim}
\cos I'=\left[1+\frac{(1-e^2)^{\frac{1}{2}}}{1+e C_f} \left(\frac{\Delta v}{v_{\rm k}}\right)S_{\theta}S_{(\phi-f)}\right]
              \left(\frac{h'^2}{h^2}\right)^{-\frac{1}{2}}.
\end{equation}

For $\Omega'$ we utilise the geometry of the problem by noting that the particle must pass through the point at which the kick is applied and that since this point also lies on the pre-kick orbit it must lie in the x-y plane.  The line through the central star and the point of application of the kick (the kick-point) thus marks the line of nodes with the nature of the kick-point as ascending or descending node determined by the sign of $\Delta v_z$ such that
\begin{equation}
\label{Omegaprim}
\Omega'=
\begin{cases}
	f	& \text{for } \theta\leq\frac{\pi}{2},\\
	f+\pi	& \text{for } \theta>\frac{\pi}{2}.
\end{cases}
\end{equation}
Note that when $\theta = \frac{\pi}{2}$, $\Omega'$ is strictly undefined since the particle orbit will remain confined to the x-y plane.  In this case $\Omega'$ may to some extent be set arbitrarily so we choose to extend the $\theta < \frac{\pi}{2}$ case to maintain consistency with our other definitions (such as Eq.~\ref{eccomegafprim} below).

In like manner from the geometry of the problem we obtain (modulo $2\pi$),
\begin{equation}
\label{eccomegafprim}
\omega'+f'=
\begin{cases}
	0	& \text{for } \theta\leq\frac{\pi}{2},\\
	\pi	& \text{for } \theta>\frac{\pi}{2}.
\end{cases}
\end{equation}

We may then use the relations
\begin{align}
\sin f' & =\frac{a'(1-e'^2)}{h'e'}\frac{\boldsymbol{R}\cdot\boldsymbol{v'}}{r},\\ \notag
\cos f' & =\frac{1}{e'}\left( \frac{a'(1-e'^2)}{r}-1 \right),
\end{align}
from \citet{murray1999} to find the values of $\omega'$ and $f'$.
Using the form of $\boldsymbol{v'}$ and $r$ and Eq.~\ref{ecceprim} these may be written as:
\begin{align}
\label{eccsinfcosf}
\sin f' & =\frac{1}{e'}\left(\frac{h'^2}{h^2}\right)^{\frac{1}{2}}
                \left( eS_f + (1-e^2)^{\frac{1}{2}} \left(\frac{\Delta v}{v_{\rm k}}\right) S_{\theta}C_{(\phi-f)}\right),\\ \notag
\cos f' & =\frac{1}{e'}\left( \frac{h'^2}{h^2}(1+eC_f)-1\right).
\end{align}

The only assumption made in the derivation of these equations is that the change in velocity may be treated as impulsive.  In physical terms this requires that the time over which the change in velocity takes place, $\Delta t$, is small in comparison to the orbital period, $P$, of the particle, such that the change in velocity due simply to the motion of the particle around its orbit during the time $\Delta t$ is small compared to the size of the kick.  In mathematical terms we may write this requirement as
\begin{equation}
 \frac{\Delta v}{v_{\rm k}} \gg \frac{|\boldsymbol{v}(t+\Delta t)-\boldsymbol{v}(t)|}{v_{\rm k}}
                                                 = 2\pi\frac{(1+eC_f)^2}{(1-e^2)^2}\frac{\Delta t}{P},
\end{equation}
where the right hand side is the fractional change in the velocity of the particle solely due to motion around its initial orbit.  \emph{Debris produced in a giant impact should always satisfy this criterion,} as the timescale for launching the debris is $\sim$minutes-hours.
Though we describe the mass $M$ as the `central body' and the mass $m$ as the `particle' we have not made any assumptions here about the relative size of $M$ and $m$ as the criterion is independent of $m$.  Any problem in which this impulsive velocity change approximation is valid may be treated using the equations presented here.

While the equations are presented in a reference frame such that $I=\Omega=\omega=0$ there is, as stated initially, no loss of generality as a result and they can easily be applied to cases in which non-zero values of $I$, $\Omega$ and $\omega$ are desired by application of simple rotations to the resulting distributions.  One caveat that should be noted with respect to rotating the distributions is that the definition of $I$ such that it only takes values in the range $[0,\pi]$, which we use here, and the related definition of $\Omega$, requires that some care be taken in such transformations.  It is easiest to unfold the $I$ distribution onto the range $[-\pi,\pi]$ before performing the transformations and then fold it back into the range $[0,\pi]$ afterwards.  To do this simply define $I'\ge0$ if $\theta \ge \frac{\pi}{2}$ and $I'<0$ if $\theta < \frac{\pi}{2}$ and always take the first condition in Eqs. \ref{Omegaprim} and \ref{eccomegafprim}.  For convenience we give equations for arbitrary initial $I$, $\omega$ and $\Omega$ in Appendix~\ref{geneq}.  Note that all of our equations assume an initially bound orbit.  The kick velocity formalism can equally be applied to initially parabolic or hyperbolic orbits, but some adjustments are necessary, in particular to account for the semi-major axis as normally defined becoming negative.

\subsection{Circular initial orbits}
\label{circparent}

To analyse some of the dependencies of the new orbital elements on the magnitude and direction of the kick velocity it is beneficial to consider the simpler case that the initial orbit of the particle is circular, such that $e=0$.  For a circular pre-kick orbit the definition of the true anomaly, $f$, becomes arbitrary and so we choose to set $f=0$; in geometric terms this means we define the particle to be on the x-axis at the time of the kick.  All dependence on $e$ and $f$ now drops out and Eqs.~\ref{aaprimecc}, \ref{ecceprim}, \ref{ecchprimh}, \ref{eccIprim} and \ref{eccsinfcosf} simplify to:
\begin{equation}
\label{aaprim}
\frac{a}{a'}=1-2\left(\frac{\Delta v}{v_{\rm k}}\right) S_{\theta}S_{\phi}-\left(\frac{\Delta v}{v_{\rm k}} \right)^2,
\end{equation}
\begin{equation}
\label{eprim2}
e'^2=1-\left(\frac{h'^2}{h^2}\frac{a}{a'}\right),
\end{equation}
\begin{equation}
\label{Iprim}
\cos I'=\left[1+\left(\frac{\Delta v}{v_{\rm k}}\right)S_{\theta}S_{\phi}\right]\left(\frac{h'^2}{h^2}\right)^{-\frac{1}{2}},
\end{equation}
\begin{equation}
\label{hhprim}
\frac{h'^2}{h^2}=1+2\left(\frac{\Delta v}{v_{\rm k}}\right)S_{\theta}S_{\phi}+\left(\frac{\Delta v}{v_{\rm k}}\right)^2(C_{\theta}^2+S_{\theta}^2S_{\phi}^2)
\end{equation}
\begin{align}
\label{sinfcosf}
\sin f' & =\frac{1}{e'}\left(\frac{h'^2}{h^2}\right)^{\frac{1}{2}} \left(\frac{\Delta v}{v_{\rm k}}\right) S_{\theta}C_{\phi} \\ \notag
\cos f' & =\frac{1}{e'}\left( \frac{h'^2}{h^2}-1\right)
\end{align}
Eqs.~\ref{aaprim}-\ref{hhprim} are equivalent to Eqs. 1-4 of \citet{jackson2012}, but note that we cast the equation for $I'$ here in a slightly different form that more intuitively handles retrograde orbits.

\subsubsection{Distribution of orbital elements}
\label{sec:circparent:aeibounds}

\begin{figure}
\includegraphics[width=\columnwidth]{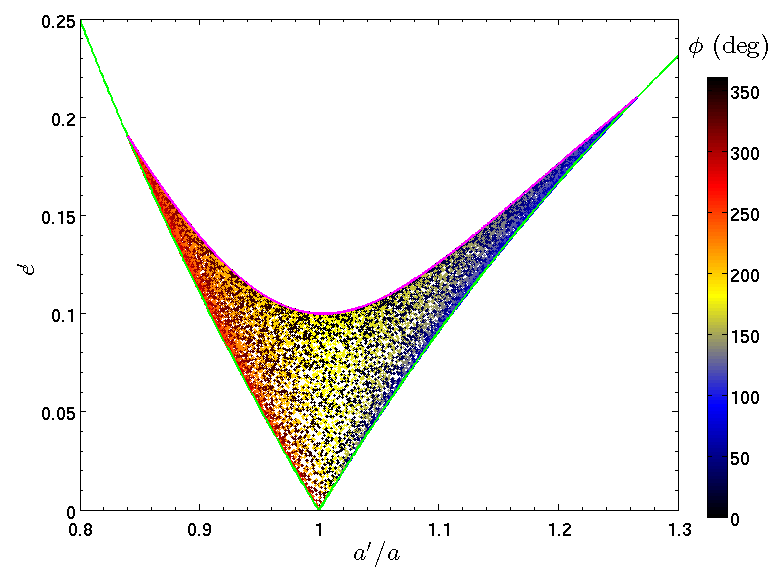}
\includegraphics[width=\columnwidth]{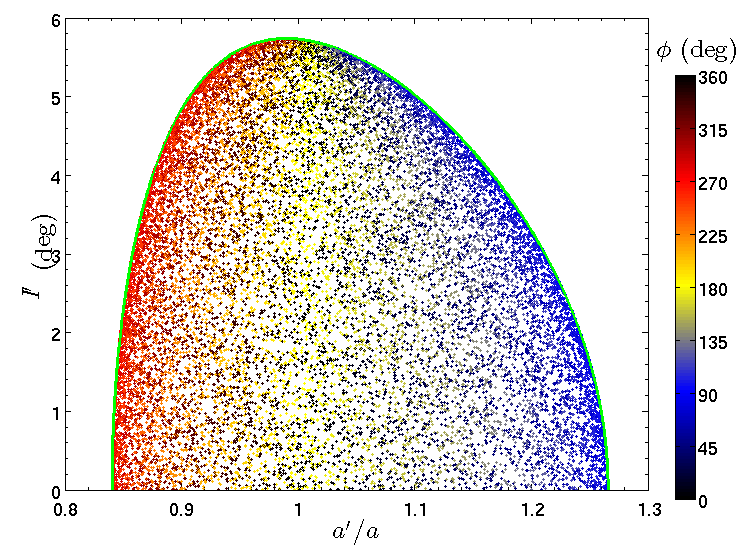}
\includegraphics[width=\columnwidth]{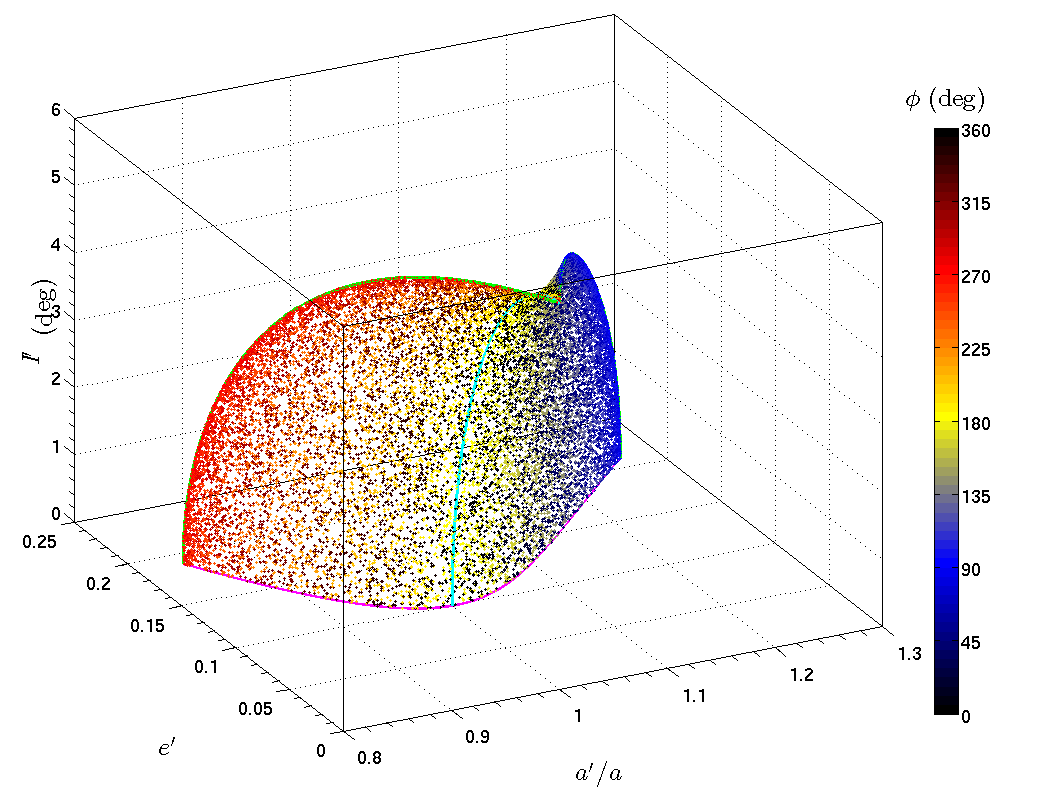}
\caption{Distributions of orbital elements for a set of 20,000 particles given a kick of magnitude $(\Delta v/v_{\rm k})=0.1$ in a random direction from an initially circular orbit.  \emph{Top}: $a-e$ distribution, \emph{middle}: $a-i$ distribution, \emph{bottom}: 3 parameter $a-e-i$ distribution.  The boundaries of the distribution are set by the new apocentre or pericentre being at the old semi-major axis distance (green lines) and the $I'=0$ contour for a $(\Delta v/v_{\rm k})=0.1$ kick (magenta line).  On the $a-e-i$ distribution we also show the line of $a'/a=1$ in cyan, which tracks close to one of the lines of principal curvature.  Points are coloured by the value of $\phi$, with black corresponding to a kick directly away from the star, yellow directly towards the star, blue in the direction of orbital motion and red against the direction of orbital motion.}
\label{aeidv01dist}
\end{figure}

Fig.~\ref{aeidv01dist} shows the distributions of semi-major axes, eccentricities and inclinations produced for a set of 20,000 particles on the same initial orbit given a kick velocity of $(\Delta v/v_{\rm k})=0.1$ and a spherically symmetric distribution of kick angles.  The key boundary of the distributions is described by the green lines, which form a distinctive `V' shape in the semi-major axis -- eccentricity distribution.  These lines are associated with the apocentre or pericentre of the new orbit being located at point at which the kick was applied (here the old semi-major axis distance), since the new orbit must pass through the point at which the kick was applied and thus the apocentre cannot be further in, nor the pericentre further out, than this distance.  These conditions on the apocentre and pericentre can be summarised by the inequality $a'(1-e') \leq r \leq a'(1+e')$.

The other boundary of the distribution, coloured magenta, is the $I'=0$ contour.  Points on this line correspond to kicks exactly in the plane of the pre-kick orbit, thus corresponding to the largest change in $a$ and $e$, and hence to how much of the `V' in semi-major axis -- eccentricity space is filled.  The height of this boundary in the upper-most panel of Fig.~\ref{aeidv01dist} is dependent on the magnitude of the kick, rising for larger kicks and falling for smaller kicks, whereas the position of the apocentre and pericentre conditions is constant.  Conversely in the middle panel the $I'=0$ contour is of course constant, whereas the height of the boundary defined by the apocentre and pericentre conditions rises with increasing kick velocity.

From the colouration of the points in Fig.~\ref{aeidv01dist} we can see that particles which have their semi-major axes reduced are those which receive kicks with $\phi > 180^{\circ}$, meaning that the component of the kick in the direction of orbital motion is opposed to the motion of the particle.  This is as expected, and indeed from the form of Eq.~\ref{aaprim} we can also see that for $a'/a < 1$ we must have $S_{\theta}S_{\phi} < 0$.  It should be noted however that this is not a sufficient condition due to the $(\Delta v/v_{\rm k})^2$ term and thus $\phi=180^{\circ}$ is not a hard boundary between decrease and increase of the semi-major axis.  The larger the magnitude of the kick the further towards higher values of $\phi$ the line of $a'/a=1$ (indicated in cyan in the lower panel of Fig.~\ref{aeidv01dist}) moves.

From Fig.~\ref{aeidv01dist} and the dependencies of Eqs.~\ref{aaprim}-\ref{sinfcosf} we can also make some simple general observations about non-spherically symmetric distributions of kick angles.  The angle $\theta=\pi/2$ at $I'=0$ and moves away from this towards $0$ or $\pi$ at higher $I'$.  So if particles are preferentially given kicks close to the plane of the original orbit, such that $\theta$ is closer to $\pi/2$, high values of $I'$ are less likely, decreasing the population of the upper parts of the middle panel of Fig.~\ref{aeidv01dist} while increasing the population of the upper parts of the upper panel.  Similarly if particles are prefentially kicked forwards (in the direction of their previous orbital motion) then the semi-major axis will be preferentially increased and the right-hand parts of both the upper and middle panels will be more strongly populated.

\begin{figure}
 \includegraphics[width=\columnwidth]{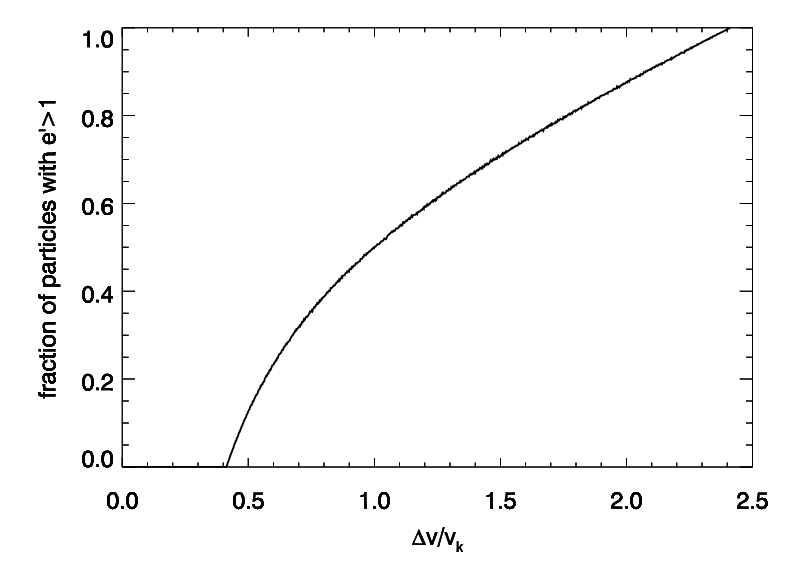}
 \caption{Fraction of particles that are put onto hyperbolic orbits ($e'>1$) as a function of the kick velocity for an isotropic distribution of kick directions.  Above $\Delta v/v_{\rm k}=\surd 2 +1$ all particles have $e'>1$.}
 \label{fig:dvfraclost}
\end{figure}

The minimum kick velocity for which an eccentricity of 1 is achievable, and thus for which particles can be ejected from the system by the kick, is $(\Delta v/v_{\rm k})= \surd 2-1$.  Conversely the maximum kick velocity for which an object can remain bound is $(\Delta v/v_{\rm k})= \surd 2+1$.  Both of these arrive as a result of the circular velocity being a factor of $\surd 2$ less than the stellar escape velocity at the same distance.  Thus a kick that increases the forward motion by a factor of $\surd 2-1$, or completely cancels the forward motion and provides an additional $\surd 2 v_{\rm k}$, are the bounding cases for escape.  Fig.~\ref{fig:dvfraclost} shows how the fraction of particles that achieve hyperbolic orbits changes as the kick velocity is varied between $\surd 2 -1$ and $\surd 2 +1$ for an isotropic distribution of kicks.

\subsubsection{Orbital distribution as a function of kick velocity}
\label{sec:circparent:orbdistdv}

\begin{figure}
\centering
 \includegraphics[width=0.9\columnwidth]{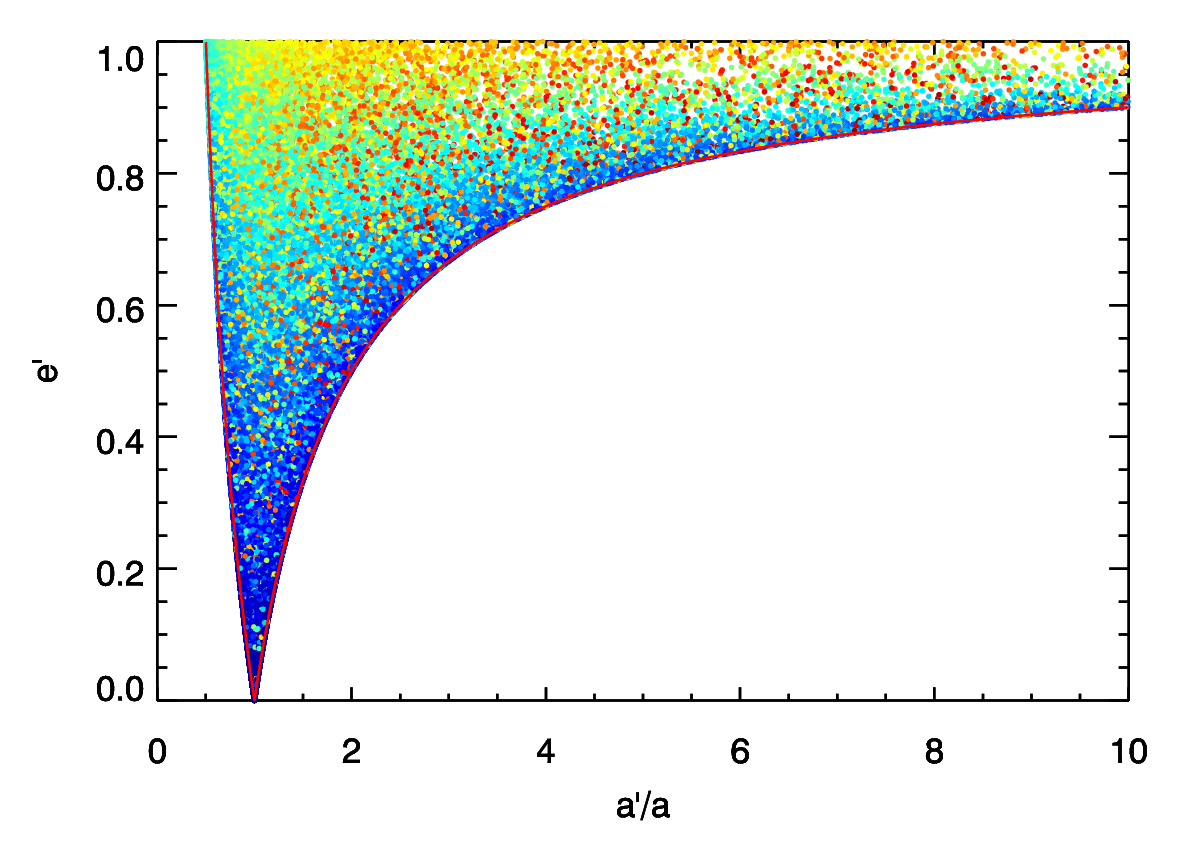}
 \includegraphics[width=0.9\columnwidth]{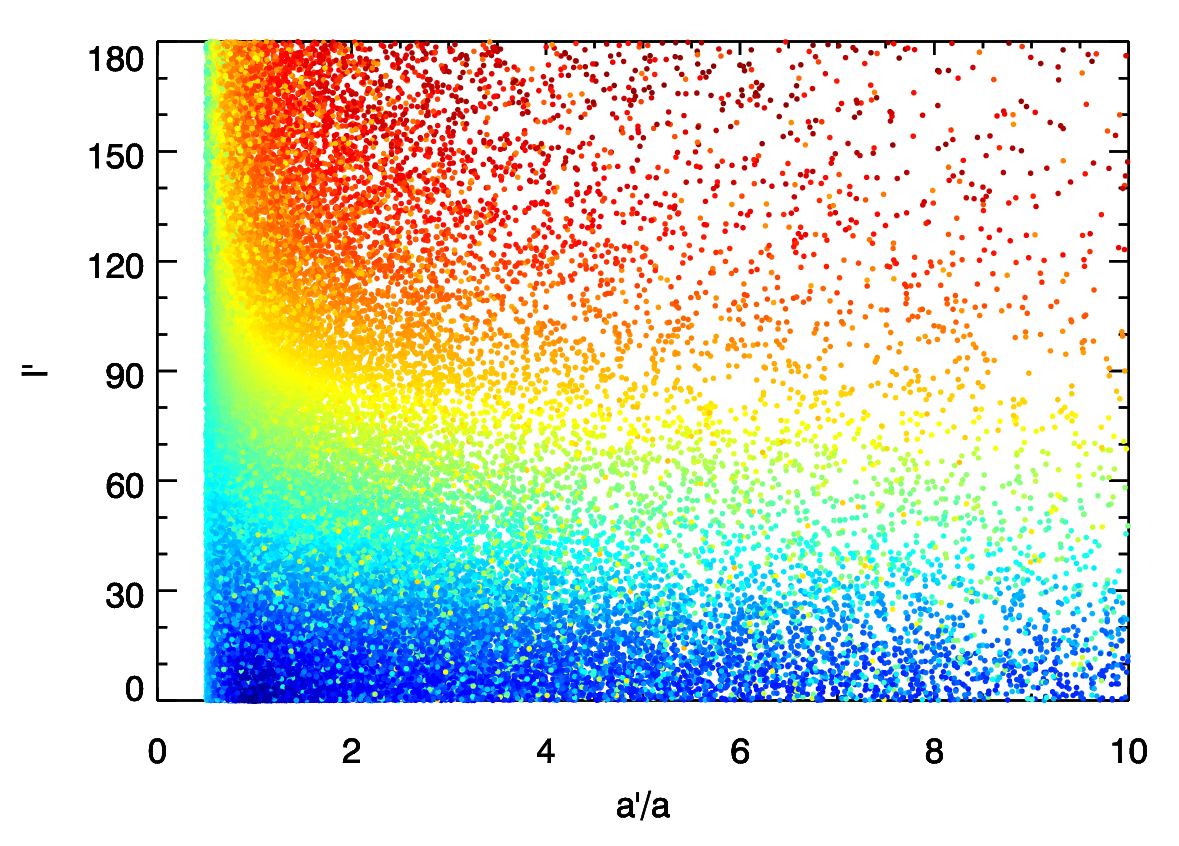}
 \includegraphics[width=0.9\columnwidth]{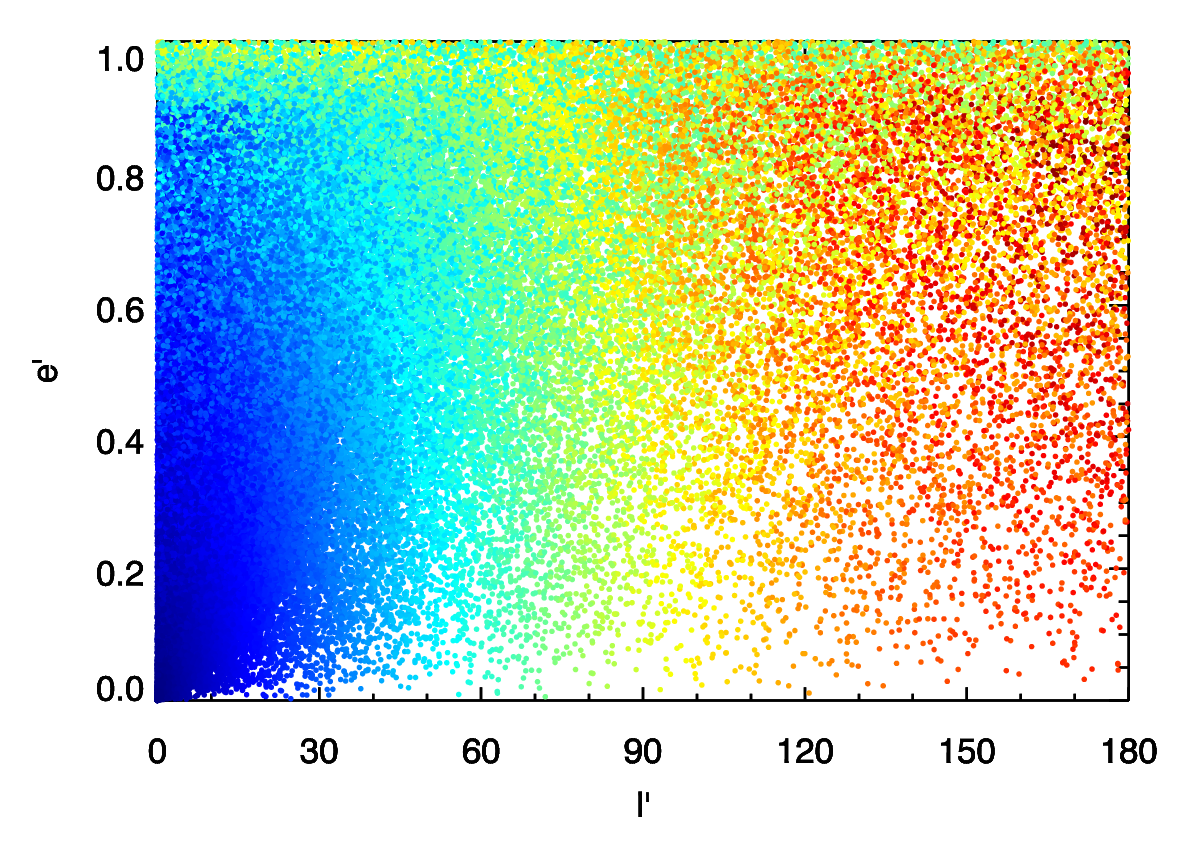}
 \includegraphics[width=0.9\columnwidth]{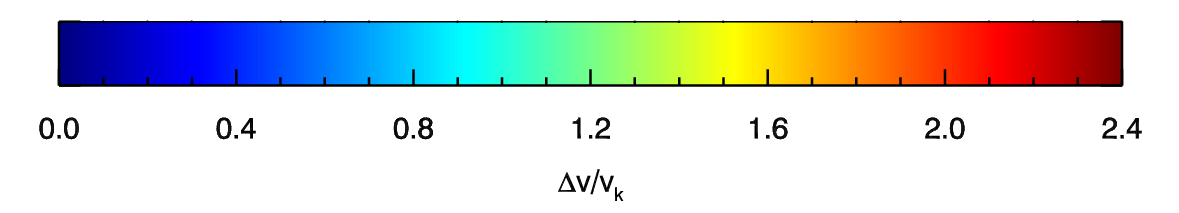}
 \caption{The orbital distributions as a function of kick velocity.  \emph{Top}: semi-major axis -- eccentricity, \emph{middle}: inclination -- eccentricity, \emph{bottom}: semi-major axis -- inclination.  100,000 particles are used with a random kick velocity chosen uniformly between $\Delta v/v_{\rm k}$ of 0 and $\surd 2 + 1$ (the maximum kick for which a body can remain bound).  On the semi-major axis -- eccentricity distribution we show the apocentre and pericentre conditions in red.}
 \label{fig:orbdistdv}
\end{figure}

In the preceding section we showed the distribution of orbital elements for a single value of the kick velocity, and discussed how the orbital elements of an individual particle depend on the kick direction (as specified by $\theta$ and $\phi$).  In a debris cloud produced by a giant impact however there will not be a single value of the kick velocity, but rather a distribution of kick velocities.  In Fig.~\ref{fig:orbdistdv} we show the distributions of the semi-major axis, eccentricity, and inclination as a function of the kick velocity, with a uniform distribution of kick velocities between the bounding values of 0 and $\surd 2 + 1$.

If we consider moving to a distribution of kick velocities in terms of the 3 parameter surface in the lower panel of Fig.~\ref{aeidv01dist} the change is that now rather than having a single surface we have a set of nested surfaces corresponding to different kick velocities.  The surfaces corresponding to the lowest kick velocities lie closest to $a'/a=1$, $e'=I'=0$, while higher velocity surfaces lie further away.  Applying this to the projections shown in Fig.~\ref{fig:orbdistdv} we can see that, since either the eccentricity or inclination of a particle must change on being kicked, the middle panel largely separates into bands of different colour as both the minimum and maximum eccentricity and inclination increases with increasing kick velocity.  In the upper panel on the other hand since the surfaces for all kick velocities touch $a'/a=1$, $e'=0$ we can see points of all colours in the lower eccentricity region, though there is a general trend towards higher velocity colours at higher eccentricities since high eccentricities are only achievable with high kick velocities and low kick velocities can only achieve low eccentricities.

The lower panel of Fig.~\ref{fig:orbdistdv} also largely separates into bands of different colour, for similar but slightly different reasons to the middle panel.  For a particle receiving a high kick velocity to obtain a relatively low inclination requires it to obtain a higher eccentricity.  For very large kicks a very low inclination can require an eccentricity greater than 1, for which the semi-major axis becomes negative and thus it does not appear in the lower panel of Fig.~\ref{fig:orbdistdv}.  As a result of the increasing fraction of particles that achieve hyperbolic orbits as the kick velocity is increased, as shown by Fig.~\ref{fig:dvfraclost}, the number of points at the high velocity end of the spectrum visible in Fig.~\ref{fig:orbdistdv} decreases significantly.  Those particles that remain bound after receiving very large kicks ($\Delta v/v_{\rm k} > \surd2$) exclusively occupy retrograde orbits.  This should be borne in mind later on, as discs containing significant numbers of particles that received very large kicks will thus possess material capable of higher collision velocities than might be otherwise expected.

\subsection{Initially eccentric orbits}
\label{eccorb}

\begin{figure}
\includegraphics[width=\columnwidth]{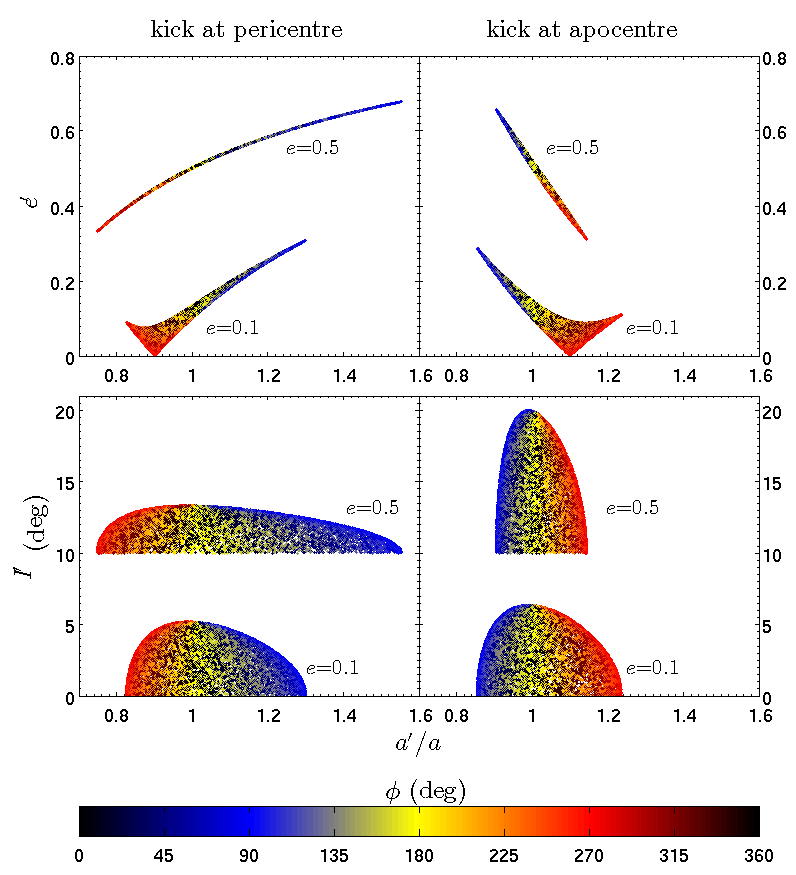}
\caption{Semi-major axis -- eccentricity distributions (\emph{upper row}) and semi-major axis -- inclination distributions (\emph{lower row}) for particles kicked at pericentre (\emph{left-hand column}) and at apocentre (\emph{right-hand column}) and two different initial eccentricities ($e=0.1$ and $e=0.5$).  An isotropic kick of magnitude $\Delta v/v_{\rm k}=0.1$ is used.  All distributions are coloured according to the value of $\phi$ as described in Fig.~\ref{aeidv01dist}.  Each distribution is annotated with the initial eccentricity.  The inclination distributions for initial eccentricity 0.5 have been offset by +10$^{\circ}$ to separate the distributions.}
\label{aedv01ecc}
\end{figure}

Now that we have analysed some of the dependencies of the new orbital elements in the simpler case of initially circular orbits it is beneficial to return to the case of initially eccentric orbits.  In Fig.~\ref{aedv01ecc} we show the semi-major axis -- eccentricity and semi-major axis -- inclination distributions for initial eccentricities of 0.1 and 0.5.

\begin{figure*}
 \begin{minipage}{\textwidth}
  \centering
    \includegraphics[width=0.325\textwidth]{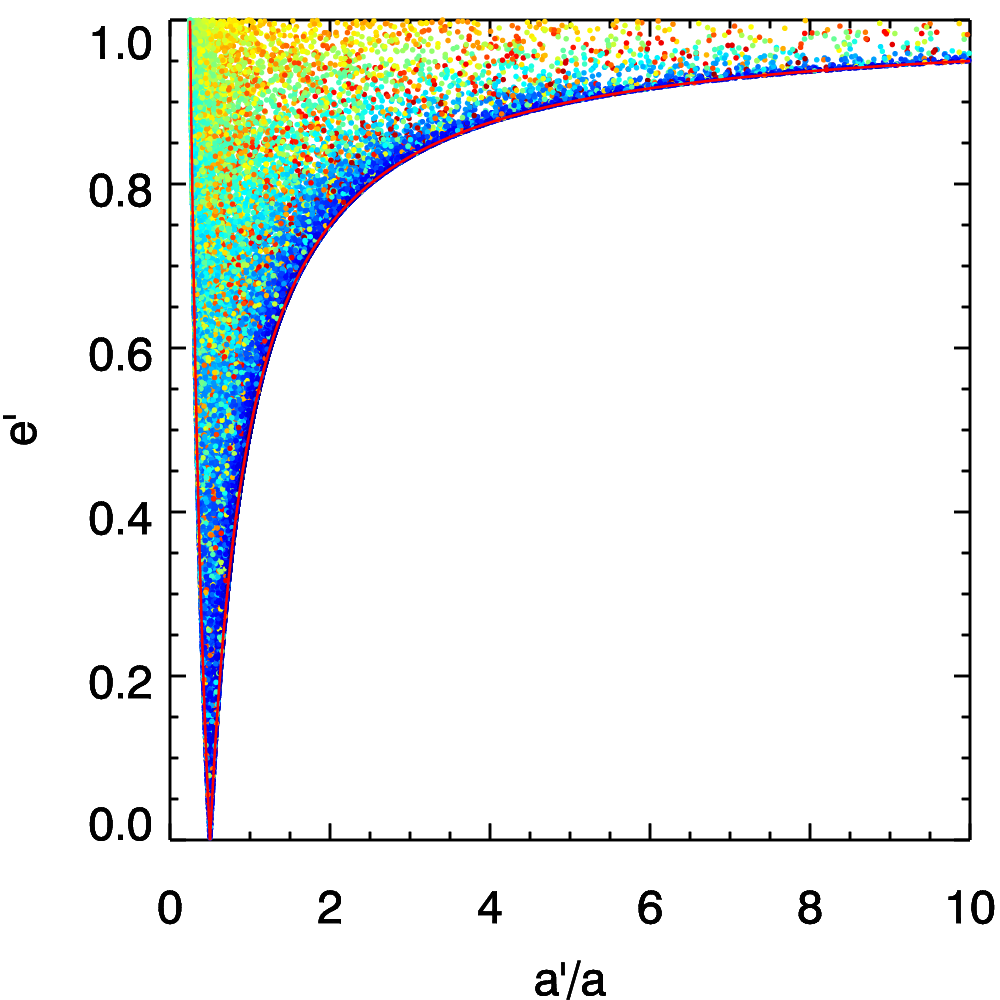}
    \includegraphics[width=0.325\textwidth]{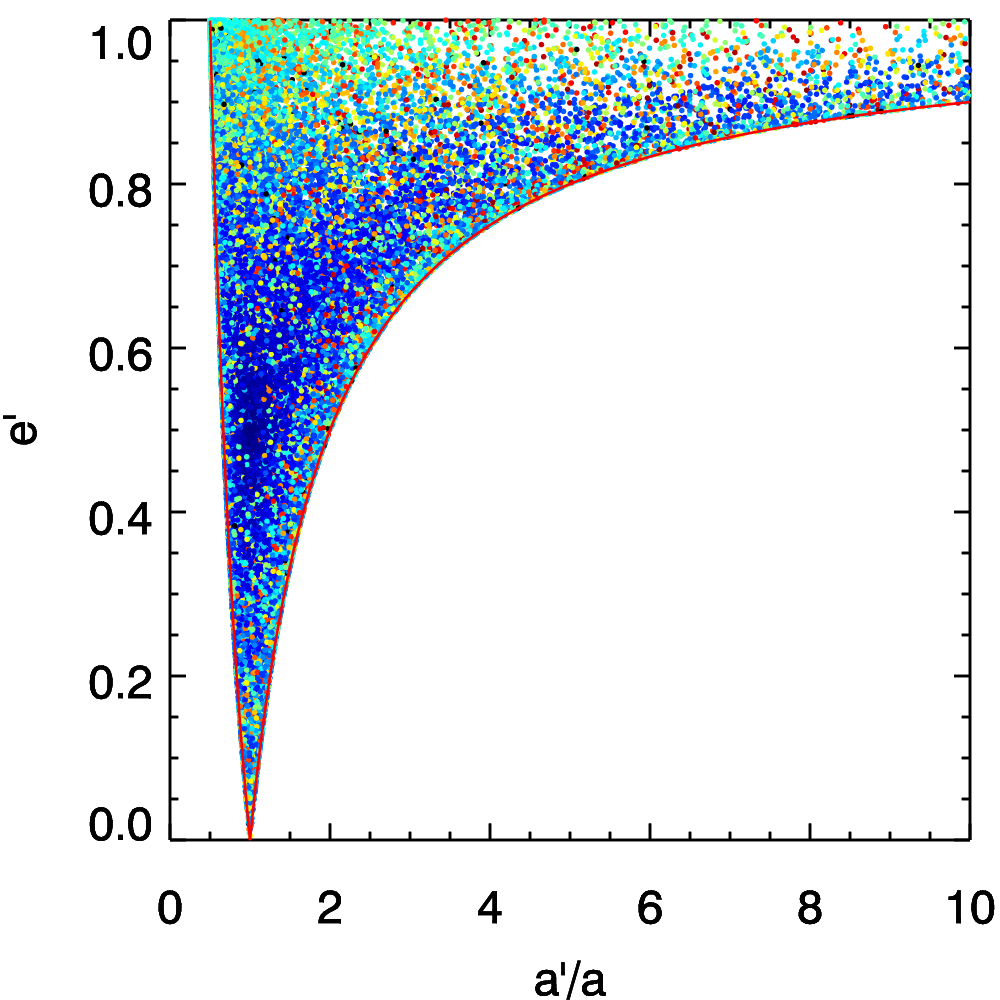}
    \includegraphics[width=0.325\textwidth]{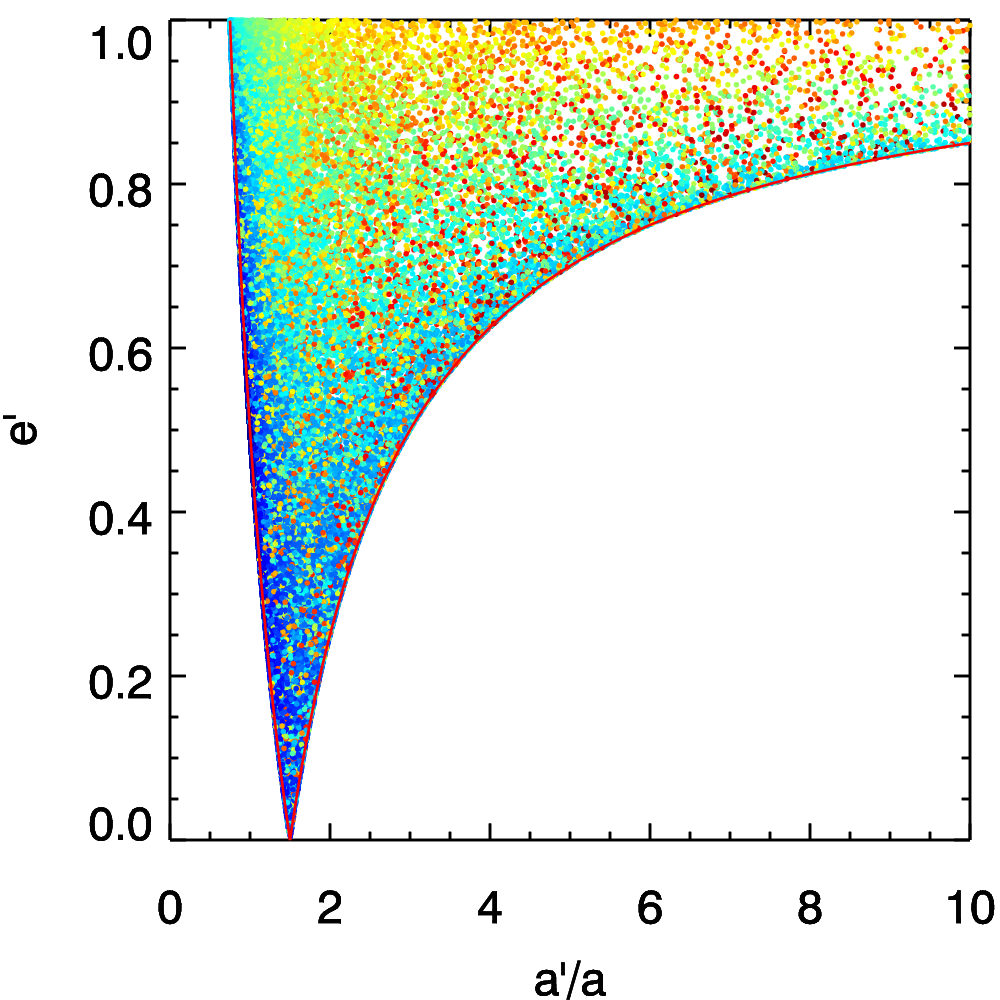}
    \includegraphics[width=0.325\textwidth]{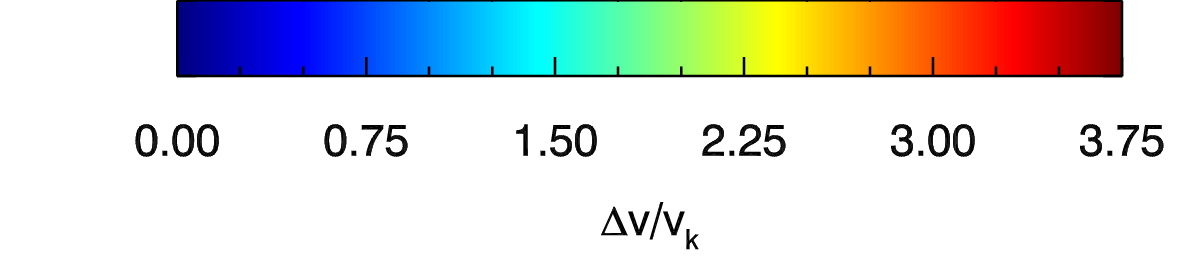}
    \includegraphics[width=0.325\textwidth]{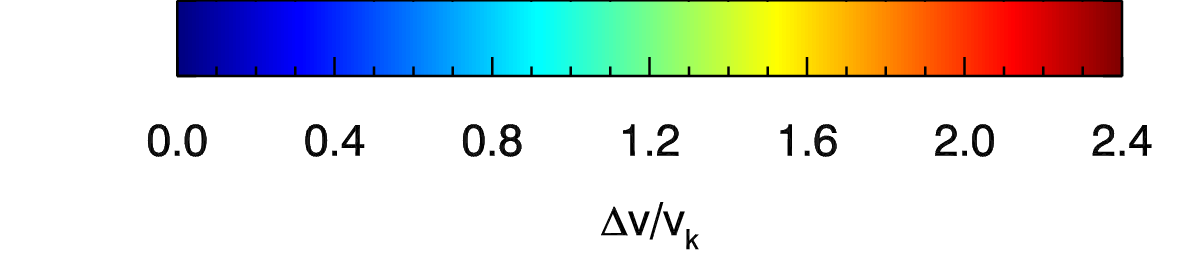}
    \includegraphics[width=0.325\textwidth]{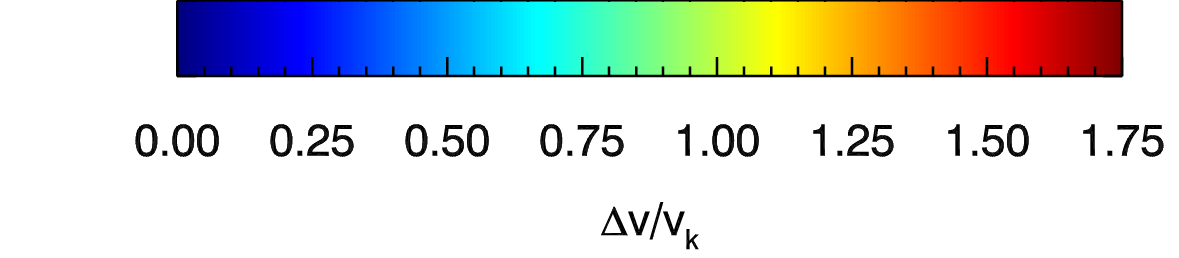}
    \caption{Semi-major axis -- eccentricity distributions for particles on an initial orbit with an eccentricity of 0.5 kicked at three different points around their orbit, \emph{left}: at pericentre, \emph{centre}: at the semi-major axis distance, \emph{right}: at apocentre.  Points are coloured by the kick velocity received with the distribution of kick velocities uniform between zero and the maximum kick a particle can receive and remain bound in each case.}
    \label{fig:aeecccomp}
 \end{minipage}
\end{figure*}

One of the first things that we can see from Fig.~\ref{aedv01ecc} is that as well as changes in the distributions for different eccentricities, the distributions also change depending on the true anomaly at which the particle is kicked, with kicks at apocentre producing a smaller range of semi-major axes, and a larger range of inclinations, than those at pericentre.  The difference between kicks at apocentre and pericentre is also more marked for higher eccentricities.  If we look at Eq.~\ref{aaprimecc} this is due to the term $(S_{(\phi-f)}+eS_{\phi})$, since when $f=0$ (pericentre) the two sines are in phase, while when $f=\pi$ they are out of phase, and as the eccentricity is increased the effect of the sines being in phase or out phase is enhanced.  While there are differences in the range of semi-major axes and inclinations between kicks at apocentre and pericentre we can see that the range of eccentricities is on the other hand very similar.

We can also expect that the minimum kick velocity at which the particle can be ejected will vary with $f$,
\begin{equation}
\label{dvminecc}
\left(\frac{\Delta v}{v_{\rm k}}\right)_{\rm{ej,min}}=\sqrt{2}\left(\frac{1+e C_f}{1-e^2}\right)^{\frac{1}{2}}
                                                - \left(\frac{1+2e C_f + e^2}{1-e^2}\right)^{\frac{1}{2}},
\end{equation}
which may also be written as
\begin{equation}
 \label{dvminecc2}
 \left(\frac{\Delta v}{v_{\rm k}}\right)_{\rm{ej,min}}=\sqrt{2}\left(\frac{r}{a}\right)^{\frac{1}{2}}-\frac{v}{v_{\rm k}},
\end{equation}
where $v$ is the velocity of the particle at $r$.  As described for circular orbits above, the maximum kick velocity that can be applied before all particles are ejected is obtained by changing the sign of the second term in Eq.~\ref{dvminecc} and \ref{dvminecc2}.  The minimum kick velocity at which particles can be ejected is lower at pericentre than at apocentre, which is perhaps slightly counterintuitive initially, but is because a particle has a higher velocity at pericentre than the circular velocity at the pericentre distance and so a smaller addition to the velocity is needed to eject it, whereas at apocentre the velocity is lower than the circular velocity at that distance.  Conversely however the maximum kick velocity at which particles can remain bound is higher at pericentre than at apocentre.  At $r=a$ the minimum kick velocity for ejection reduces back to $\Delta v/v_{\rm k}=\surd 2 -1$ as in the circular case.

The change in the distribution as a function of true anomaly can be further seen in Fig.~\ref{fig:aeecccomp}, which highlights the changing position of the outer bounding `V', centred at the location of the kick-point, which varies in distance as we change the true anomaly at which the kick is applied.  Fig.~\ref{fig:aeecccomp} also shows how the `V' of the apocentre and pericentre conditions always forms an absolute outer boundary even if the kick is not large enough for the distribution to reach one or both conditions in a particular instance, as is the case for the initial $e=0.5$ distributions in Fig.~\ref{aedv01ecc}.

The apparent change in the shape of the inclination distribution for initially eccentric orbits is less dramatic, maintaining the same domed shape as in the middle panel of Fig.~\ref{aeidv01dist}, albeit stretched or compressed.  The inclination follows the opposite trend with $f$ to the semi-major axis, with higher inclinations achievable for apocentre kicks than for pericentre kicks.  This we can understand intuitively since the inclination of the orbit must be set by the ratio of the vertical component of the kick velocity to the total velocity in the old orbit plane (the initial orbital velocity plus the planar component of the kick), and since at apocentre the orbital velocity is lowest, it follows that this ratio can be larger here.

\section{Observing dusty debris}
\label{sec:obsdisc}

The figures in Section~\ref{orbeq} give us an overview of where particles that receive different kicks will end up, both in terms of orbital elements and spatially, and shows us how particles concentrate near the progenitor orbit.  We now want to ask, if we observe the aftermath of a real giant impact, what are we likely to see?

Real giant impacts release debris with a distribution of kick velocities, and whereas in the figures in Section~\ref{orbeq} we used a uniform distribution simply to demonstrate the range of outcomes, we must now ask what this distribution is.  For the debris released by the Moon-forming giant impact \citet{jackson2012} found that the velocity distribution was well fit by a truncated Gaussian.  This was preferred over a power law due to a tail of particles at high velocities, a feature also found in many of the simulations of \citet{leinhardt2012}.  If we correct for the energy required by to escape Earth's gravity, and also weight by mass to account for the finding of \citet{leinhardt2012} that there are variations in the velocity distribution with the mass of the debris fragment, the velocity distribution of the Moon-formation debris is well fit by a Gaussian with mean zero and standard deviation 5.20 km~s$^{-1}$.  We expect the velocity dispersion to scale with the escape velocity of the progenitor body \citep[e.g.][]{leinhardt2012}, so different values of the standard deviation will correspond to different masses for the progenitor.  For progenitors of a constant density the escape velocity scales as $M^{1/3}$, and so since the orbital velocity scales as $r^{-1/2}$ to obtain the same dispersion in terms of $\Delta v/v_{\rm k}$ at different orbital distances we should rescale the mass as $r^{-3/2}$.

\subsection{Establishment of the disc}
\label{sec:asymlifetime}

\begin{figure}
 \includegraphics[width=0.5\columnwidth]{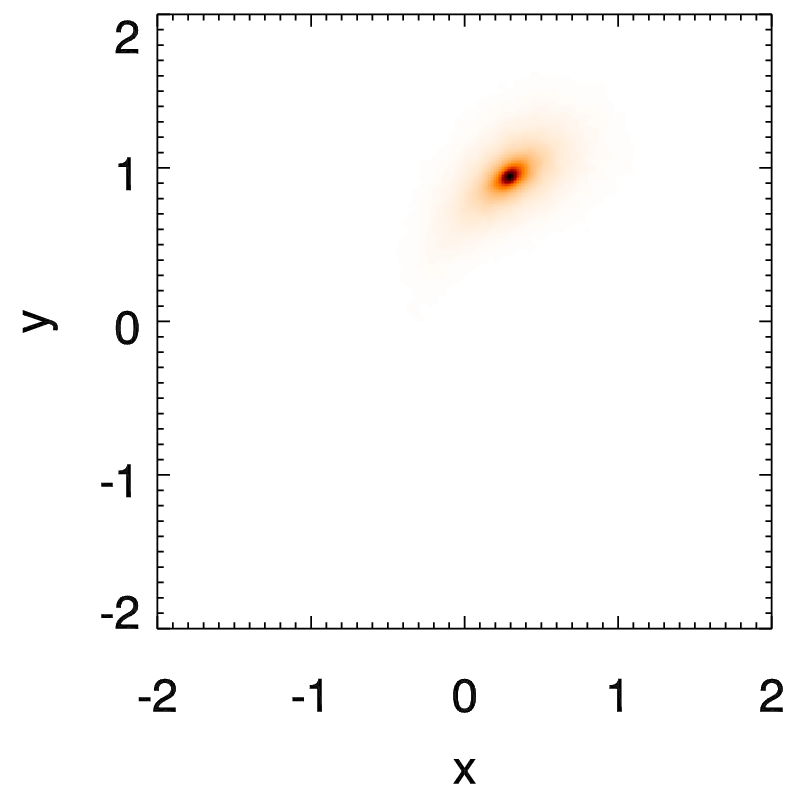}
 \includegraphics[width=0.5\columnwidth]{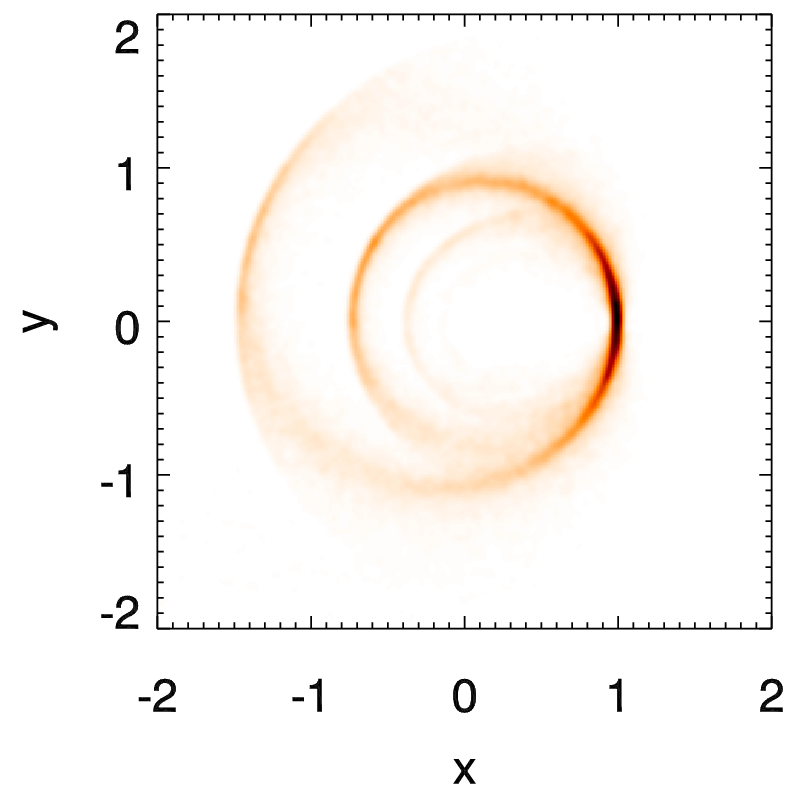}
 \includegraphics[width=0.5\columnwidth]{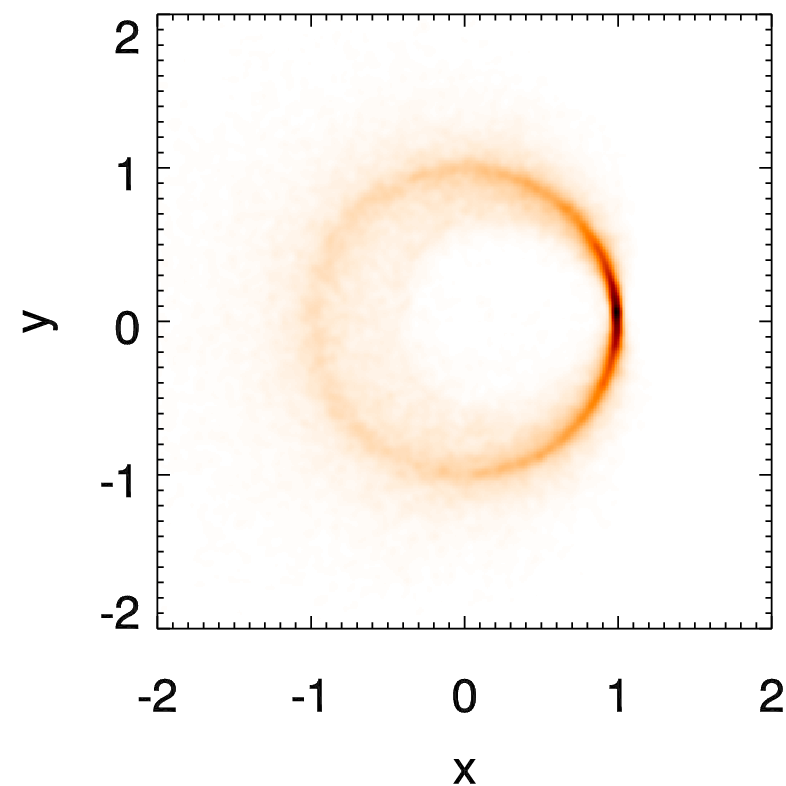}
 \includegraphics[width=0.5\columnwidth]{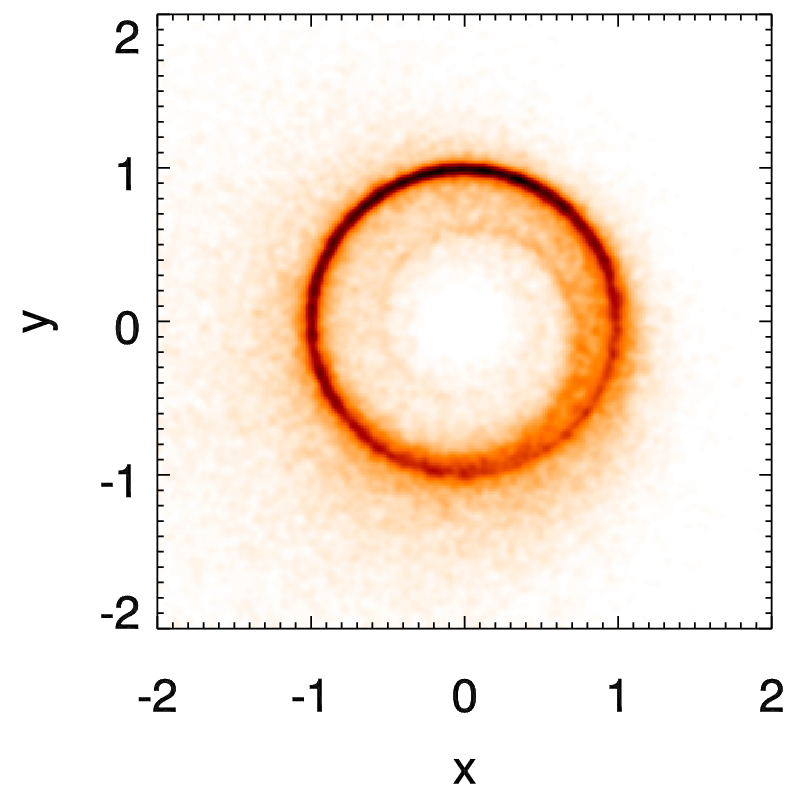}
 \includegraphics[width=\columnwidth]{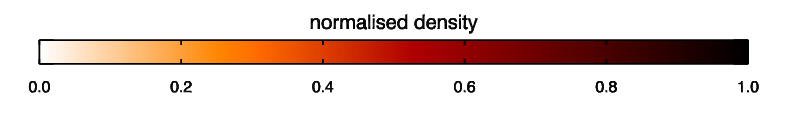}
 \caption{Phases of the dynamical evolution of a giant impact debris disc created at (1,0) at t=0 from a progenitor on a circular orbit.  \emph{Top left}: Appearance after 0.2 orbits.  \emph{Top right}: After 2 orbits.  \emph{Bottom left}: 200 orbits.  \emph{Bottom right}: 10,000 orbits.  A velocity dispersion of $\sigma_v/v_{\rm k}=0.3$ is used.  The effects of precession due to a Jupiter mass planet interior to the disc at 0.2 semi-major axis units are included.  All images are normalised individually and have a gaussian smoothing with FWHM 0.05 semi-major axis units applied.}
 \label{fig:earlyphases}
\end{figure}

It takes the asymmetric disc a short time to become established and smooth following the impact event.  Immediately after the impact the debris forms an expanding clump that follows the progenitor object (top left panel of Fig.~\ref{fig:earlyphases}).  As the clump moves around the orbit it shears out as a result of the gradient in orbital period and velocity across the clump from faster moving particles interior to the progenitor to slower moving particles exterior to the progenitor.  Although the rate at which the clump shears is dependent on the velocity dispersion of the debris, it is a rapid process and the clump will typically have been dispersed after only 1 orbit.  Even for a debris cloud with a very tight velocity dispersion this will take no more than a few orbits.  The `clump' phase is thus unlikely to last more than around 500 years at 50~AU.

After the clump phase we enter the `spiral' phase.  Here the initial debris clump has been sheared out into a spiral structure, but there is still enough coherence for there to be a visible spiral pattern with coils composed of debris particles on similar orbits.

The spiral structure of the spiral phase would not be detectable for some edge-on configurations and at lower resolutions, but high resolution images close to face-on would detect variations in the dust density.  The shearing process will continue and gradually coil the spiral tighter and tighter until it has been completely smeared out and the coils have merged.  The shearing again has some dependence on the velocity dispersion of the debris (and on the resolution of observations), but will have been completed after 100 orbits for even very tight velocity dispersions, and after around 50 orbits in most cases.

Once the spiral has become so tightly coiled the structure is no longer visible we enter a phase with a smooth, but still strongly asymmetric, disc (bottom left of Fig.~\ref{fig:earlyphases}).  The most important feature of the disc in this phase, which first appears in the spiral phase, is the phenomenon which we refer to as the \emph{collision-point}.  This arises from the requirement that the new orbit of a particle must pass through the point at which the kick is applied that is responsible for the apocentre and pericentre conditions.  Since for giant impact debris, the point at which the kick is applied is the point at which the collision occurs, and this is the same for all of the debris, the collision-point forms a nexus for the orbits of all of the debris fragments.  In all of the figures in this Section the collision takes place on the $x$-axis.

This smooth asymmetry also has a finite lifetime, and will eventually smear out into an axisymmetric structure (bottom right of Fig.~\ref{fig:earlyphases}).  The presence of other massive bodies in the system, or potentially even self-gravity of the debris itself, will gradually cause both the longitudes of ascending node, $\Omega$, and arguments of pericentre, $\omega$, of the orbits to precess.  For a particle with an orbital period, $T$, and semi-major axis, $a$ (with the ratio of planet and particle semi-major axes, $a_{\rm p}/a$, given by $\alpha$), the precession period is 
\begin{equation}
t_{\rm precess}=4 T \left[ \alpha b^1_{3/2}(\alpha) \frac{M_{\rm p}}{M_*} \right]^{-1}
\label{eq:precession}
\end{equation}
where $t$ is the time since the starting point, $M_{\rm p}$ and $M_*$ are the planet and star masses, and $b^1_{3/2}(\alpha)$ is a Laplace coefficient (see e.g. \citealt{murray1999, wyatt1999}).  To first order in $\alpha$, $b^1_{3/2}(\alpha)\approx3\alpha$.  Note that the argument of pericentre and the ascending node precess in opposite directions.

Around a sun-like star with a Jupiter mass planet at one fifth of the orbital distance of the debris (e.g. 10 and 50 AU) this leads to a precession period of around 33,000 debris orbital periods ($\sim$10Myr at 50 AU).  The debris distribution will have a range of semi-major axes and the precession period varies quite rapidly with the semi-major axis ratio.  Including the dependence of the orbital period the precession period varies as $a^{-7/2}$, leading to a factor of 2 difference in precession period for debris with orbits differing by only 20 per cent in semi-major axis.  If we return to the semi-major axis distributions in Figs.~\ref{aeidv01dist}, \ref{fig:orbdistdv} and \ref{aedv01ecc} we can see that such variations in semi-major axis appear even for small kicks.  As such we expect variations of factors of 2 or more in precession rate between different particles such that one precession period at the progenitor semi-major axis will be sufficient for the asymmetry to have been washed out.

In addition to possible massive planets elsewhere in the system, interactions with the progenitor body itself will also lead to precession of the debris orbits.  This is not easy to quantify with an equation like Eq.~\ref{eq:precession}, because the semi-major axis ratio for many of the particles will go to 1 and the expansions on which such equations rely fail to converge.  The typical precession periods induced by the co-orbital progenitor body are however similar in magnitude to that calculated above for a distant Jupiter.

The spatially confined collision-point itself is shorter lived, lasting only a few tenths of a precession period, but there is still visible asymmetry after the collision-point is no longer a meaningful concept, as we can see from the bottom panels of Fig.~\ref{fig:earlyphases}.  This sets a typical lifetime for the collision-point of a few thousand debris orbits ($\sim$1 Myr at 50 AU), and a typical timescale for achieving total axisymmetry of a few tens of thousands of debris orbits.

Once the collision-point has begun to be smeared out through precession the asymmetry in the small, visible, dust grains may also be further washed out as a result of collisional diffusion.  At early times when the collision-point is dominant however collisional diffusion should not be significant, due to both the dominance of the initial impact in setting the velocity dispersion, and the dominance of the collision-point in the collision rate amongst the debris (see Section~\ref{sec:collevol}).

There is therefore roughly a factor of 50-100 in lifetime between both the clump phase and spiral phase, and the spiral phase and the asymmetric disc.  As such, while it is not beyond the realms of possibility for the aftermath of a giant impact to be observed in the spiral phase, it is much less likely than observing it in the asymmetric phase.  The clump phase however has a lifetime over 3 orders of magnitude shorter than the asymmetric disc, and so observing a system during this phase is very unlikely.  The most likely phase in which to observe a giant impact debris disc is of course probably the axisymmetric phase (though this depends on collisional lifetimes, see Section~\ref{sec:collevol}), however other than transience arguments based on system age \citep[e.g.][]{wyatt2007} it may be difficult to distinguish a giant impact debris disc in this state from other cold discs.  As such we focus on the smooth asymmetric phase.

All of the phases of evolution discussed above also of course apply to hot debris produced by impacts in the inner planetary system, such as that released by the Moon-forming impact as studied by \citet{jackson2012}.  Since the orbital timescales in the inner system are much faster the asymmetric phases are much shorter lived than for impacts occuring in the outer planetary system.  As such observing a system in an asymmetric state after an impact occuring in the inner planetary system would be less likely.

\subsection{Disc morphologies}
\label{sec:obsdisc:morph}

\begin{landscape}
\begin{figure}
\centering
  \includegraphics[width=0.41\textwidth]{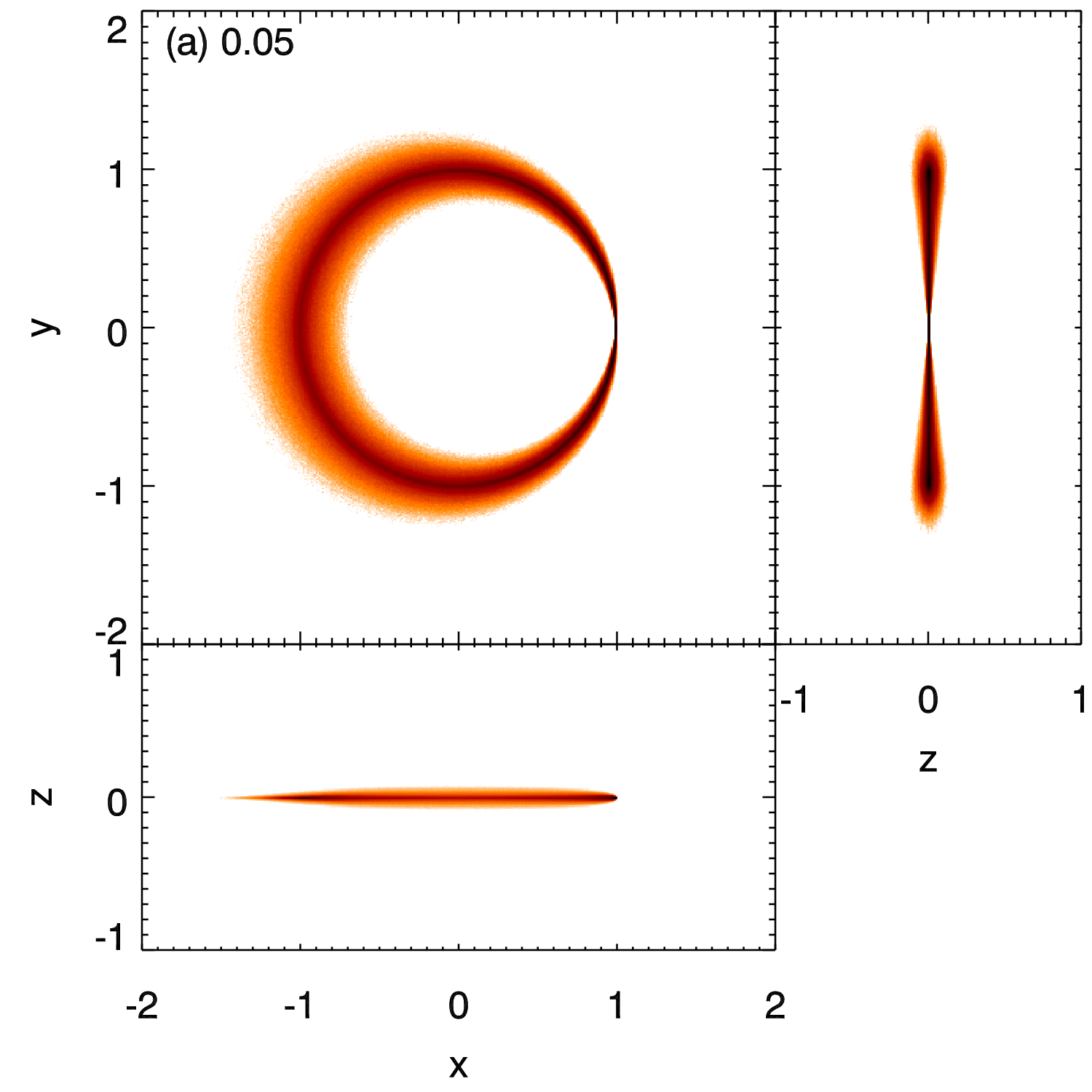}
  \includegraphics[width=0.41\textwidth]{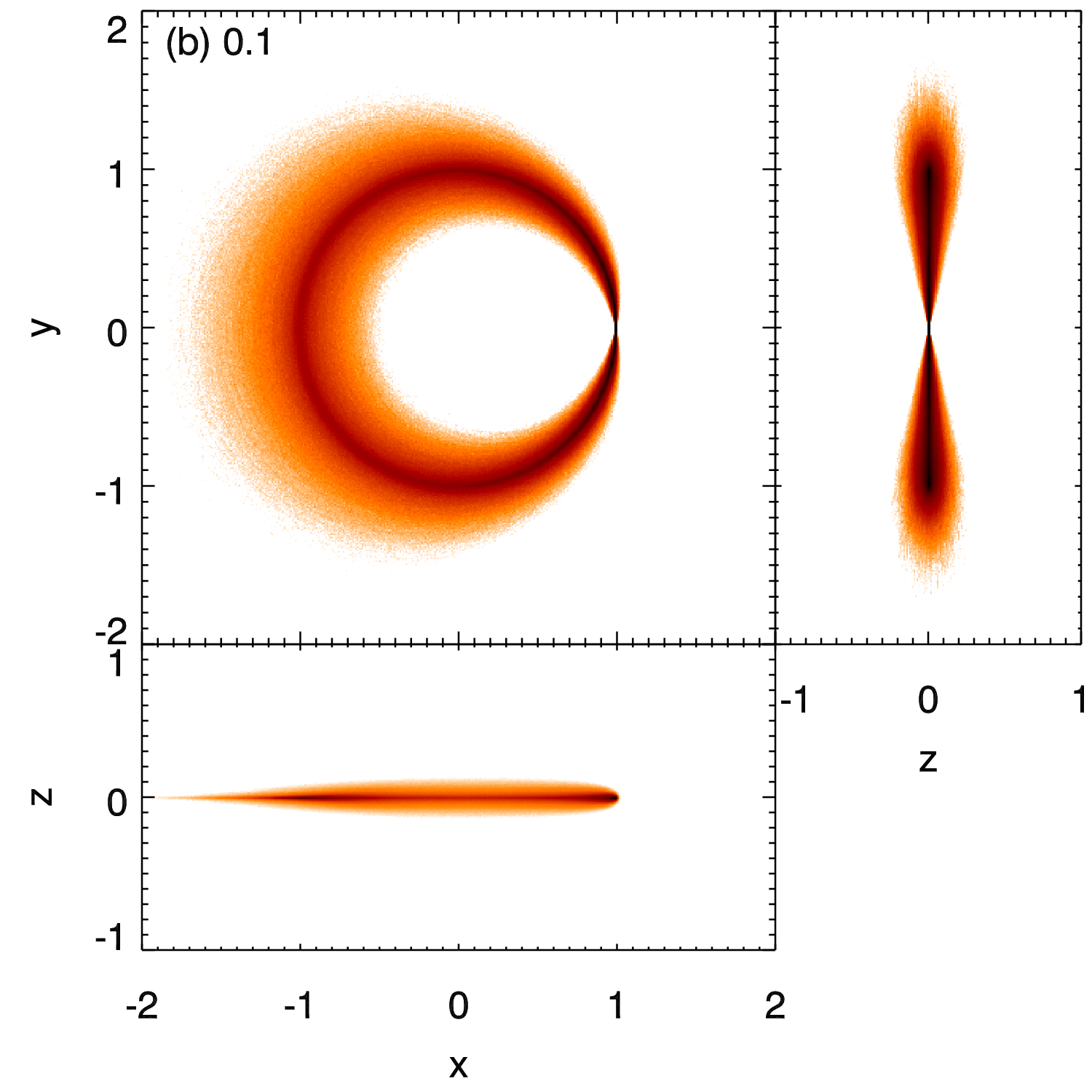}
  \includegraphics[width=0.41\textwidth]{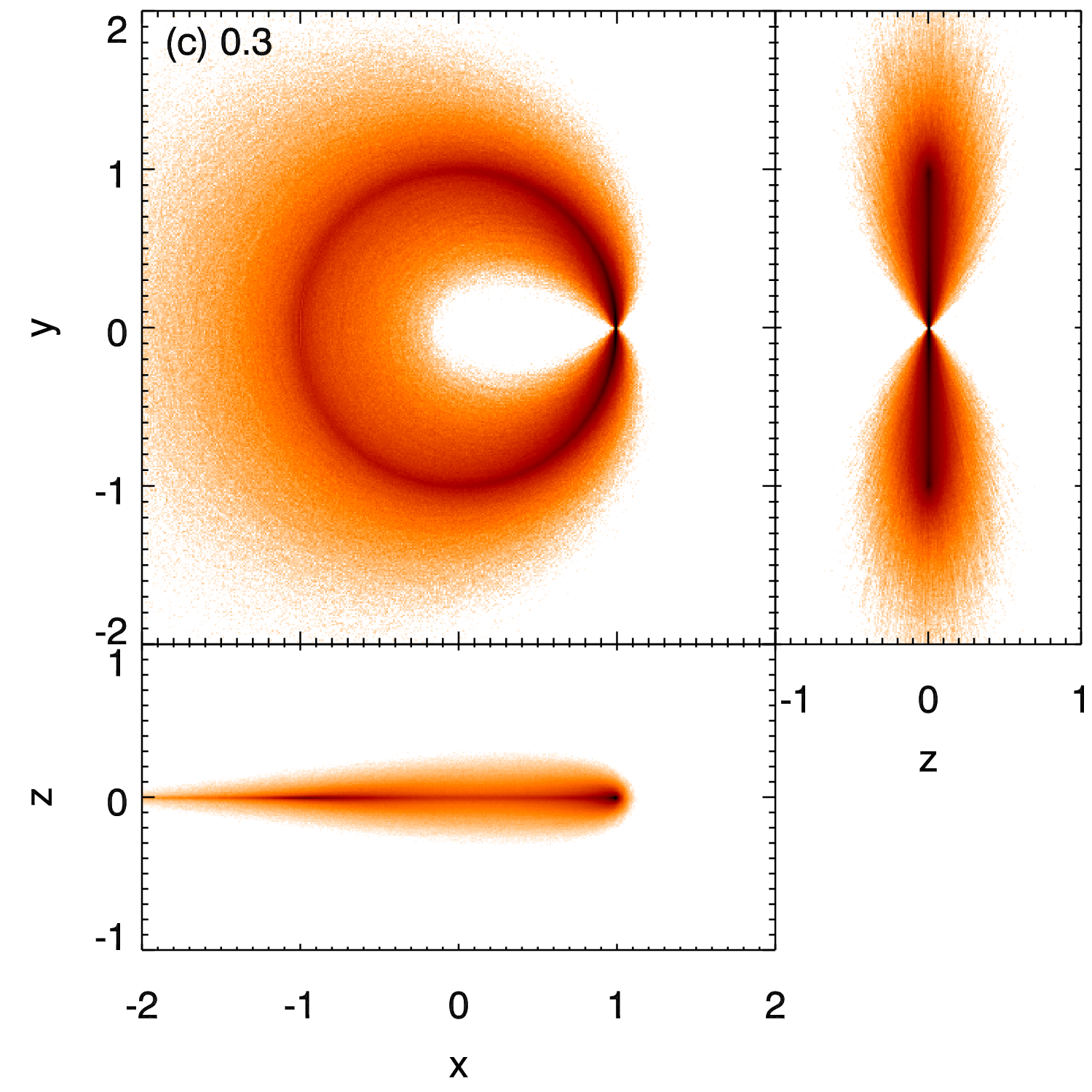}
  \includegraphics[width=0.075\textwidth]{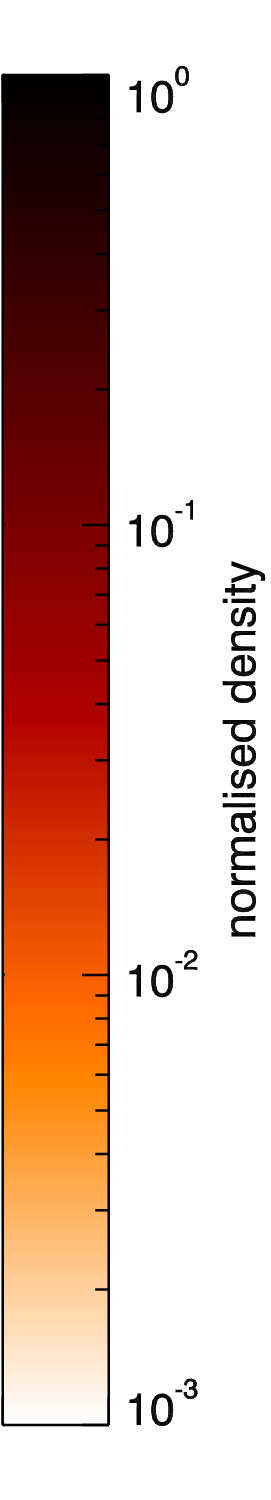}
  \includegraphics[width=0.41\textwidth]{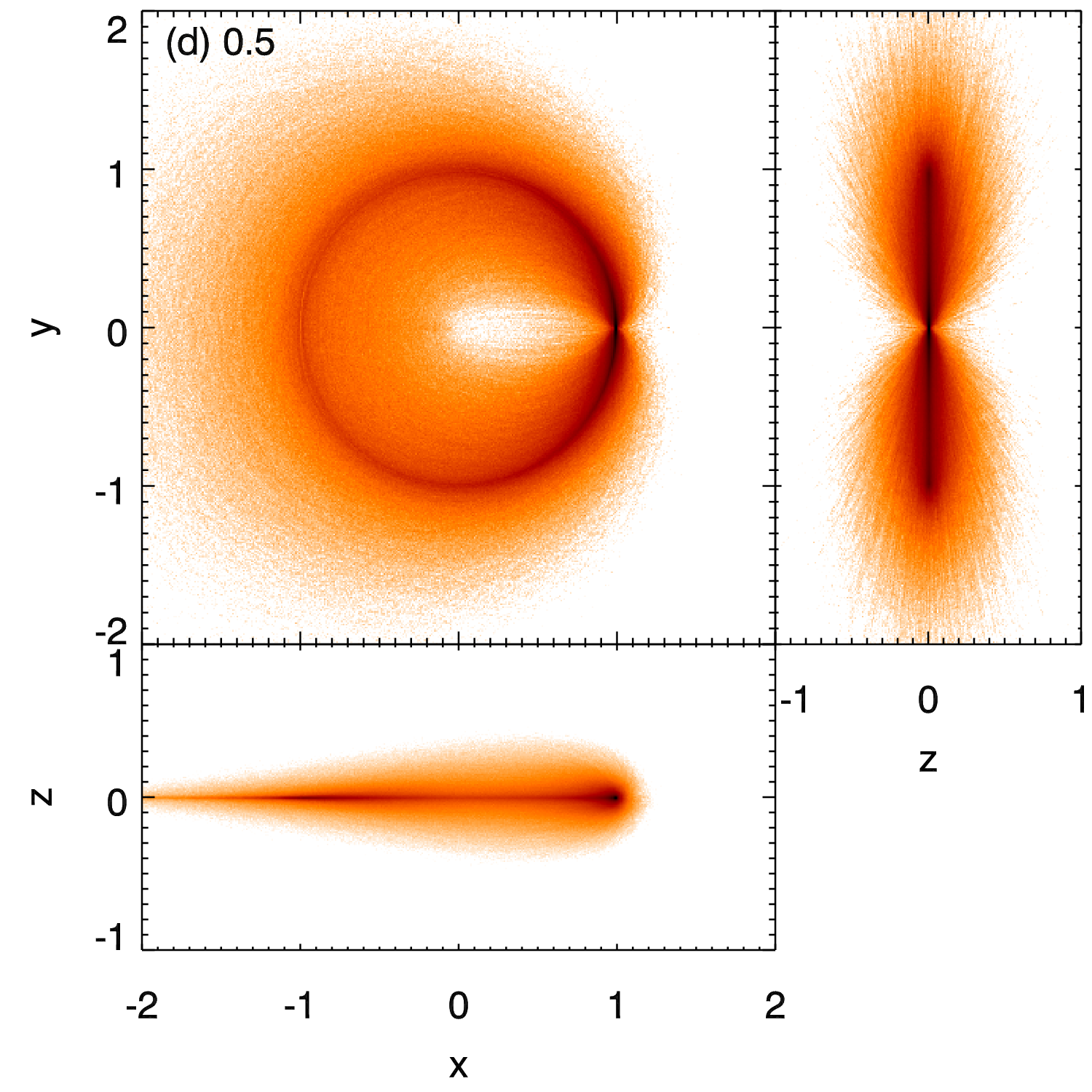}
  \includegraphics[width=0.41\textwidth]{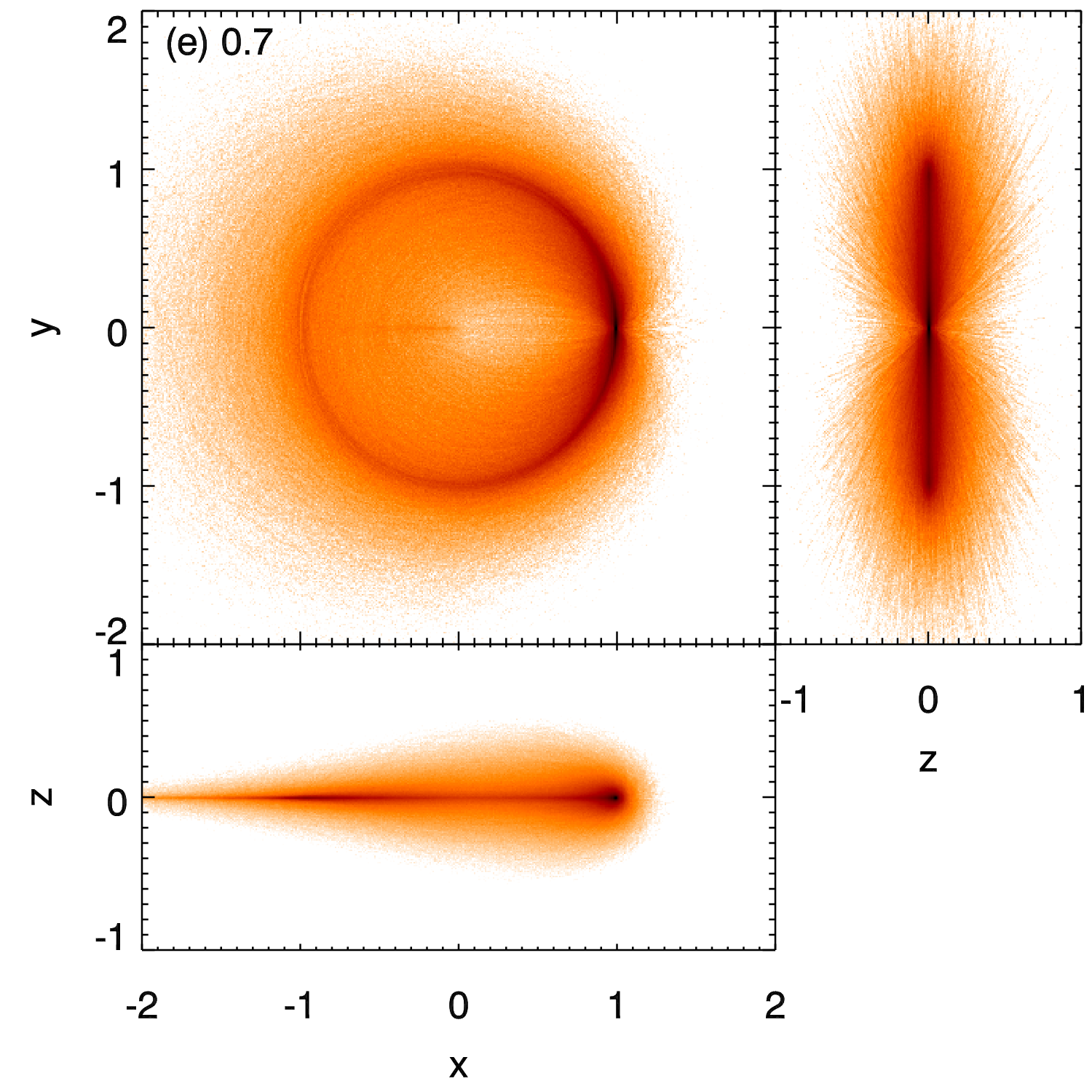}
  \includegraphics[width=0.41\textwidth]{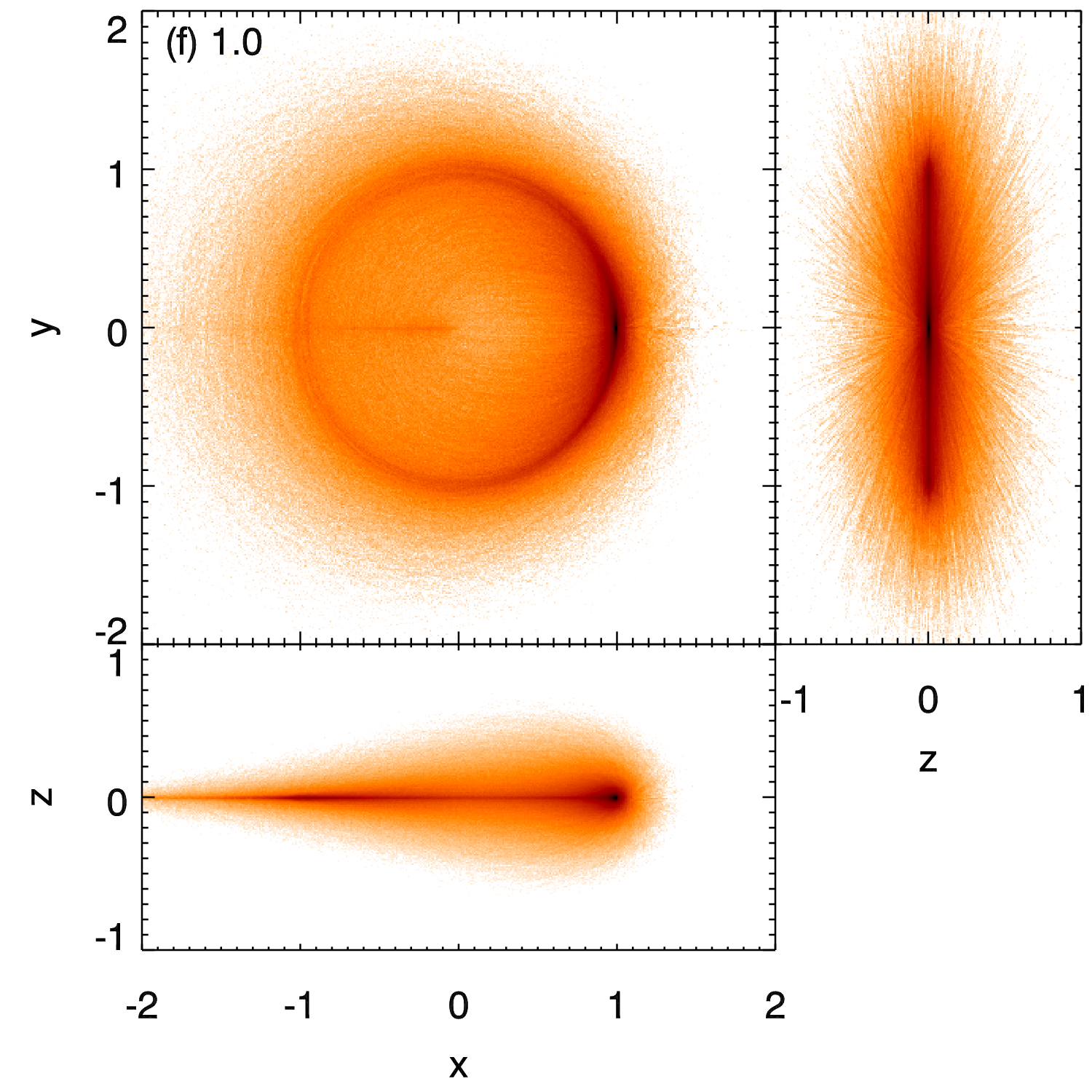}
  \includegraphics[width=0.075\textwidth]{densdvlogcolbar2.png}
  \caption{Dust density for debris produced by impacts with gaussian distributions of kick velocity of varying widths.  Each set of 3 panels shows the same disc face-on and in two edge-on configurations, as seen from the right and bottom of the face-on image respectively.   From left to right and top to bottom the triples correspond to velocity dispersions with standard deviations, $\sigma_v/v_{\rm k}=$ 0.05, 0.1, 0.3, 0.5, 0.7 and 1, as labelled.  All images are normalised to a maximum of 1.  The axes are in units of the progenitor semi-major axis, which here is 50~AU.  However discs produced at different semi-major axes will look very similar.}
  \label{fig:discimages}
\end{figure}
\end{landscape}

In Fig.~\ref{fig:discimages} we show a selection of images of collisional debris in the smooth asymmetric phase.  The resulting debris structure is almost entirely dependent on magnitude of the velocity kick (and so on the mass of the progenitor), and so we show images produced using velocity dispersions with different standard deviations, $\sigma_v$.  The velocity dispersions used ($\sigma_v/v_{\rm k}=$ 0.05, 0.1, 0.3, 0.5, 0.7 and 1), may be compared to the mass scale Ceres - Pluto - Moon - Mars - Earth, which at 50AU from the Sun corresponds to $\sigma_v/v_{\rm k}=$ 0.056, 0.13, 0.27, 0.55, and 1.24.

The images in Fig.~\ref{fig:discimages} are simple density maps that illustrate the density that one would expect in dust grains large enough not to be significantly affected by radiation pressure.  Millimetre-size and larger grains will follow these distributions, while smaller grains closer to the radiation blow-out size of $\sim$1~$\mu$m will deviate.  We discuss the distribution of small, blow-out size, grains in Section~\ref{sec:blowout}.  In observations there will be other effects on top of these density maps that will depend on the wavelength of observation and how that compares with the peak of the dust emission spectrum.  Most observational effects will tend to enhance the brightness of the dust interior to the progenitor, but generally not in a simple $1/r^2$ fashion due to wavelength dependencies.  Unlike Fig.~\ref{fig:earlyphases}, in Fig.~\ref{fig:discimages} and later figures we use a logarithmic colour scale to better bring out the structure of the disc.

Fig.~\ref{fig:discimages} has been produced by generating a cloud of 10$^5$ particles launched from their progenitor in accordance with the equations in Section~\ref{orbeq} and then following the dynamical evolution for 100 orbits, to the end of the spiral phase, including the gravitational effect of the appropriate progenitor body, with the \textsc{Mercury} N-body integrator\footnote{see \citet{chambers1999}}.  The particles are then spread around their orbits by randomisation of the mean anomaly to produce a smoother image.  This process can result in a slight streakiness around the edges of the image where the orbits are sparse.

All images in Fig.~\ref{fig:discimages} are generated using progenitor semi-major axes of 50~AU around a 1~$M_{\odot}$ star, however a disc with the same $\sigma_v/v_{\rm k}$ at a different orbital distance will look very similar.  The only influence that can change noticeably between different semi-major axes is the fractional Hill Radius of the progenitor, $R_{\rm Hill}/a$, which determines how much the collision-point is puffed out in the radial direction owing to scattering of particles during the progenitor's passage through the collision-point.  Moving from 50~AU to 10~AU or 250~AU while maintaining the same $\sigma_v/v_{\rm k}$, $R_{\rm Hill}/a$ will only change by a factor of $\sim$~$\surd 5$ which has little effect on Fig.~\ref{fig:discimages}.  Variation with stellar mass is even less important since if we maintain the same $\sigma_v/v_{\rm k}$ at the same orbital distance while varying the stellar mass $R_{\rm Hill}/a$ will only vary as $M_*^{1/6}$.

Since the collision-point will be either the ascending or descending node of the particle orbit (as determined relative to the progenitor orbit), all of the debris fragments will thus also share the same line of nodes, and on the opposite side of the star from the collision-point there will be a line we call the \emph{anti-collision line} through which all of the orbits will also pass.  The alignment of the line of nodes is responsible for the bow-tie like appearance of many of the edge-on views shown in the $y-z$ panels of Fig.~\ref{fig:discimages}, because particles that receive smaller kicks will lie closer to the progenitor orbit.  At high values of $\sigma_v/v_{\rm k}$ significant amounts of material is put onto polar orbits and the bow-tie structure is less apparent, however the anti-collision line begins to become visible at the left hand side of the face-on image as a result of the enhanced surface density.

As we move from lower to higher $\sigma_v$ in Fig.~\ref{fig:discimages} the disc becomes progressively broader at locations away from the collision-point, and more dominated by the collision-point.  At moderate to large values of $\sigma_v$ the wide breadth of the ring away from the collision-point means that the region interior to the orbit of the progenitor can be characterised by a cavity between the star and the collision-point with dust filling the rest of the region.  Above $\sigma_v=1$ the trend towards increasing dominance of the collision-point over the rest of the disk continues, but more slowly, such that it is difficult to distinguish a Neptune/super-Earth scale impact ($\sigma_v=$2.5-3) from an Earth scale impact by visual inspection, particularly in a face-on view.  The disc continues to become thicker in the $z$ direction, but again this is a slow increase.  The growing number of particles of retrograde orbits as $\sigma_v$ is increased will make collisions within the debris distribution more violent and accelerate the collisional evolution, but this does not affect the visible morphological strongly.

The increasing dominance of the collision-point as we increase $\sigma_v$ is due to the fact that as the kick velocity is increased the volume of space accessible to the debris increases, as we can see from Fig.~\ref{aedv01ecc} since the range of semi-major axes, eccentricities and inclinations all increase.  Away from the collision-point there is thus a much larger volume of space through which the debris can be spread at higher $\sigma_v$.  All of the particles must still pass through the collision-point however, and so the density contrast between the collision-point and the rest of the disc increases.

\begin{figure*}
 \begin{minipage}{\textwidth}
  \includegraphics[width=0.315\textwidth]{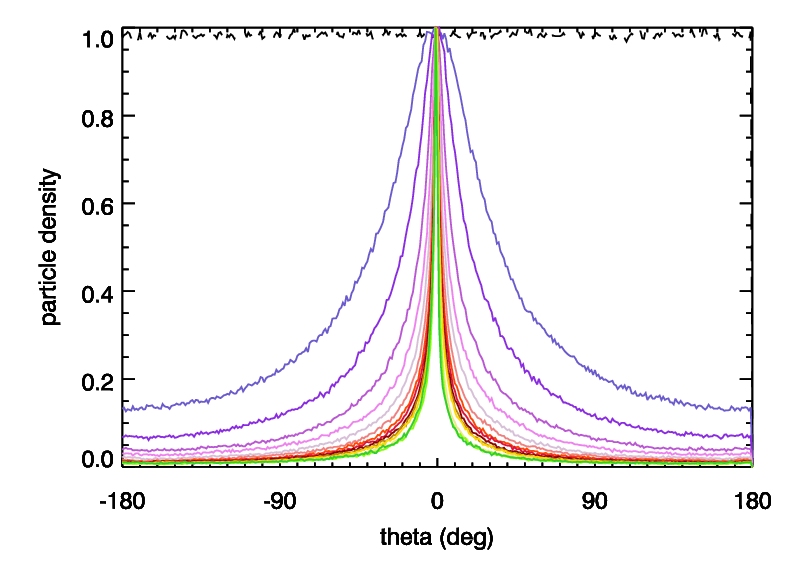}
  \includegraphics[width=0.315\textwidth]{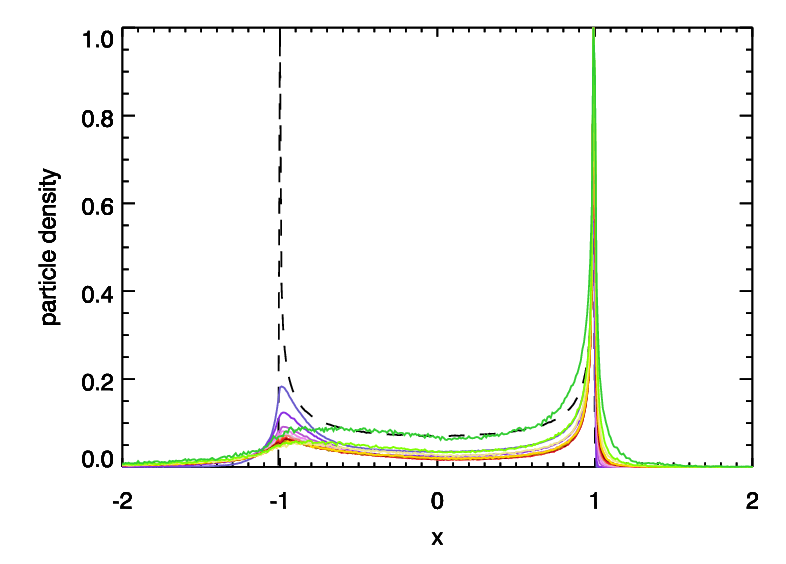}
  \includegraphics[width=0.315\textwidth]{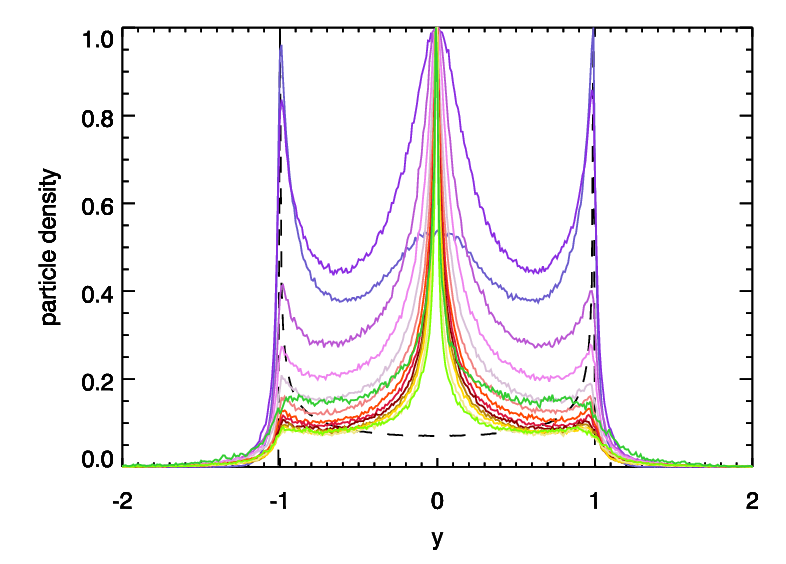}
  \includegraphics[width=0.045\textwidth]{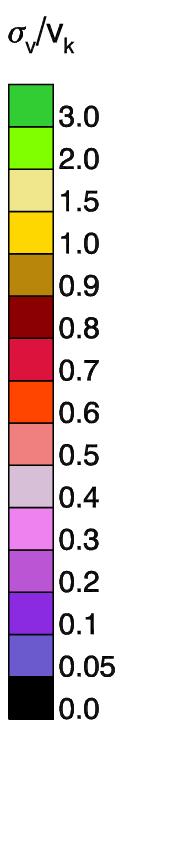}
  \includegraphics[width=0.315\textwidth]{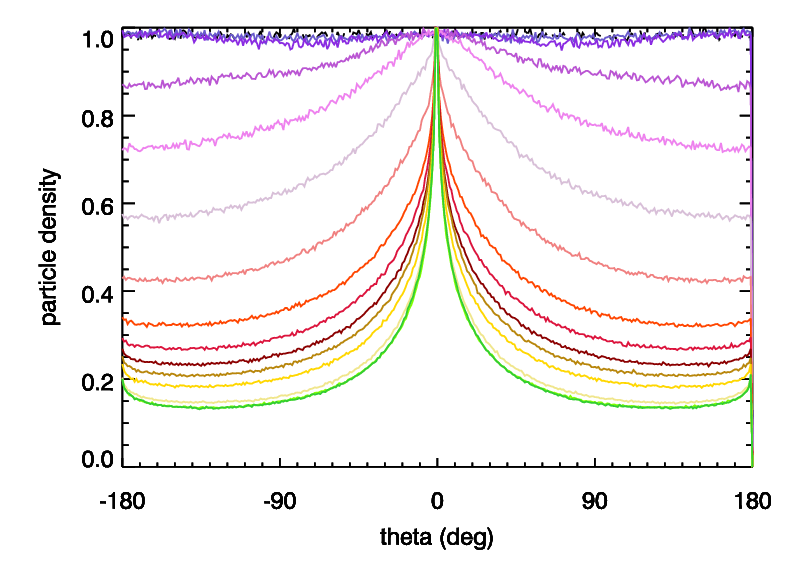}
  \includegraphics[width=0.315\textwidth]{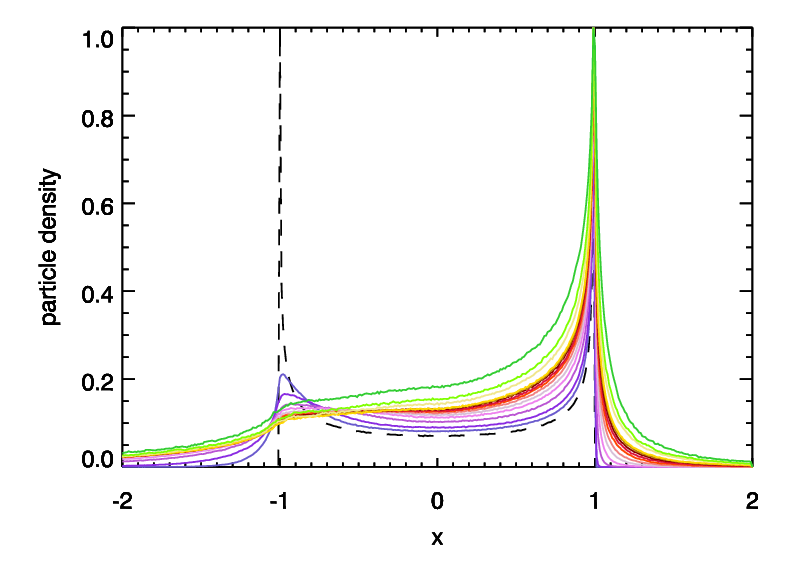}
  \includegraphics[width=0.315\textwidth]{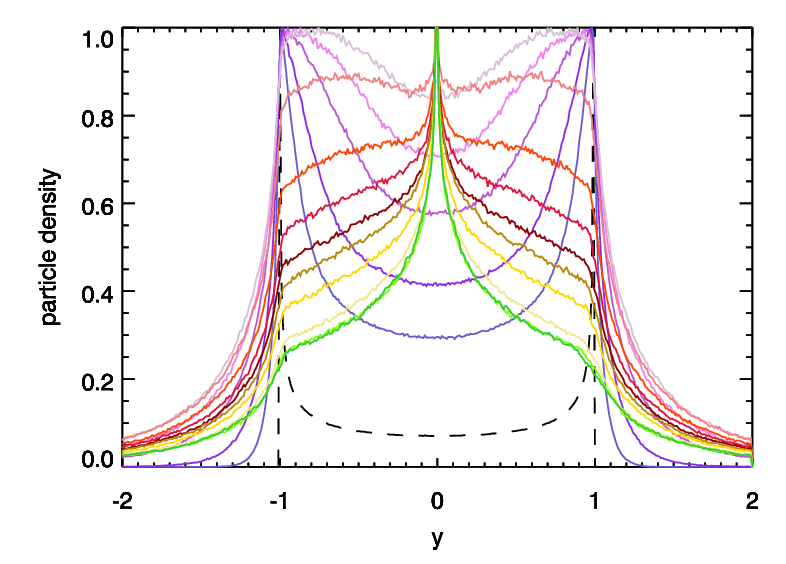}
  \includegraphics[width=0.045\textwidth]{cutscolbar2.png}
  \caption{\emph{Top row}: Particle density along line cuts through the images in Fig.~\ref{fig:discimages}.  At \emph{left} the cut is around a ring of radius 1 in the $x-y$ plane, at \emph{centre} it is along $z=0$ in the $x-z$ plane, and at \emph{right} it is along $z=0$ in the $y-z$ plane.  \emph{Bottom row}: Particle density integrated over one dimension of the images in Fig.~\ref{fig:discimages}.  At \emph{left} the integral is over radius, and at \emph{centre} and right \emph{right} it is over $z$.  Within each frame all curves are normalized to a peak value of 1.}
  \label{fig:linecuts}
 \end{minipage}
\end{figure*}

\subsubsection{Dust density variations}
\label{sec:densvar}

To supplement Fig.~\ref{fig:discimages}, in Fig.~\ref{fig:linecuts} we show how the particle density varies across the ring, both by taking the density along a line cut through the images in Fig.~\ref{fig:discimages} and by integrating over one of the dimensions, and how this changes with $\sigma_v$.  These cuts and integrals illustrate even more clearly how the collision-point becomes progressively more dominant as $\sigma_v$ is increased.  In particular the transition from a two horned profile to a three horned one and finally to a centrally peaked profile in the $y-z$ images is illustrated very clearly.  The exact shape of the density variation along a line cut will of course vary with the resolution of observations, however the integrals will be much more robust.

\begin{figure}
 \includegraphics[width=\columnwidth]{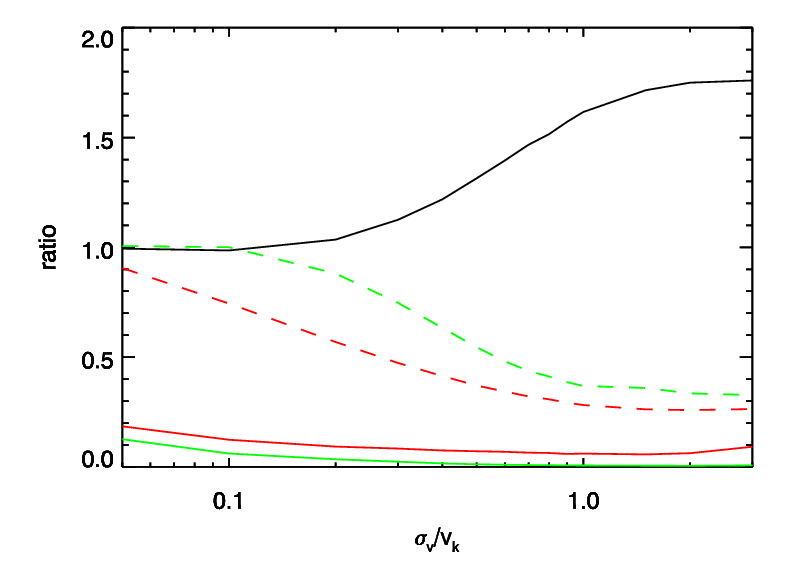}
 \caption{Ratios of different quantities from Figs.~\ref{fig:discimages} and \ref{fig:linecuts} as a function of the kick velocity.  \emph{Black}: The ratio of the integral of the right hand side of the face-on images over the left hand side ($+x/-x$).  \emph{Red}: Ratio of the ansae values in the $x-z$ plane (centre column of Fig.~\ref{fig:linecuts}), \emph{solid} the ratio of peaks from the line cut (upper frame), \emph{dashed} ratio of the sums within 0.15 of -0.95 and 0.95 in the lower frame. \emph{Green}: collision-point vs. anti-collision line ratio with respect to $\theta$ (left column of Fig.~\ref{fig:linecuts}), \emph{solid} trough-to-peak ratio in the upper frame, \emph{dashed} ratio of the sums within 20 degrees of the anti-collision line and collision-point.}
 \label{fig:discratios}
\end{figure}

In Fig.~\ref{fig:discratios} we then show how a selection of potentially observable diagnostic ratios between different parts of the disk vary with $\sigma_v$, with a view to using observations to constrain $\sigma_v$ and so the progenitor mass.  We can see that at low $\sigma_v$ the ratio between the ansae in the $x-z$ plane (red lines) fall rapidly with increasing $\sigma_v$ before flattening out at higher $\sigma_v$.  The ratio between the collision-point and anti-collision line (green lines) behaves similarly, though the ratio constructed from the integrals flattens out more slowly.  The ratio between the two halves of the disk in the $x-y$ plane (black line) however is rather flat at lower $\sigma_v$ before rising at intermediate values and then levelling off once more.

\subsubsection{Discs with eccentric progenitors}
\label{sec:obsdiscecc}

\begin{figure*}
 \begin{minipage}{\textwidth}
 \centering
  \includegraphics[width=0.41\textwidth]{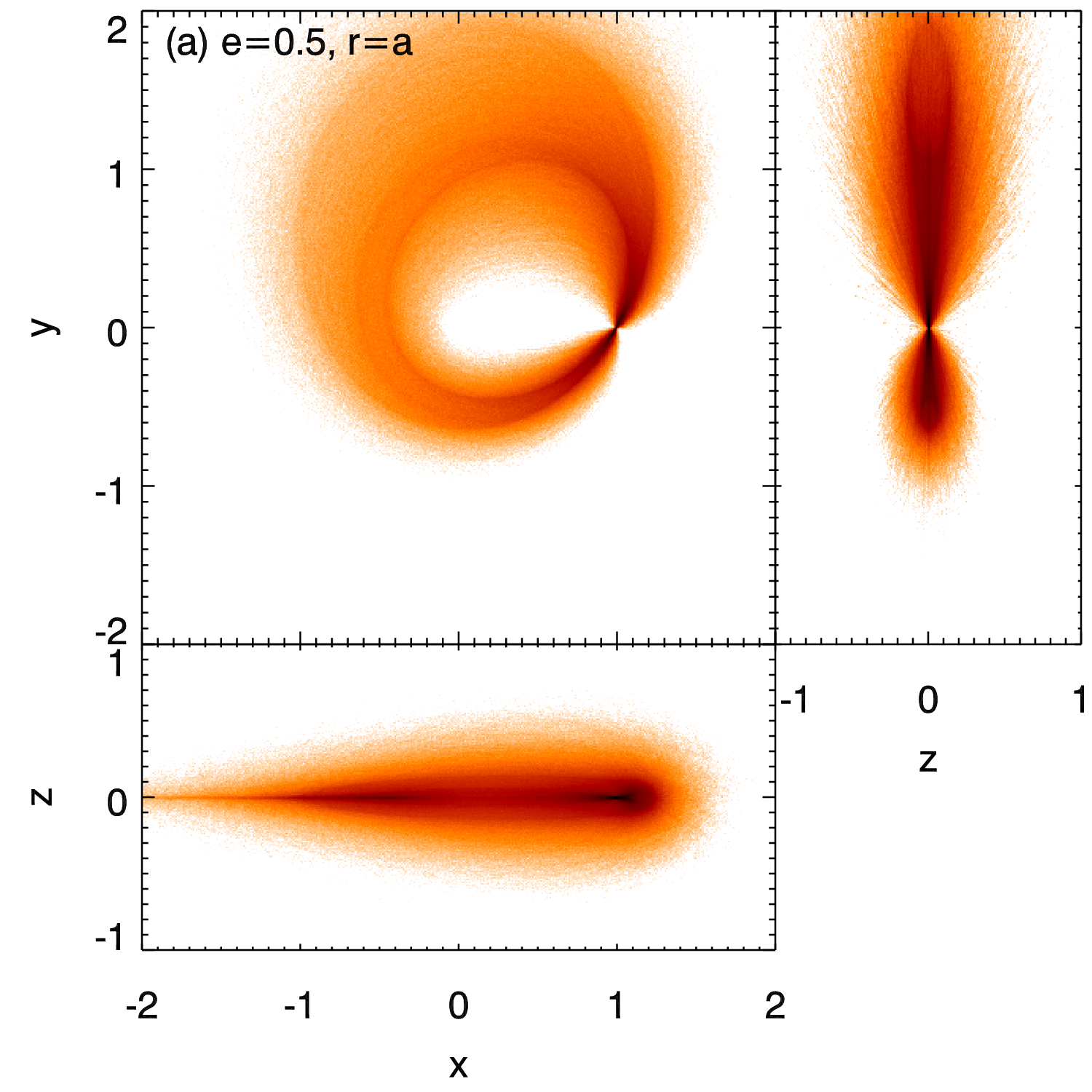}
  \includegraphics[width=0.41\textwidth]{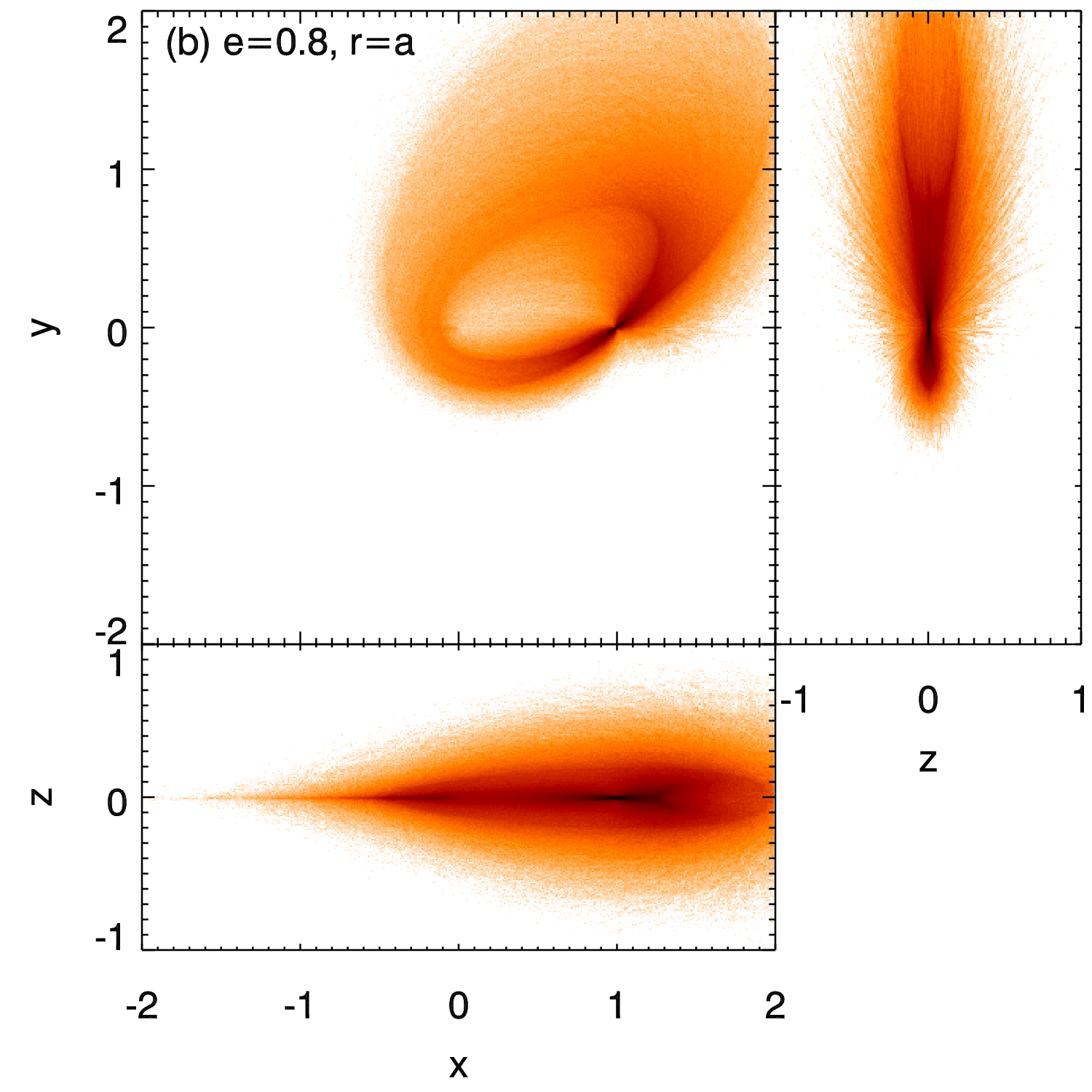}
  \includegraphics[width=0.41\textwidth]{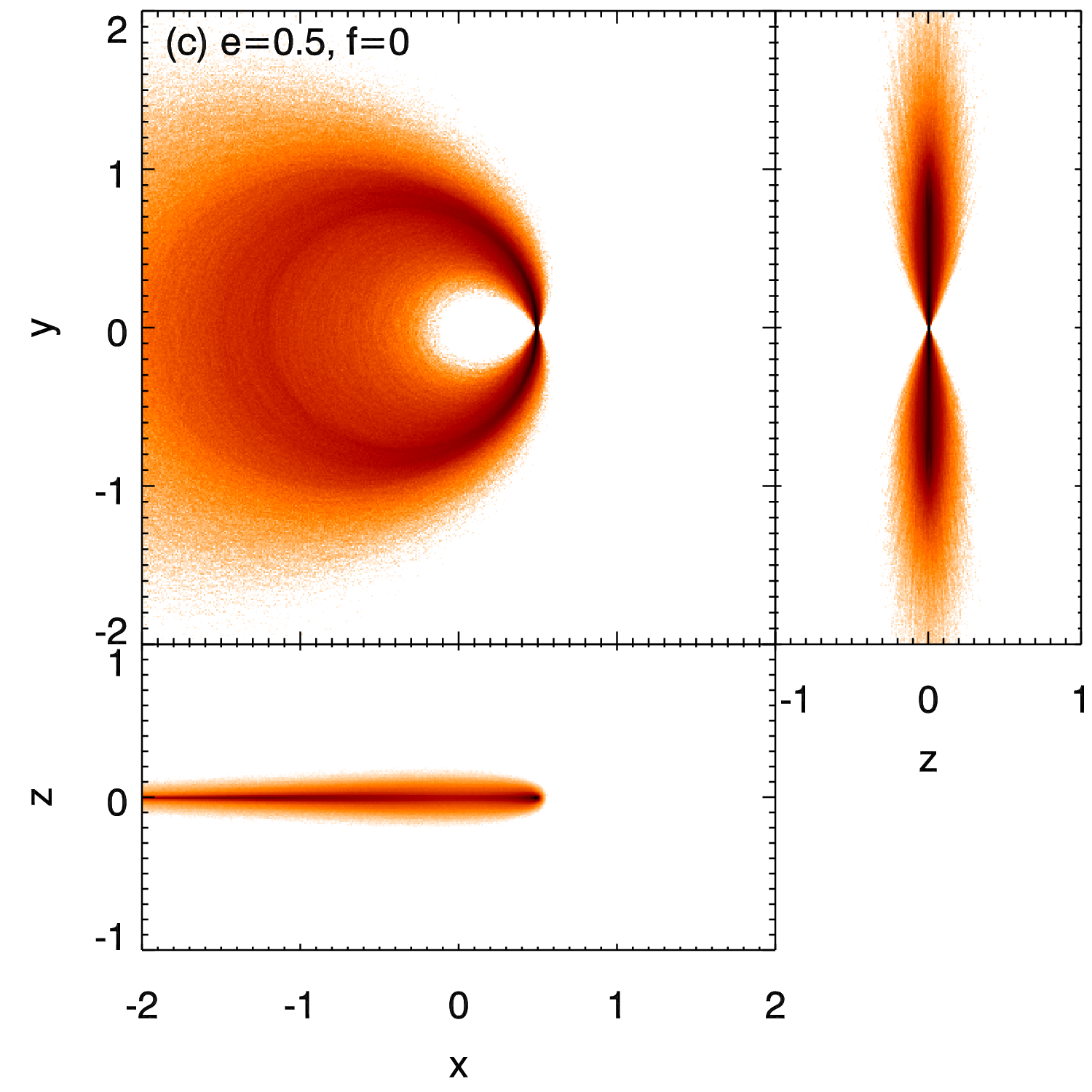}
  \includegraphics[width=0.41\textwidth]{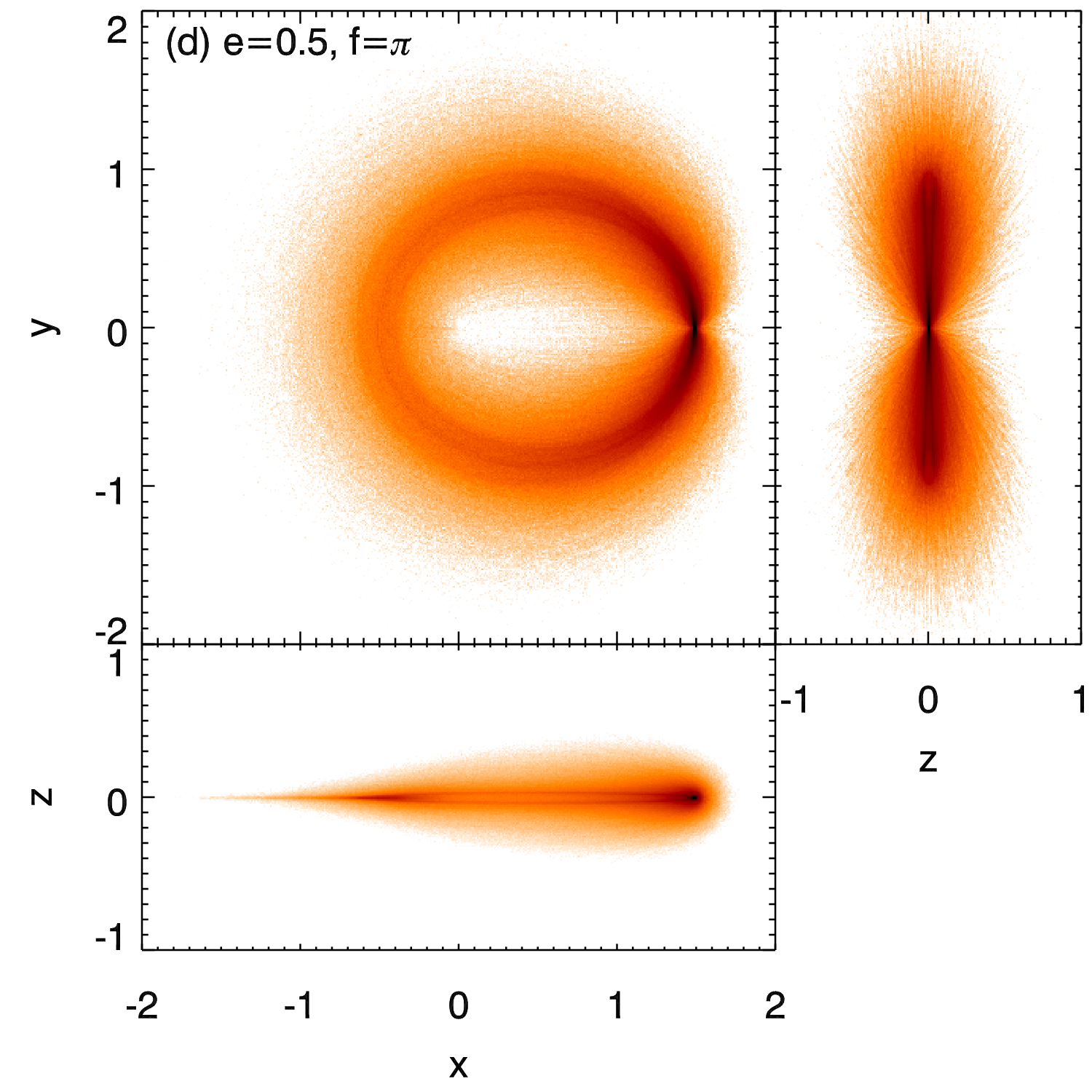}
  \includegraphics[width=\textwidth]{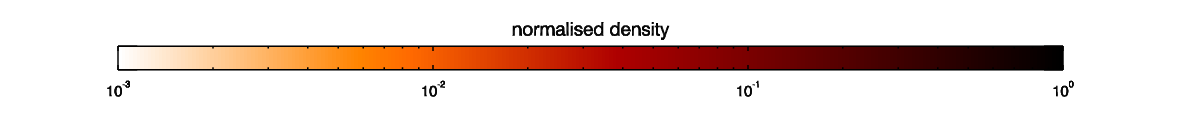}
  \caption{Dust density for debris produced by impacts involving progenitors on initially eccentric orbits.  The eccentricity of the progenitor and location of the impact on the orbit are indicated at the top left of each set of frames.  A velocity dispersion of $\sigma_v/v_{\rm k}=$ 0.3 is used in all cases.  All images are normalised to a maximum of 1.}
  \label{fig:eccdiscimages}
 \end{minipage}
\end{figure*}

All of the discs in Fig.~\ref{fig:discimages} are the result of impacts involving progenitors on circular orbits, however as we saw in Section~\ref{eccorb} the eccentricity of the progenitor can have a significant effect on the resulting distributions of orbital elements.  In Fig.~\ref{fig:eccdiscimages} we show dust density maps like those of Fig.~\ref{fig:discimages} but now for non-zero progenitor eccentricities.

The primary effect of a non-zero progenitor eccentricity is to introduce additional sources of asymmetry in the disc. Firstly the disc is now centred around the elliptical orbit of the progenitor.  Additionally, since particles spend more time near apocentre than near pericentre, the dust density is also enhanced near the apocentre of the progenitor orbit.  This additional asymmetry increases in strength as the progenitor eccentricity is increased and interacts with the asymmetry due to the collision-point.

Since the collision-point asymmetry results in higher dust density near the collision-point, and the eccentric progenitor asymmetry results in higher dust density near the apocentre of the progenitor orbit, the interaction between the two can be either constructive or destructive depending on whether the impact occurs nearer to the pericentre or apocentre of the progenitor orbit.  The extreme cases are the impact occuring exactly at pericentre or apocentre as we can see in the lower row of Fig.~\ref{fig:eccdiscimages} where we can see that the density asymmetry is much stronger for the impact occuring at apocentre than at pericentre.  The opposite is true for the asymmetry in how dispersed the disc material is at a given location, since material is dispersed over a wider range of radii at the apocentre of the progenitor orbit, and opposite the collision-point.  We can also see that in the general case where the impact does not occur at the apocentre or pericentre that the bow-tie shape of the disc when viewed down the line of nodes is now lopsided rather than symmetrical.

In addition to introducing further asymmetry into the disc, since the kick velocity at which material can be ejected depends on the true anomaly at which it is kicked (equations \ref{dvminecc} and \ref{dvminecc2}) the amount of debris material that is lost immediately after the impact depends on the true anomaly at which it occurs.  For example in Fig.~\ref{fig:eccdiscimages} in the bottom right panel just under 1 per cent of material is placed onto hyperbolic orbits, whereas in the bottom left panel over 7 per cent achieves escape.

\subsection{Detectability of the dust}
\label{sec:detectability}

In addition to considering the lifetime of a disk in the asymmetric state we must also consider the brightness of the dust emission, and its detectability.  The brightness of the dust emission is determined by several factors, and will vary depending on the wavelength at which we observe it.  If we consider simply the bolometric luminosity of the dust to keep the situation as simple as possible the most important factors are the mass of debris in the disk, its orbital radius, and the size distribution of the debris.  The characteristic orbital radius of the disk is simply set by the orbital distance of the progenitor body.

The mass of debris in the disk is also dependent on the progenitor.  At the most basic level the mass of debris clearly cannot exceed the mass of the progenitor.   Completely pulverising a body requires extremely energetic impacts with impact velocities many times the escape velocity of the body \citep[e.g.][]{leinhardt2012,stewart2012}.  When we are considering massive bodies at large orbital radii the orbital velocities are smaller than the escape velocity of the body, rendering such violent impacts impossible.  While small bodies may be able to participate in catastrophic impacts and shed large fractions of their mass as debris, larger ones will be restricted to less violent regimes, with commensurately lower debris production.  For the hit and run and partial accretion/erosion regimes of \citet{leinhardt2012,stewart2012}, within which we would expect impacts involving large bodies to fall, the typical debris production is around 3-5 per cent of the colliding mass (S. Stewart, private communication).  For massive bodies we thus make a relatively conservative estimate that the debris mass is 3 per cent of the mass of the progenitor, and would expect typical variations of around a factor of 2.

The final factor in determining the brightness of the dust, the size distribution, is the least well constrained.  Considering a differential size distribution $n(D)dD \propto D^{-\alpha}dD$ \citet{leinhardt2012} find that $\alpha=3.85$ provides the best fit to their simulations.  However there is a considerable degree of uncertainty in this, particularly for the low energy impacts that are most likely for massive bodies at large orbital distances.  A size distribution with $\alpha=3.5$ is expected to occur in a steady state, self-similar, collisional cascade \citep*[][]{dohnanyi1969,tanaka1996}, which is similar to the value found by \citet{leinhardt2012} and has other useful properties, in particular with consideration to the future evolution of the disc brightness.  As such we adopt a size distribution with $\alpha=3.5$ here.  Having adopted a slope for the size distribution we must also set the upper and lower bounds of the distribution.  The lower bound will be set by the removal of small dust grains by radiation pressure.  The blow-out size, $D_{\rm bl}$, is given by $D_{\rm bl}=0.8 (L_*/L_{\odot})(M_{\odot}/M_*)(2700 {\rm kg m^{-3}}/\rho) \mu$m \citep{wyatt2008} and will be around a micron in most cases (it can be significantly different for collisions occuring predominantly at the periapse of an eccentric orbit, e.g. \citealt{wyatt2010}).  The upper bound is much more difficult to constrain and so we consider a wide range of possible values.

\begin{figure}
 \includegraphics[width=\columnwidth]{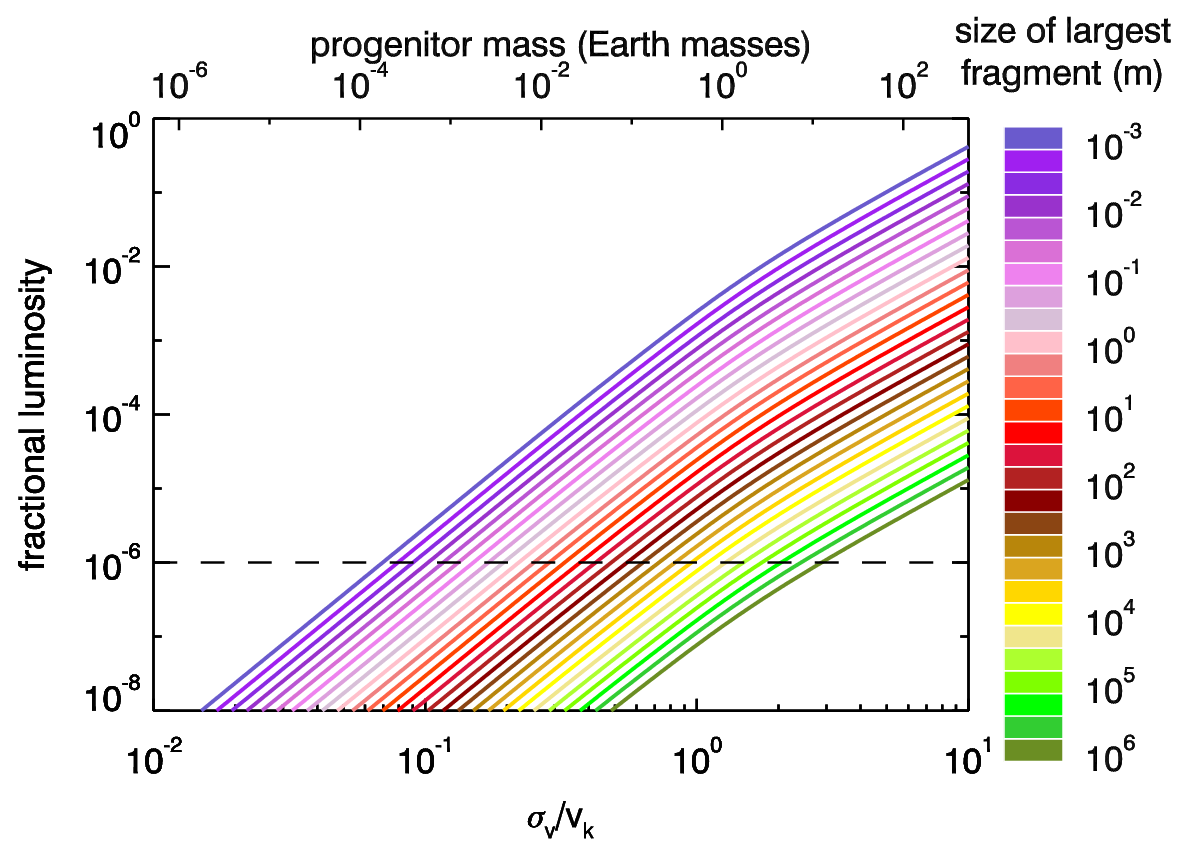}
 \caption{Initial fractional luminosity of impact debris versus $\sigma_v$ and progenitor mass for an impact occuring at 50AU from a Sun-like star assuming 3 per cent of the progenitor mass is released as debris.  The black dotted line represents the approximate detection limit.  The relationship between $\sigma_v$ and progenitor mass assumes a constant density equal to that of Earth.}
 \label{fig:fraclum}
\end{figure}

For a size distribution with power law slope of $\alpha=3.5$ the fractional luminosity of the debris is related to the mass of the debris by
\begin{equation}
 f=0.37r^{-2}D_{\rm bl}^{\frac{1}{2}}D_{\rm max}^{\frac{1}{2}}M_{\rm tot}
 \label{eq:fraclum}
\end{equation}
where $D_{\rm max}$ is the size of the largest objects in km, $r$ is the radius of the disk in AU, $D_{\rm bl}$ is the radiation blow-out size in $\mu$m and $M_{\rm tot}$ is the mass of the debris in $M_{\oplus}$ \citep{wyatt2008}.  In Fig.~\ref{fig:fraclum} we use this to show how the initial fractional luminosity varies with different values of $\sigma_v$ and $D_{\rm max}$ for an impact at 50AU from a sun-like star.  As we increase the progenitor mass, then if a fixed percentage of the progenitor mass is released as debris the debris mass and thus fractional luminosity also increases.  Through the expected proportionality of $\sigma_v$ to the escape velocity we can also relate the fractional luminosity to $\sigma_v$.  Here we assume that the density of the progenitor remains constant at that of Earth in order to maintain a simple relation between $\sigma_v$ and progenitor mass ($\sigma_v \propto M_{\rm prog}^{1/3}$), but note that in general we expect the density to increase as the mass increases, such that $\sigma_v$ should increase slightly faster with mass than our $M_{\rm prog}^{1/3}$.

Although the mass released as debris is kept as a fixed percentage of the progenitor mass in Fig.~\ref{fig:fraclum}, the debris mass that is retained in the disk increases more slowly at higher $\sigma_v$.  This is because at higher kick velocities a larger fraction of particles are lost (see Fig.~\ref{fig:dvfraclost}), and so in broader velocity distributions that contain a larger fraction of high velocity particles more material will be ejected and not form part of the disk.  This is the cause of the slight curvature of the lines above $\sigma_v/v_{\rm k} \approx 0.5$.

Note that for Fig.~\ref{fig:fraclum} we assume that the $r$ in Eq.~\ref{eq:fraclum} is the semi-major axis of the progenitor (here 50AU).  Since the disc is radially broad away from the collision-point, especially for higher values of $\sigma_v$, we can expect that this will not be quite right because of the extra contribution from material closer to the star.  This can increase the brightness by up to 20-30 per cent, but the effect on Fig.~\ref{fig:fraclum} is fairly small.

As we decrease the size of the largest objects in the distribution the surface area of the debris increases, and thus so does the fractional luminosity.  A reasonable estimate for the level at which a cold debris disk can be detected is a fractional luminosity of $10^{-6}$, shown by the dotted black line in Fig.~\ref{fig:fraclum} \citep[e.g.][]{eiroa2013}.  We can thus compare the fractional luminosity curves to this detection level and determine whether we would be able to detect the debris from a giant impact involving a progenitor of a given mass for a certain size of the largest objects in the resulting debris.  For a Lunar mass progenitor (0.012$M_{\oplus}$) the largest objects would need to be smaller that 10m in size, while for an Earth mass progenitor the largest objects can be up to 10~km size.   For an object less massive than Ceres ($\sim10^{21}$kg) even if the largest objects were mm-size the debris would not be detectable at the 3 per cent release level.

Although large bodies will only be able to take part in low energy impacts, small bodies may be able to participate in more violent encounters.  This would release larger fractions of the progenitor mass as debris and raise all of the curves in Fig.~\ref{fig:fraclum}, potentially by a factor of up to 30 (corresponding to total destruction).  Impacts involving small bodies may thus still be detectable, but only if they are highly destructive (i.e.~a large fraction of the progenitor is converted into debris).  Since the width of the velocity dispersion is proportional to the escape velocity of the progenitor we can suggest that destructive impacts are possible for $\sigma_v/v_{\rm k}\la0.1$.  At larger values of $\sigma_v/v_{\rm k}$ we expect smaller variations in the height of the curves in Fig.~\ref{fig:fraclum}, typically around a factor of 2.

While the size of the largest objects in the size distribution is a poorly constrained parameter, we may reasonably expect that it decreases as we decrease the size of the progenitor.  As such we can expect that rather than following a single line in Fig.~\ref{fig:fraclum} we should move up the lines as we move to smaller progenitor masses, meaning that the fractional luminosity will vary more shallowly with progenitor mass and $\sigma_v$.

Finally we should note that, as we have hinted earlier, the quantity described in Fig.~\ref{fig:fraclum} is the \emph{initial} fractional luminosity, and this will subsequently evolve with time.  After their initial production the debris fragments will continue to collide with one another and shatter, gradually grinding down until they reach the blow-out size and are removed.  The mass of the disk and its fractional luminosity will thus diminish over time.  A feature of the size distribution we have adopted is that the timescale of the collisional evolution is determined by the size of the largest objects.  We discuss the collisional evolution in detail in Section~\ref{sec:collevol}.

\section{Collisional Evolution}
\label{sec:collevol}

As we noted earlier, after the debris is released in the generating giant impact it will then continue to evolve through mutual collisions within the disc of debris as well as dynamically through interactions with the gravity of massive bodies in the system.  If we follow an individual debris fragment this fragment will experience collisions with other members of the debris distribution at a rate
\begin{equation}
 R_{\rm col}=n \sigma v_{\rm rel},
\end{equation}
where $\sigma$ is the cross-sectional area of the fragment for collision (which may be larger than the physical cross-sectional area due to gravitational focussing), and $n$ and $v_{\rm rel}$ are the \emph{locally} calculated number density of fragments and their relative velocity with respect to the fragment we are following.  That $n$ and $v_{\rm rel}$ are locally calculated is of key importance.  In any disc we can expect that there will be radial gradients in the disc properties such that the collision rate may vary with orbital distance, however in the case of a highly asymmetric disc like those we are dealing with here we must also account for azimuthal variations in the disc quantities.

\begin{figure}
 \includegraphics[width=\columnwidth]{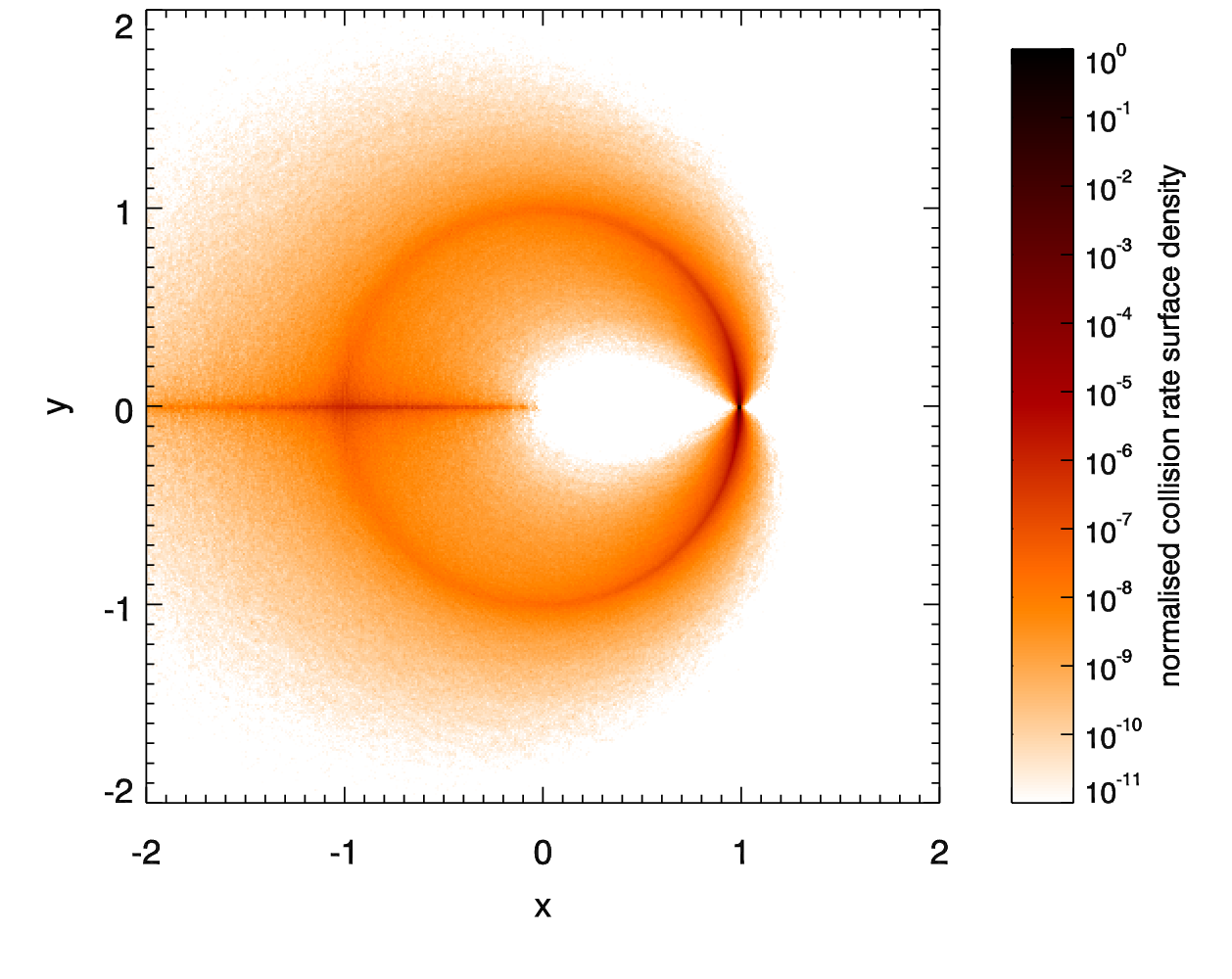}
 \caption{Image showing the normalised rate of collisions within a disk produced by an impact with $\sigma_v/v_{\rm k}=0.3$. Produced using the same data as the bottom left panel of Fig.~\ref{fig:discimages}}
 \label{fig:collmap}
\end{figure}

In Fig.~\ref{fig:collmap} we show a map of the collision rates across the disc for debris produced by an impact with $\sigma_v/v_{\rm k}=0.3$.  To construct this map the collision rate has been calculated individually for each particle in the image by determining the density and relative velocity in the immediate vicinity of the particle, the image is then a two-dimensional histogram weighted by the individual particle collision rates and thus represents the collision rate surface density.

By comparing with its counterpart in Fig.~\ref{fig:discimages}(c) we can see that the surface density of collisions largely follows the particle surface density, but with a much stronger variation.  This is as would be expected since the collision rate for each individual particle varies in proportion to the density, so the density of collisions should vary roughly as the square of the particle density.  The collision-point and anti-collision line in particular are much more prominent because of the substantially increased density at these locations.  The effect for the anti-collision line is particularly noticeable considering that it could not be seen at all in Fig.~\ref{fig:discimages}(c).  This is because the particle surface density (which is what is shown by Fig.~\ref{fig:discimages}) does not increase significantly at the anti-collision line, since the disc is thinner vertically, but nonetheless the \emph{volume} density does.  At the collision-point on the other-hand both the surface and volume densities increase substantially.  In addition the collision-point and anti-collision line receive a further minor enhancement due to the fact that as all of the particle orbits cross here, the relative velocities of the particles are also higher.  Our naming of the collision-point intentionally foreshadowed its dominant role in the collisional evolution of the disc.

When integrated over the whole of the disc the effect of the asymmetry is to dramatically increase the collision rate by several orders of magnitude over the collision rate that would be calculated for an axisymmetric disc.  As a result the disc will evolve significantly faster collisionally than an axisymmetric disc would be expected to.

\subsection{Evolving the cascade}
\label{sec:collevol2}

\begin{figure*}
\begin{minipage}{\textwidth}
 \includegraphics[width=\textwidth]{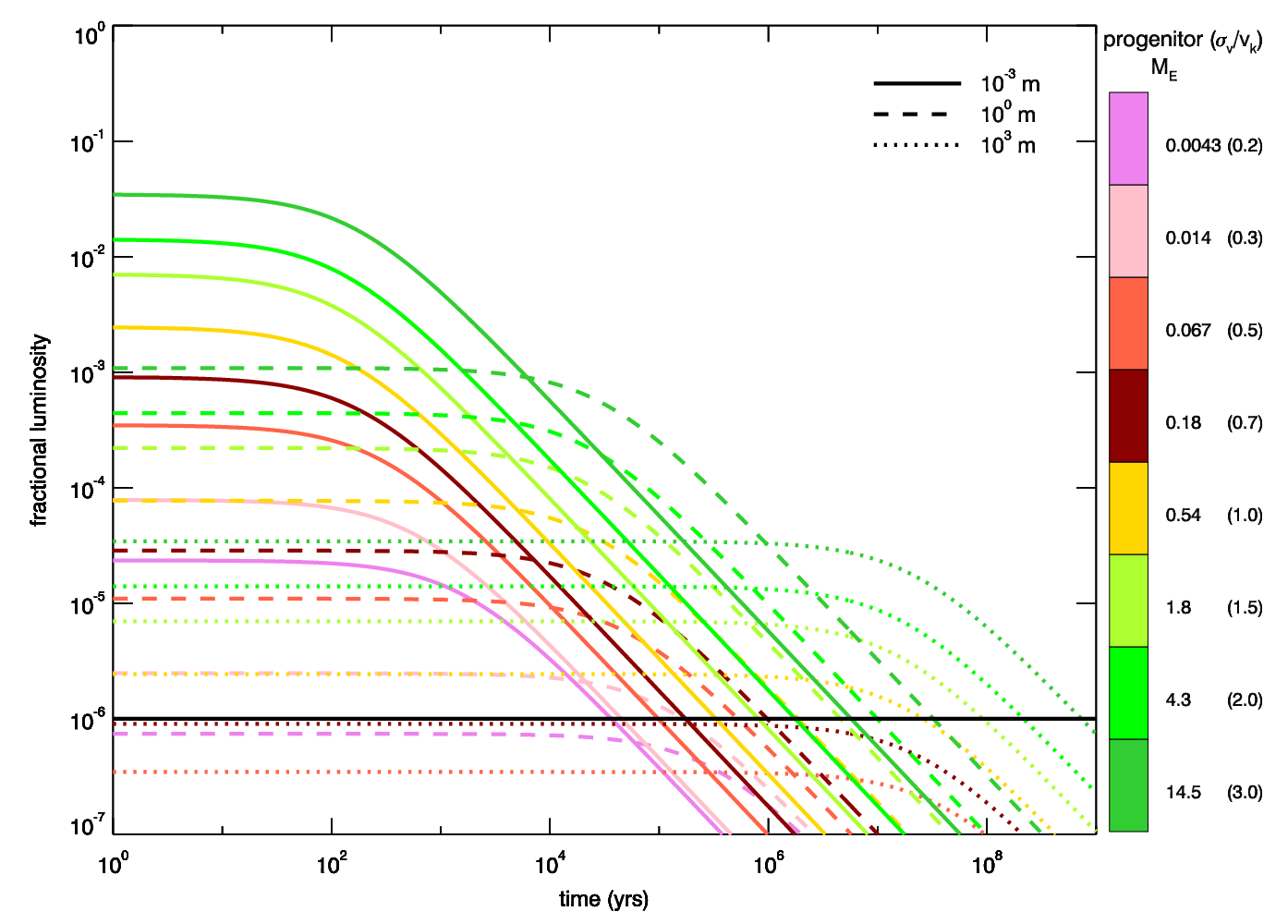}
 \caption{Temporal evolution of the fractional luminosity of a disc produced by a giant impact occuring at 50AU from a Sun-like star, releasing debris equal to 3 per cent of the progenitor mass, for a selection of velocity dispersions and largest fragment sizes.  The black line indicates the approximate detection limit.  The colour scale gives the mass of the progenitor (in $M_{\oplus}$), with the corresponding value of $\sigma_v/v_{\rm k}$ in parentheses.  The line style corresponds to the value of $D_{\rm max}$, as indicated.}
 \label{fig:fracevol}
 \end{minipage}
\end{figure*}

Now that we know what the collision rate is within the disc we can proceed to evolve the mass of the collisional cascade, and thus its luminosity.  One of the advantages of the $n(D)dD \propto D^{-3.5}dD$ size distribution that we have adopted is that the mass of the cascade is dominated by the largest objects in the distribution.  While the surface area, and thus luminosity, of the disc is dominated by the smallest objects it is the break-up of the largest objects that replenishes the supply of small dust and so the lifetime of the largest objects sets the evolution timescale of the whole disc.

To determine the lifetime of the largest objects we need to know $D_{cc}(D_{\rm max})$, the size of the smallest object that is capable of colliding catastrophically with an object of size $D_{\rm max}$.  To calculate this we utilise the velocity dependent dispersal threshold of \citet{stewart2009}.  The rate at which the largest objects experience catastrophic collisions is then $R_{cc}=n\sigma_{cc}(D_{\rm max})v_{\rm rel}$ where $\sigma_{cc}(D_{\rm max})$ is the catastrophic collision cross-section for an object of size $D_{\rm max}$, given by
\begin{equation}
\sigma_{cc}=\int_{D_{cc}(D_{\rm{max}})}^{D_{\rm{max}}} n(D) \left( \frac{D_{\rm{max}}+D}{2} \right) ^2 dD,
\label{eq:sigmacc}
\end{equation}
neglecting gravitational focussing.  The lifetime of the largest objects, $\tau$, is then simply $1/R_{cc}$, and will increase over time as the mass of the debris distribution decreases and thus the normalisation of $n(D)$ falls. The mass of the disc then evolves as
\begin{equation}
 m(t+\delta t)=m(t)\frac{1}{1+\delta t /\tau}.
 \label{eq:massevol}
\end{equation}

Solving these equations for a selection of velocity dispersions/initial disc masses, and sizes of the largest debris fragments, we obtain Fig.~\ref{fig:fracevol}.   This shows us that provided that the disc is initially detectable it will in general remain so for at least $10^5$yr at 50AU and typically for the whole $\sim10^6$yr duration of the asymmetric phase.  In order to get a single curve for the evolution of each distribution we calculate, we integrate the collision rate around each particle orbit and then average over all particles to obtain a mean value of $R_{cc}$.  Note that we neglect the effects of dynamical evolution since we are focussed on a single dynamical phase and during the asymmetric phase dynamical effects such as re-accretion will have a negligible effect on the disc mass aside from fragments put onto initially hyperbolic orbits, which are already accounted for.  As such the luminosity evolution in Fig.~\ref{fig:fracevol} beyond around $10^6$yr will be faster than in reality, since once the disc has been symmetrised the rate of collisional evolution will slow down.  During the period that the disc remains asymmetric however the evolution in Fig.~\ref{fig:fracevol} will be accurate.

Note that the collisional model we use, assuming that the mass flux is given by the ratio of the total mass and the catastrophic collision timescale of the largest objects (Eq.~\ref{eq:massevol}), is an approximation.  A more complex model, such as that of \citet{kobayashi2010}, which explicitly accounts for a distribution of collision outcomes integrated across the whole collisional cascade, may produce a mass flux that is higher than our simple model.  Compared to a model such as that of \citet{kobayashi2010} we may thus slightly overestimate the detectable lifetime of the disc for a specific value of $D_{\rm max}$.

In addition, as stated in Section~\ref{sec:detectability}, our size distribution slope of $\alpha=3.5$ is also an approximation.  Furthermore, although $\alpha=3.5$ is the slope of an idealised self-similar cascade, in reality variation of the strength of debris objects with size, and removal of small dust grains by radiation forces, can cause deviations from this \citep*[e.g.][]{wyatt2011}.

In light of the orders of magnitude uncertainty in the value of $D_{\rm max}$, and the uncertainties in the impact energies required for catastrophic collisions ($Q_D^*$), the errors introduced by the approximations in our simple model are comparatively small.  As such, while the limitations of our simple collisional model should be borne in mind, for the purposes of this study its utility and ease of understanding outweighs its limitations.

\subsection{Blow-out grains}
\label{sec:blowout}

In general it is assumed that within the main debris disc the disc luminosity is dominated by bound grains.  Dust grains with sizes of $\la$1$\mu$m are strongly influenced by radiation pressure and thus do not follow the same orbits as the larger grains from which they are produced.  These grains are put onto highly eccentric, or hyperbolic, orbits and thus are swept outwards from the main disc where they can dominate the luminosty at very large distances as a `halo'.  A number of bright debris discs have been observed to have such extended haloes \citep[e.g.][]{kalas2005b,fitzgerald2007,kalas2013}, which are most readily detectable at short wavelengths due to the small sizes of the dust grains.  In addition to these broad haloes it is also possible for unbound grains to contribute significantly to the emission of the main disc if they are produced at a high enough rate.

The strength of the radiation pressure force on a dust grain can be quantified by the parameter $\beta$, which is the ratio of the radiation pressure and the stellar gravity.  Since both radiation pressure and gravity fall off as $r^{-2}$ the value of $\beta$ is constant for a given dust grain regardless of its distance from the central star and we can think of the effect of a non-zero $\beta$ as being to modify the effective mass of the star as seen by the dust grain to $(1-\beta)M_*$.  Any particle with a $\beta>1$ will thus always be blown out, while dust grains originating from a parent population on a circular orbit will be removed if $\beta > 0.5$.  Dust grains originating from eccentric orbits however can both be removed at lower values of $\beta$, and remain bound at higher values, depending on where around the eccentric orbit they are produced.

\begin{figure}
 \includegraphics[width=\columnwidth]{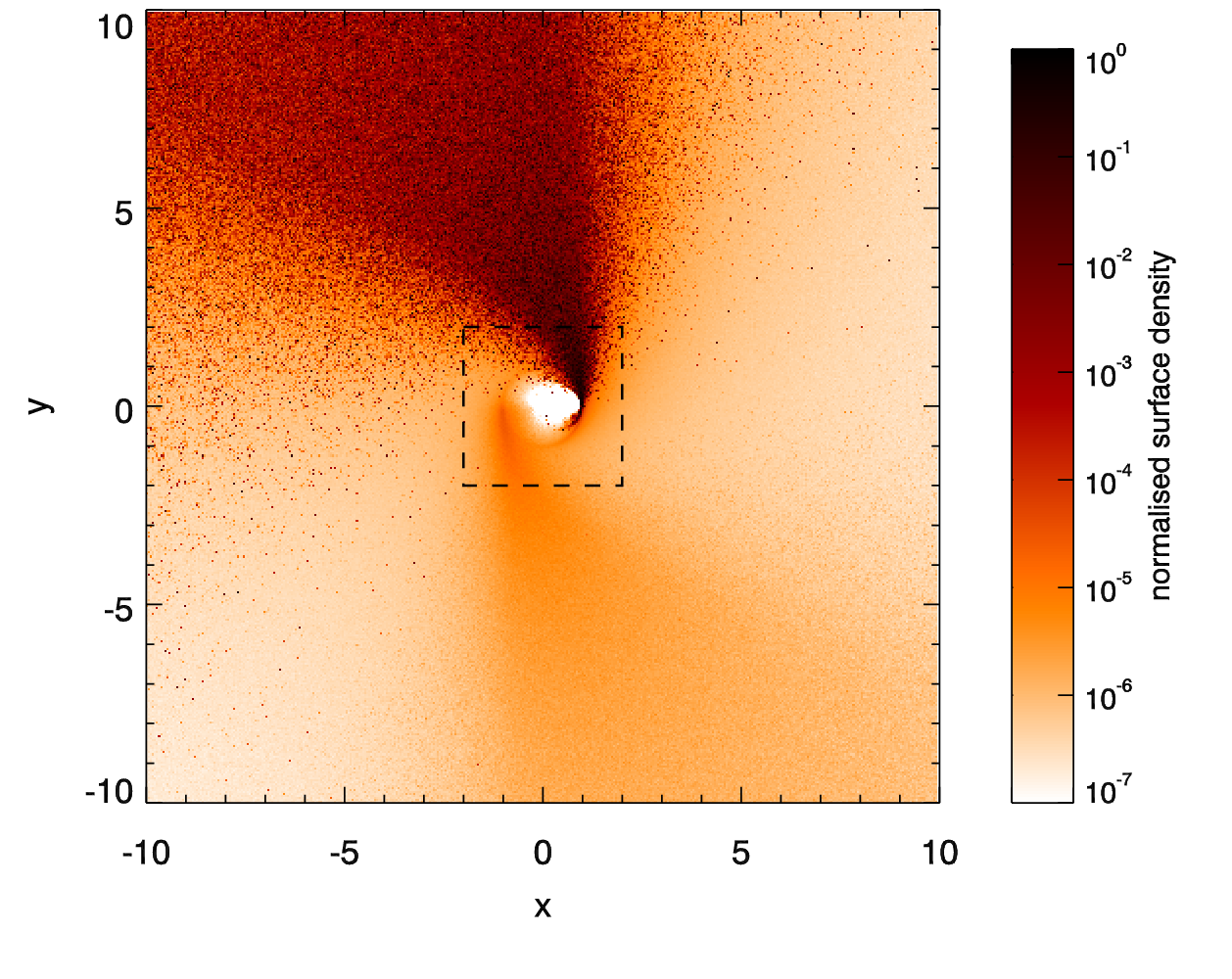}
 \includegraphics[width=\columnwidth]{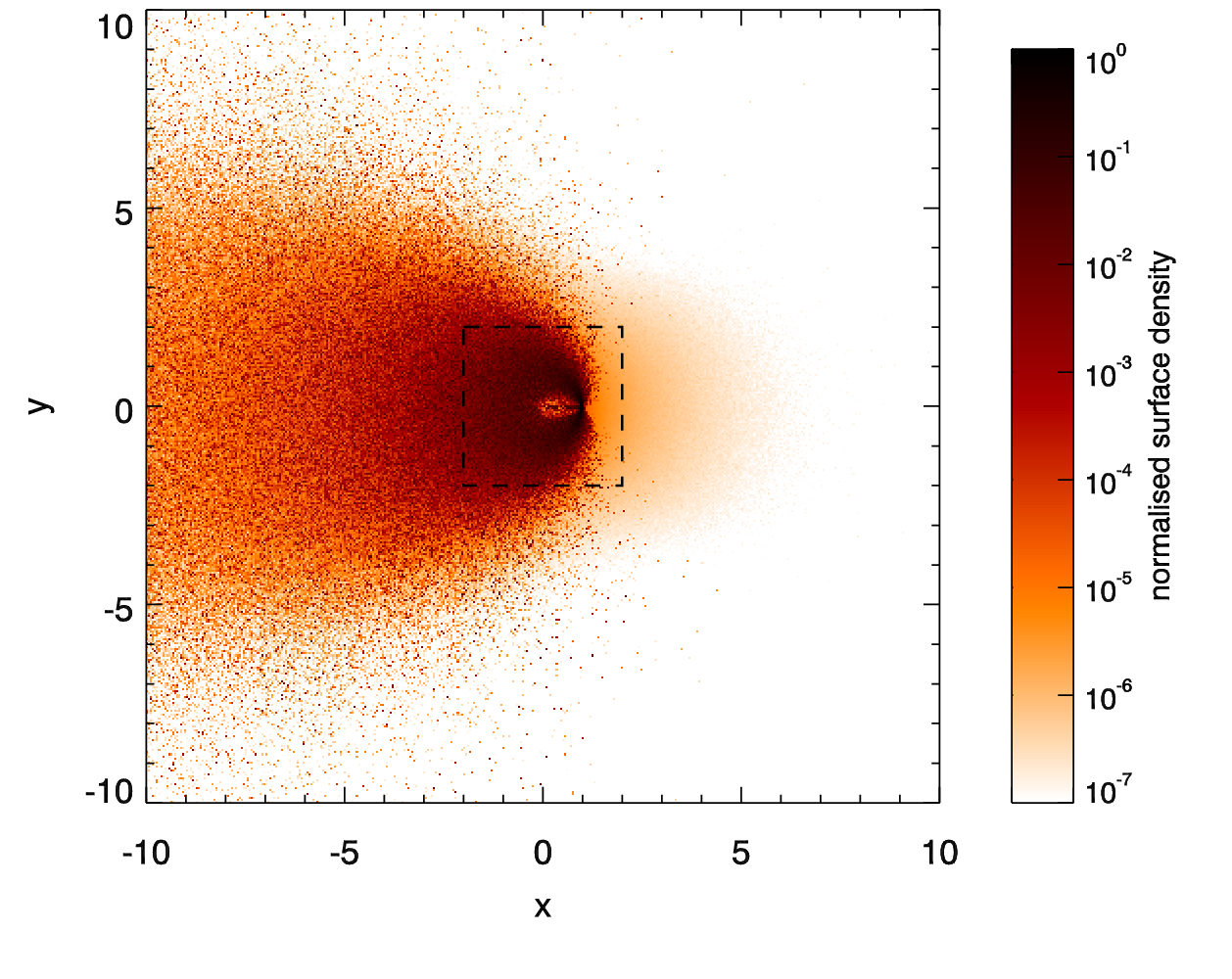}
 \caption{The distribution of small dust grains for a disc produced by an impact with $\sigma_v/v_{\rm k}=0.3$.  \emph{Top}: unbound grains, \emph{bottom}: bound grains.  In both images $\beta$ is uniformly distributed between 0 and 0.99.  Note the zoomed out scale in comparison with previous figures, the dashed box indicates the region covered by earlier images.  The corresponding distribution of large grains can be found in Fig.~\ref{fig:discimages}(c).}
 \label{fig:blowout}
\end{figure}

The overall effect is that dust grains with $\beta \ga 0.5$ are generally not present in stable, bound orbits, but rather are produced in collisions involving larger dust grains and then blown out.  As a result the distribution of these small dust grains, which are most visible in observations at shorter wavelengths, is strongly dependent on the collision rate and we can think of Fig.~\ref{fig:collmap} as being a map of their production sites.  Since the collision-point strongly dominates the collision rate it follows that it will also be the dominant site for the production of blow-out grains.

In Fig.~\ref{fig:blowout} we show the distribution of small dust grains produced for a disc with $\sigma_v/v_{\rm k}=0.3$, and a uniform distribution of $\beta$ between 0 and 0.99.  In the upper image we show the dust grains that are unbound leaving the system on hyperbolic orbits.  This image is constructed by integrating the unbound grains forward from their starting locations for 10 orbital periods of the progenitor.  The dominance of the collision-point in the production of the grains results in the appearance of a broad `jet' like structure emanating from the collision-point in the direction of orbital motion.  There is also a secondary jet emanating from the anti-collision line, since this is the location with the second highest collision rate, however this is much fainter than the primary jet.  These broad jets roughly span the region between a parabolic orbit with pericentre at (1,0) or (-1,0), the orbits expect for a grain with $\beta=0.5$, and a straight line at $x=\pm1$, the path expected for a grain with $\beta=1.0$.

The lower panel of Fig.~\ref{fig:blowout} shows the dust grains that remain bound for the same distribution of $\beta$ as the upper panel.  For the bound grains the image is constructed by randomising the mean anomaly of the dust grains.  These grains have a distribution that is symmetric about the line of nodes (the $x$-axis), since they trace complete eccentric orbits.  The distribution of the bound grains is also a lot more concentrated toward the inner region in which the grains originate than the unbound grains as would be expected, however they still show a significantly longer tail extending out in the negative $x$ direction away from the collision-point due to the increased eccentricity of the dust grains.  Note that there is considerable uncertainty in the relative normalisation between the bound and unbound dust grains owing to uncertainties in the lifetime of the bound grains.  The distribution of large grains, which are the parents of the small grains found in Fig.~\ref{fig:blowout} can be found in Fig.~\ref{fig:discimages}(c).

The uniform distribution of $\beta$ used in Fig.~\ref{fig:blowout} is almost certainly not realistic, however the true distribution of $\beta$ is heavily dependent on the uncertain outcomes of high velocity collisions between small dust grains at the collision-point.  We can use the results of the uniform distribution to inform us as to the behaviour expected if we change the distribution.  If we weight the distribution toward low values of $\beta$ the jet of unbound grains from the collision-point will be denser toward the parabola centred at (1,0), while the bound grains will be even more heavily concentrated toward the parent disc.  On the other hand if we weight the distribution toward high values of $\beta$ the jet of unbound grains will be denser toward the $x=1$ line while the tail of bound grains at large negative $x$ will become more prominent.  If we allow values of $\beta$ higher than 1 this results in anomalous hyberbolae that curve away from the star at their point of origin \citep{krivov2006}, this would broaden the jet to the right of the $x=1$ line.

While the relative normalisation between the bound and unbound dust grains is rather uncertain we can still make inferences about what would be seen in observations.  By comparing the upper and lower panels of Fig.~\ref{fig:blowout} we can see that the density of the bound grains falls by around four orders of magnitude along the negative $x$-axis between the location of the progenitor ring and the edge of the image.  In contrast the density of the jet of unbound grains falls off much more slowly.  As such we can expect that even if the bound grains dominate close to the parent ring there will be a distance at which the unbound grains become dominant.  In addition both the tail of bound grains and the jet of unbound grains lie mostly in the upper left quadrant, so whatever the normalisation between the two populations the halo will be concentrated in this region.

\subsection{CO gas}
\label{sec:CO}

Debris discs are generally thought of as gas free systems, however a number of debris discs have now been observed to possess carbon monoxide (CO) gas, such as Beta Pictoris \citep{roberge2000, troutman2011}, 49 Ceti \citep{roberge2013} and HD21997 \citep{kospal2013}.  In some cases this may be remnants of the primordial gas from the protoplanetary stage.  For Beta Pictoris in particular however this does not seem to be the case as there are stringent upper limits on the mass of hydrogen present \citep{freudling1995,lecavelier2001} and it is instead suggested that the CO is secondary gas produced by planetesimals containing CO ice.  Here we are considering that the parent bodies contain $\sim$1-10 per cent by mass of CO ice, as for solar system comets (though the exact value is unimportant for our model), and that this can be released during collisions.

CO is photodissociated on fairly short timescales by interstellar ultraviolet radiation.  \citet*{visser2009}, for example, find 120-170 years for the lifetime of CO molecules in the absence of shielding, depending on the radiation field used.  At the large orbital distances where we are likely to find bodies containing substantial quantities of CO ice this lifetime is shorter than the orbital timescale and so the distribution of CO will be strongly influenced by the distribution of the production regions.

As a result of photodissociation the distribution of CO will trace the distribution of its production sites in the same way as the small, blow-out, grains.  However here instead of being blown out from the original orbits of the parent particles the CO gas will continue to follow the same orbit since it is minimally affected by radiation pressure.

\begin{figure}
 \includegraphics[width=\columnwidth]{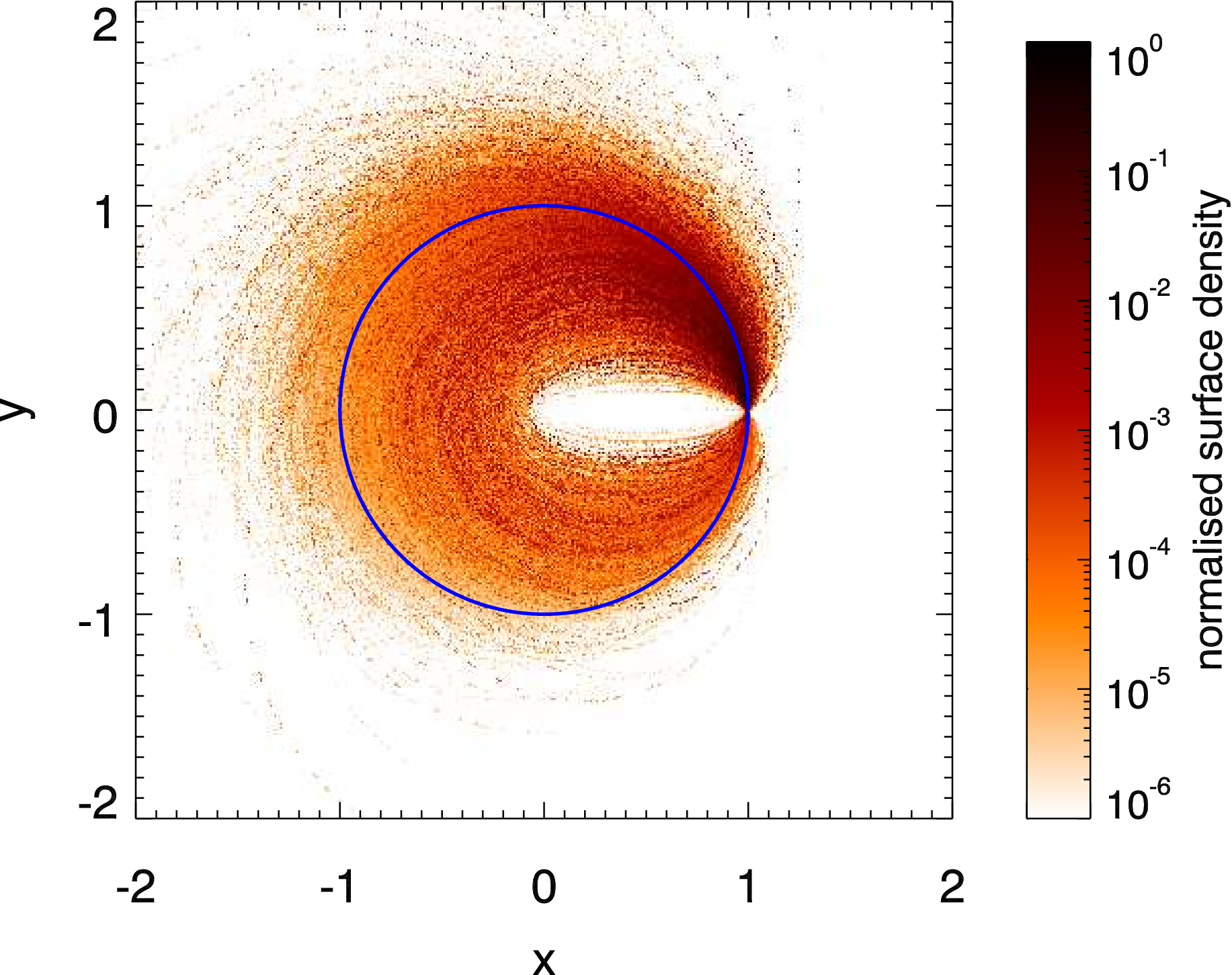}
 \caption{The distribution of CO for a disc produced by an impact with $\sigma_v/v_{\rm k}=0.3$ and with a CO decay time of 1/10 of the progenitor orbital period}
 \label{fig:COdecay}
\end{figure}

In Fig.~\ref{fig:COdecay} we implement the dissociation as a simple exponential decay of the CO density as it moves away from the production site (with the timescale here being 1/10 of the progenitor orbital period).  As for the blow out grains of Section~\ref{sec:blowout} we may think of the collision map of Fig.~\ref{fig:collmap} as a map of the production sites.  This simple exponential decay assumes that CO gas is released promptly after a destructive collision and that the lifetime of the CO is constant.

As with the small dust grains of Section~\ref{sec:blowout} the effect of the CO being concentrated more strongly at the locations it is produced is to enhance the asymmetry of the disc above that seen in the larger debris fragments traced by Figs.~\ref{fig:discimages} and \ref{fig:eccdiscimages}.  The constant decay time also has the effect that the CO reaches further around the disc at smaller orbital distances, since the orbital period here is shorter.  As a result the CO is concentrated slightly interior to the orbit of the progenitor.  Interestingly CO reaching to smaller orbital distances than dust is found in HD21997 by \citet{kospal2013}, where the inner edge of the CO disc is at $<$26~AU, while the inner edge of the dust disc is at $\sim$55~AU.  As the CO in the HD21997 disc is optically thick however the decay times are likely too long for a similar effect to that described here to cause this difference in inner edge distances.

In reality there are effects that may not be captured in the simple assumptions above.  Although at smaller orbital distances the orbital period is shorter and so the CO reaches further around the orbit in Fig.~\ref{fig:COdecay} the temperatures to which the debris is subjected will also be higher.  This can have the effect that any CO contained in large debris fragments on these orbits outgases and is lost very early in the disc lifetime such that at later times CO is absent from short orbits.  In addition the photodissociation timescale of 120-170 years is that in the absence of shielding, but self-shielding within the CO gas can be important even at fairly low column densities, significantly increasing the photodissociation lifetime \citep[e.g.][]{visser2009}.  As the gas is then photodissociated and the density falls, so the self-shielding decreases and the photodissociation will rate accelerate.  It is also possible that in addition to an initial burst of CO release after destructive collisions there may be a slower release of CO from debris fragments that did not undergo catastrophic collisions, but rather had cratering events that exposed new surfaces to space.  Such continued production would also have the effect of apparently prolonging the CO lifetime.  Despite these potential complications however, the overall picture of CO being produced largely at sites of high collision rate and subsequently decaying as it moves around the orbit is unchanged.

\section{Beta Pictoris}
\label{sec:betapic}

Having set out the theoretical basis for debris discs produced by giant impacts at large distances from their host star it is highly desirable to have a comparison with a real system.  We are fortunate to have an appropriate system with which to compare in the young (~12Myr, \citealt{zuckerman2001}), nearby (19.44pc, \emph{Hipparcos}), A-star Beta Pictoris.

\begin{figure}
 \includegraphics[width=\columnwidth]{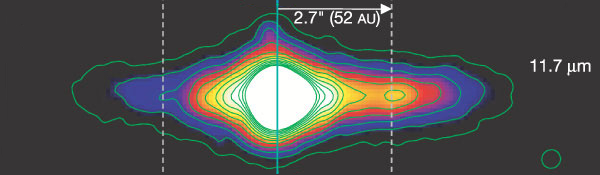}
 \caption{11.7$\mu$m image of Beta Pictoris (Fig.~1d of \citealt{telesco2005}).  The image has been rotated 58$^{\circ}$ counter-clockwise with respect to the sky position-angle; Northeast is to the left and Southwest is to the right.  The small circle at bottom right indicates the T-ReCS point-source FWHM. The vertical solid line is at the star (centre) and the vertical dotted lines are at 52 AU from the star.  The contours (in units of 0.01 mJy pixel$^{-1}$) are 10, 17, 26, 36, 45, 56, 68, 80, 93, 108, 123, 140.  The brightest (inner) colours correspond to the highest-numbered contour levels.  Note that the star is heavily saturated in this image.  There is a strong left-right asymmetry, peaked at a separation of 52~AU from the star.}
 \label{fig:telesco2005}
\end{figure}

Beta Pictoris is a very well studied system and several complex structures and asymmetries have been discovered in its edge-on disc.  At large distances from the star the extent of the disc has been found to be asymmetric, reaching 1450AU in the Southwest extension but 1835AU in the Northeast extension \citep[][]{larwood2001}.  Closer to the star the disc is observed to be warped \citep[e.g.][]{golimowski2006}.  Most pertinent to our models here are the observations of \citet{telesco2005} which revealed a large brightness asymmetry in the mid-infrared at a projected separation of about 50AU from the star (see Fig.~\ref{fig:telesco2005}).  New observations in the sub-mm with the Atacama Large Millimetre Array (\textsc{alma}) by Dent et al. subm. also reveal a similar asymmetry in the sub-mm continuum and in emission from CO gas.

\citet{telesco2005} suggested two alternative hypotheses for the origin of the prominent asymmetry in the mid-infrared, production from larger planetesimals trapped in resonance with a planet (not the planet Beta Pictoris b imaged by \citealt{lagrange2009, lagrange2010}, rather one farther from the star), as studied in the context of Vega by \citet{wyatt2003}, or the very recent ($\sim$50~yr ago) break up of a $\sim$100~km body.  The main issue with the suggestion of collisional break-up of a large asteroid/comet is the very short timescale and thus the low likelihood of catching the system in its current state.  Here we suggest that a larger impact event involving a planetary scale body avoids this problem since the asymmetry in the debris disc produced persists for much longer.  That is, the clump could be the aftermath of a collision that occured up to $\sim$1Myr ago.

As we saw in Figs.~\ref{fig:discimages}, \ref{fig:linecuts}, \ref{fig:discratios} and \ref{fig:eccdiscimages} if we look at a debris disc generated by a giant impact edge on, with the collision-point to one side of the star the result is a large brightness asymmetry between the two sides of the disc.  In the context of Beta Pictoris as the Southwest side of the disc is brighter in the mid-infrared the collision-point would be on the Southwest side.  Although the projected separation of the clump from the star is 52AU this does not mean that the impact must have occured at 52AU from the star, rather this sets a minimum orbital distance of 52AU since if the impact occured further out but not at the ansa of its orbit the collision-point could also appear at 52AU.  This is one of the key features of CO observations that can be obtained with \textsc{alma}, as CO emission is line emission and it is thus possible to obtain velocity information.  Having line of sight velocity information gives access to the structure along the line of sight and makes it possible to, for example, determine the true orbital distance of the asymmetry.  The \textsc{alma} observations of Dent et al. reveal, amongst other things, that the true orbital distance of the CO is $\sim$85~AU.

Considering the orbital distance of $\sim$85~AU from Dent et al., the orbital period at this distance from Beta Pictoris is 590 years. If the mean lifetime of CO is $\sim$120-170 years this implies that CO produced at the collision-point will have gone through at least one decay time, and thus decayed by at least 63 per cent, over the course of moving from the Southwest side of the disc to the Northeast side.  This does not automatically translate into a brightness asymmetry of the same magnitude, since the CO will continue to decay as it moves from the Northeast side back to the Southwest side.   Nonetheless we can see how this leads to the expectation that the asymmetry in CO should be larger than that in the larger parent grains that dominate the emission at longer wavelengths (particularly in the sub-mm).

In addition to considerations of CO, mid-infrared and mm-emission a giant impact model can also produce consistency with the asymmetry at very large scales observed by \citet{larwood2001} in optical scattered light.  This asymmetry is in the opposite sense to the asymmetry at longer wavelengths, with the Northeast side of the disc observed to extend around 25 per cent further than the Southwest side.  It is at these short wavelengths that the jet of blow-out grains we described in Section~\ref{sec:blowout} will be most visible, and if this jet is oriented in a Northeasterly direction it would result in the Northeast side of the disc halo being brighter, and thus observed to larger distances.  This jet orientation requires that the giant impact occurs at a larger distance from the star than 52AU, such that the jet is at an angle to the line of sight, consistent with the $\sim$85~AU distance revealed by Dent et al.  In addition, since \citet{olofsson2001} showed that the sense of orbital motion in the disc is toward us in the Southwest, this jet orientation requires that the collision-point is closer to us along the line of sight than the star.

In our giant impact model the collision-point is stationary on orbital timescales, only moving on much longer timescales as a result of precession.  As such if the clump in the Beta Pictoris disc is indeed the result of giant impact it should be stationary.  This is a difference with a resonance model similar to that of \citet{wyatt2003}, and indeed to the original suggestion of \citet{telesco2005}, as in both of those models the clump will move.  In the case of the original \citet{telesco2005} model it would move with the Keplerian velocity at its orbital distance, while in the resonance model it would move on the faster orbital timescale of the interior planet driving the resonance.  \citet{li2012} re-imaged the mid-infrared clump and found tentative evidence for motion.  If confirmed this would rule out a giant impact model, but at present the evidence is inconclusive.  Nonetheless further observations over decade timescales will enable us to definitively either detect or rule out motion of the clump, helping us to determine its origin.

We have not here attempted to conduct a full statistical modelling of Beta Pictoris.  Rather we have shown qualitatively that debris released by a giant impact is broadly capable of reproducing the large scale asymmetries observed in the Beta Pictoris disc in the mid-infrared by \citet{telesco2005}, in the CO/sub-mm by Dent et al. subm., and in scattered light by \citet{larwood2001}.

\section{Conclusions}
\label{conclusion}

Planetary scale, giant, impacts have occurred in the outer reaches of our own solar system, and it is not unreasonable to expect similar impacts to occur in the outer reaches of other planetary systems.  These large impacts release substantial quantities of debris that will go into orbit around the host star and produce an, initially, high asymmetric disc.

The behaviour of this giant impact debris is governed by a set of equations that we described in Section~\ref{orbeq}.  The key result of these equations is the existence of the collision-point, a fixed point in space at the location at which the originating giant impact takes place and through which all of the debris must pass.  The collision-point is of paramount importance in determining the appearance and evolution of the debris disc, and is what produces the strong asymmetry in these discs.

We have studied the morphologies of debris discs generated by giant impacts and shown how the character of the disc varies depending on the mass of the progenitor body (and its orbital distance).  At the same orbital distance a debris disc produced by an impact involving a more massive progenitor is broader both radially on the side of the disc opposite the collision-point, and vertically.  In addition a debris disc produced by an impact involving a more massive progenitor is more strongly dominated by the collision-point due to the effect that while all material must pass through the collision-point, elsewhere the material is more dispersed than in a disc originating from a less massive progenitor.

The lifetime of the asymmetry due to the collision-point is determined by precession caused by other bodies in the system, a typical lifetime however is a few thousand orbits.  The long orbital periods at large distances from the parent star translates this into timescales of $\sim$1Myr.

The eccentricity of the orbit of the progenitor body interacts with the asymmetry present due to the collision-point.  Depending on the location of the impact around the orbit this can enhance or reduce the asymmetries in the brightness and radial extent of the disc at the collision-point and opposite it.  In general the complexity of the disc structure is increased for eccentric progenitors.

We have also studied the collisional evolution of the asymmetric discs generated by giant impacts.  In the second meaning of its name the collision-point dominates the collisional evolution of the disc.  As such material whose distribution depends strongly on the location at which it is produced, such as small dust grains that are strongly influenced by radiation pressure, and CO, demonstrate even stronger asymmetries, focussed on the collision-point.  In addition the highly asymmetric debris disc produced by a giant impact evolves much faster collisionally than an equivalent axisymmetric disc.  Nonetheless for a disc that is detectable initially the expected detectable lifetime is typically at least as long as the lifetime of the asymmetry.  As such it is reasonable to expect that we can observe asymmetric discs resulting from impacts between Moon-size and larger bodies at large distances ($\sim$50AU) from their host star.

We applied our model of giant impact debris discs to the debris disc around the star Beta Pictoris and demonstrated that it is capable of broadly reproducing the asymmetry observed in the mid-infrared by \citet{telesco2005}, in CO/sub-mm by Dent et al. subm., and in scattered light by \citet{larwood2001}.  A more detailed analysis would be required however to determine if this is the best model for the disc, and if so what the system parameters are.

If debris discs generated by giant impacts are found in the outer reaches of extrasolar planetary systems, for example if this is shown to be the best model for Beta Pictoris, this has important implications for planet formation models.  The occurrence of giant impacts could imply that rocky/icy bodies routinely grow to large, planetary, sizes at substantial distances from their host star.

\section*{Acknowledgements}
\label{acknowledgements}
APJ is supported by an STFC postgraduate studentship, AB acknowledges the support of the ANR-2010 BLAN-0505-01 (EXOZODI),  MCW acknowledges the support of the European Union through ERC grant number 279973 and DV acknowledges the support of the European Union through ERC grant number 320964.  The authors thank Cathie Clarke for valuable discussions in the course of preparing this manuscript, and the referee for thought provoking comments.

{\footnotesize
\bibliographystyle{myMNRAS}
\bibliography{largeradiiGIs}
}

\appendix
\onecolumn
\section{Equations for arbitrary initial $I$, $\Omega$ and $\omega$}
\label{geneq}

As described in Section~\ref{orbeq} our equations can be applied to situations in which non-zero initial values of $I$, $\Omega$ and $\omega$ are desired by application of rotations to the resulting distributions.  As this is quite a common situation to which the equations would be applied we give below the appropriate transformation of $\theta$ and $\phi$, and also equations for $I'$, $\Omega'$ and $\omega'$ that incorporate the rotations directly, which might in particular be convenient for some programming applications.

Moving into a frame in which $I, \Omega, \omega \ne 0$ introduces a new Cartesian basis, which would now be the preferred basis for the calculations.  This can be related easily to our original definition of $\theta$ and $\phi$ by defining a new $\theta_1$ and $\phi_1$, which are spherical polar angles defined relative to the Cartesian basis of the $I, \Omega, \omega \ne 0$ frame.  We can then relate $\theta, \phi$ and $\theta_1, \phi_1$ by
\begin{align}
\cos(\theta) & = C_{\theta_1}C_I - S_{\theta_1}S_I S_{\beta}, \\ \notag
\tan(\phi)   & = \frac{S_{\theta_1}(S_{\beta}C_I C_{\omega} - C_{\beta}S_{\omega}) + C_{\theta_1}S_I C_{\omega}}
                        {S_{\theta_1}(S_{\beta}C_I S_{\omega} + C_{\beta}C_{\omega}) + C_{\theta_1}S_I S_{\omega}}.
\end{align}
where we define $\alpha=\omega+f$ and $\beta=\phi_1-\Omega$.

For $a'$, $e'$ and $f'$ we can then simply use these definitions of $\theta$ and $\phi$ in terms of $\theta_1$ and $\phi_1$ in the equations of Section~\ref{orbeq} and obtain the correct results while describing the kick velocity with respect to the Cartesian basis of the $I, \Omega, \omega \ne 0$ frame.  Although they will be calculated correctly $I'$, $\Omega'$ and $\omega'$ will however still be defined relative to the orbit reference frame rather than the $I, \Omega, \omega \ne 0$ frame.  The values in the $I, \Omega, \omega \ne 0$ frame can of course be determined by rotation of the resulting distributions, but for convenience we give equations for $I'$, $\Omega'$ and $\omega'$ in terms of $\theta_1, \phi_1$ that incorporate the rotations directly as

\begin{equation}
\label{Igen}
\cos(I') = \left[ C_I + \frac{\sqrt{1-e^2}}{1+eC_f} \left( \frac{\Delta v}{v_{\rm k}} \right) S_{\theta_1} \left(C_{\alpha}S_{\beta}-S_{\alpha}C_{\beta}C_I\right) \right]
                    \left(\frac{h'^2}{h^2}\right)^{-\frac{1}{2}},
\end{equation}
\begin{equation}
\label{Omegagen}
\tan(\Omega')=\frac{S_{\Omega}S_I + \frac{\sqrt{1-e^2}}{1+eC_f} \left( \frac{\Delta v}{v_{\rm k}} \right) \bigl[ C_{\theta_1}\left(S_{\Omega}C_{\alpha}+C_{\Omega}S_{\alpha}C_I\right) - S_{\theta_1}C_{\phi_1}S_I S_{\alpha} \bigr]}
                            {C_{\Omega}S_I + \frac{\sqrt{1-e^2}}{1+eC_f} \left( \frac{\Delta v}{v_{\rm k}} \right) \bigl[ C_{\theta_1}\left(C_{\Omega}C_{\alpha}-S_{\Omega}S_{\alpha}C_I\right)  - S_{\theta_1}C_{\phi_1}S_I S_{\alpha} \bigr]},
\end{equation}
and
\begin{align}
\label{omegafprimgen}
\sin (\omega' + f') & = \frac{S_{\alpha}S_I}{S_{I'}},\\ \notag
\cos (\omega' + f') & = \frac{1}{C_{\Omega'}}\left(C_{\Omega}C_{\alpha} - S_{\Omega}S_{\alpha}C_I
                                           + S_{\Omega'}S_{\alpha}\frac{S_I C_{I'}}{S_{I'}} \right).
\end{align}
Note that these equations do not include any mechanism preventing the inclination being negative.

\section{Comparison to the Gauss planetary equations}
\label{gausscomp}

For simplicity we will only give the example of a purely radial kick ($\theta=\pi/2$, $\phi=f$) here, but the same logic applies to kicks in other directions.  Let us first consider the change in the semi-major axis.  For a purely radial kick Eq.~\ref{aaprimecc} simplifies to
\begin{equation}
 \frac{a}{a'}=1-\left(\frac{\Delta v}{v_{\rm k}}\right)^2 - \frac{2}{\sqrt{1-e^2}}\left(\frac{\Delta v}{v_{\rm k}}\right)e S_f.
 \label{aaprimrad}
\end{equation}
Now let us consider only a small change in the semi-major axis, such that $a'=a+\Delta a$ and  $\Delta v/v_{\rm k}$ is also small, allowing us to neglect the term in $\Delta v^2$.  The above then becomes
\begin{equation}
 \frac{\Delta a}{a'} = \frac{2}{\sqrt{1-e^2}}\left(\frac{\Delta v}{v_{\rm k}}\right)e S_f.
 \label{deltaaprim}
\end{equation}
We can also think of $\Delta v$ as a change in the specific momentum of the particle, and allow this to be introduced over a small time $\Delta t$ by a force $R$, rather than exactly instantaneously.  So we can write $\Delta v$ as $R \Delta t$, and Eq.~\ref{deltaaprim} becomes
\begin{equation}
 \frac{1}{a'}\frac{\Delta a}{\Delta t} = \frac{2 a^{1/2}}{\sqrt{G(M+m)(1-e^2)}}e R S_f,
\end{equation}
where we have also substituted $v_{\rm k}=\sqrt{G(M+m)/a}$.  If we now take the limit as $\Delta a$ and $\Delta t$ become infinitesimally small, and note that $a' = a+da \approx a$, this becomes
\begin{equation}
 \frac{da}{dt} = \frac{2 a^{3/2}}{\sqrt{G(M+m)(1-e^2)}}e R S_f,
\end{equation}
which is simply Gauss' equation for the rate of change of the semi-major axis under the action of a radial force, as expressed by e.g. \citet{burns1976} or \citet{murray1999}.

The eccentricity can be treated in exactly the same way.  Firstly notice that in the case of a purely radial kick $h'^2/h^2=1$ and so we can write Eq.~\ref{ecceprim} as
\begin{equation}
 e'^2 = 1 - (1-e^2) \left(1-\left(\frac{\Delta v}{v_{\rm k}}\right)^2 - \frac{2}{\sqrt{1-e^2}}\left(\frac{\Delta v}{v_{\rm k}}\right)e S_f\right).
\end{equation}
Following exactly the same procedure of considering a small change $e'=e+\Delta e$, with $\Delta v = R \Delta t$ and neglecting non-linear terms in small quantities we obtain
\begin{equation}
 \frac{\Delta e}{\Delta t} = \sqrt{\frac{a(1-e^2)}{G(M+m)}} R S_f,
\end{equation}
which again taking the limit as $\Delta e$ and $\Delta t$ become infinitesimally small becomes
\begin{equation}
 \frac{de}{dt} = \sqrt{\frac{a(1-e^2)}{G(M+m)}} R S_f,
\end{equation}
Gauss' equation for the rate of change of the eccentricity.

We can thus see that in the case of a small change, which is the assumption under which the Gauss planetary equations are derived, we can relate the kick equations directly to the planetary equations.  The differences between the kick formalism and the planetary equations become important when we consider large changes for which the non-linear terms are not negligible.  Perhaps the most obvious example of a simple difference and its importance is that if we consider a 3-dimensional force ${\bf F}=R{\bf \hat{e}}_r + T {\bf\hat{e}}_T + N{\bf \hat{e}}_N$, with ${\bf \hat{e}}_N$ the basis vector normal to the orbit, and ${\bf \hat{e}}_T$ in the plane of the orbit perpendicular to the radial, the full planetary equation for the semi-major axis is
\begin{equation}
 \frac{da}{dt} = \frac{2 a^{3/2}}{\sqrt{G(M+m)(1-e^2)}}\left[e R S_f + T(1+e C_f)\right].
\end{equation}
The lack of a ${\bf \hat{e}}_N$ term illustrates that a force perpendicular to the plane of the orbit cannot change the semi-major axis, whereas in the kick formalism a perpendicular kick can change the semi-major axis (though the term linear in $\Delta v$ is zero).  We can understand this by considering that to induce a large velocity kick in the vertical we need to have a large force act (or do so for a longer time) and the \emph{direction of that force must be constant}.  In the formalism of the planetary equations the definition of ${\bf \hat{e}}_N$ means that if we start applying a force in the ${\bf \hat{e}}_N$ direction the plane of the orbit will begin to rotate, and thus \emph{so will the definition of ${\bf \hat{e}}_N$}.  So if we want to induce a larger velocity kick in the vertical the definition of the direction of the force will change over the time of application and gain an apparently radial component.

\label{lastpage}
\end{document}